\documentclass[fleqn,usenatbib,useAMS]{mnras}

\usepackage{graphicx}	
\usepackage{amsmath}	
\usepackage{amssymb}	
\usepackage{multicol}        
\usepackage{pdflscape}	
\usepackage{float}
\usepackage[export]{adjustbox}

\usepackage{wasysym}  
\usepackage{enumitem}  
\setlist[enumerate]{label={\arabic*.}}

\usepackage{hyperref}
\newcommand{\cloudy}{{\footnotesize CLOUDY}\xspace}
\newcommand{\beagle}{{\footnotesize BEAGLE}\xspace}

\newcommand{\mappings}{{\footnotesize MAPPINGS\,V}\xspace}

\newcommand{\bpass}{{\footnotesize BPASS}\xspace}
\newcommand{\sevn}{{\footnotesize SEVN}\xspace}
\newcommand{\parsec}{{\footnotesize PARSEC}\xspace}
\newcommand{\galaxev}{{\footnotesize GALAXEV}\xspace}

\newcommand{\CB}{{\footnotesize C\&B}\xspace}

\newcommand{\JWST}{\textit{JWST}\xspace}
\newcommand{\GLS}{{\footnotesize GALSEVN}\xspace}
\newcommand{\wmbasic}{{\footnotesize WM-BASIC}\xspace}
\newcommand{\powr}{{\footnotesize PoWR}\xspace}
\newcommand{\spectres}{{\footnotesize SpectRes}\xspace}
\newcommand{\stars}{{\footnotesize STARS}\xspace}
\newcommand{\sphinx}{{\footnotesize SPHINX}\xspace}
\newcommand{\ares}{{\footnotesize ARES}\xspace}
\newcommand{\simpl}{{\footnotesize SIMPL}\xspace}

\newcommand{\Msun}{\hbox{M$_{\rm{\odot}}$}\xspace}

\newcommand{\tprime}{\hbox{$t^\prime$}\xspace}
\newcommand{\hii}{\hbox{H\,{\sc ii}}\xspace}

\newcommand{\zsun}{\hbox{${Z}_{\odot}$}\xspace}
\newcommand{\zpsun}{\hbox{${Z}_{\odot}^0$}\xspace}

\newcommand{\zism}{\hbox{$Z_\mathrm{ISM}$}\xspace}

\newcommand{\nh}{\hbox{$n_{\mathrm{H}}$}\xspace}

\newcommand{\xid}{\hbox{$\xi_{\rm{d}}$}\xspace}

\newcommand{\CO}{\hbox{C/O}\xspace}
\newcommand{\OH}{\hbox{O/H}\xspace}

\newcommand{\COsol}{\hbox{(C/O)$_\odot$}\xspace}

\newcommand{\logoh}{\hbox{12 + log(\OH)}\xspace}

\newcommand{\Uav}{\hbox{$\langle U\rangle$}\xspace}

\newcommand{\Te}{\hbox{$T_{\mathrm{e}}$}\xspace}
\newcommand{\logU}{\hbox{$\log \Uav$}\xspace}
\newcommand{\Nhdot}{\hbox{$\dot{N}_\mathrm{H}$}\xspace}
\newcommand{\Nheidot}{\hbox{$\dot{N}_\mathrm{He\,I}$}\xspace}
\newcommand{\Nheiidot}{\hbox{$\dot{N}_\mathrm{He\,II}$}\xspace}
\newcommand{\fbin}{\hbox{$f_{\mathrm{bin}}$}\xspace}
\newcommand{\mchar}{\hbox{$m_{\mathrm{char}}$}\xspace}
\newcommand{\Muv}{\hbox{$M_{\mathrm{UV}}$}\xspace}
\newcommand{\Mdotacc}{\hbox{$\dot{M}_{\mathrm{acc}}$}\xspace}
\newcommand{\Mdotedd}{\hbox{$\dot{M}_{\mathrm{Edd}}$}\xspace}
\newcommand{\Ledd}{\hbox{$L_{\mathrm{Edd}}$}\xspace}
\newcommand{\fedd}{\hbox{$f_{\mathrm{Edd}}$}\xspace}
\newcommand{\Lacc}{\hbox{$L_{\mathrm{acc}}$}\xspace}
\newcommand{\Mc}{\hbox{$M_{\mathrm{c}}$}\xspace}
\newcommand{\rmax}{\hbox{$R_{\mathrm{max}}$}\xspace}
\newcommand{\Lx}{\hbox{$L_{\mathrm{X}}$}\xspace}
\newcommand{\fa}{\hbox{$f_{\mathrm{A}}$}\xspace}
\newcommand{\ea}{\hbox{$\epsilon_{\mathrm{A}}$}\xspace}
\newcommand{\Ltotx}{\hbox{$L_{\mathrm{X}}^\mathrm{tot}$}\xspace}
\newcommand{\Vs}{\hbox{$V_{\mathrm{s}}$}\xspace}
\newcommand{\rsn}{\hbox{$r_{\mathrm{SN}}$}\xspace}
\newcommand{\Mares}{\hbox{$M_{\mathrm{ARES}}$}\xspace}
\newcommand{\effsn}{\hbox{$\varepsilon_{\mathrm{SN}}$}\xspace}
\newcommand{\effw}{\hbox{$\varepsilon_{\mathrm{W}}$}\xspace}

\newcommand{\civ}{\hbox{C\,\textsc{iv}\,$\lambda1549$}\xspace}

\newcommand{\heii}{\hbox{He\,\textsc{ii}\,$\lambda1640$}\xspace}

\newcommand{\oiii}{\hbox{O\,\textsc{iii}]\,$\lambda1664$}\xspace}

\newcommand{\ciiid}{\hbox{[C\,\textsc{iii}]$\lambda1907$+C\,\textsc{iii}]$\lambda1909$}\xspace}
\newcommand{\ciii}{\hbox{C\,\textsc{iii}]\,$\lambda1908$}\xspace}

\newcommand{\nevopt}{\hbox{[Ne\,\textsc{v}]\,$\lambda3426$}\xspace}

\newcommand{\oiiopt}{\hbox{[O\,\textsc{ii}]$\lambda3727$}\xspace}

\newcommand{\heiiopt}{\hbox{He\,\textsc{ii}\,$\lambda4686$}\xspace}
\newcommand{\hb}{\hbox{H$\beta$}\xspace}
\newcommand{\oiiiopt}{\hbox{[O\,\textsc{iii}]\,$\lambda5007$}\xspace}

\newcommand{\oiopt}{\hbox{[O\,\textsc{i}]\,$\lambda6300$}\xspace}
\newcommand{\ha}{\hbox{H$\alpha$}\xspace}
\newcommand{\niiopt}{\hbox{[N\,\textsc{ii}]\,$\lambda6584$}\xspace}

\newcommand{\oiiioptc}{\hbox{[O\,\textsc{iii}]\,$\lambda4363$}\xspace}

\newcommand{\lhei}{\hbox{He\,\textsc{i}}\xspace}

\newcommand{\loiii}{\hbox{O\,\textsc{iii}]}\xspace}

\newcommand{\lciii}{\hbox{C\,\textsc{iii}]}\xspace}
\newcommand{\lciv}{\hbox{C\,\textsc{iv}}\xspace}

\newcommand{\loiiopt}{\hbox{[O\,\textsc{ii}]}\xspace}

\newcommand{\lheii}{\hbox{He\,\textsc{ii}}\xspace}
\newcommand{\lniiopt}{\hbox{[N\,\textsc{ii}]}\xspace}

\usepackage[T1]{fontenc}
\usepackage{ae,aecompl}

\usepackage{newtxtext,newtxmath}

\usepackage{xspace}
\usepackage[dvipsnames]{xcolor}
\usepackage{orcidlink}

\title[Nebular emission from young stellar populations]{Nebular emission from young stellar populations including binary stars}

\author[M. Lecroq et al.]
{Marie~Lecroq$^{\orcidlink{0009-0008-2198-9651}}$,$^1$\thanks{E-mail: lecroq@iap.fr} 
Stéphane~Charlot$^{\orcidlink{0000-0003-3458-2275}}$,$^1$ 
Alessandro~Bressan$^{\orcidlink{0000-0002-7922-8440}}$,$^{2,3}$ 
Gustavo~Bruzual$^{\orcidlink{0000-0002-6971-5755}}$,$^4$ 
\newauthor  Guglielmo~Costa$^{\orcidlink{0000-0002-6213-6988}}$,$^{7,3,5,6}$    
Giuliano~Iorio$^{\orcidlink{0000-0003-0293-503X}}$,$^{5,6,3}$ 
Mario~Spera$^{\orcidlink{0000-0003-0930-6930}}$,$^{2,8,9}$ 
Michela~Mapelli$^{\orcidlink{0000-0001-8799-2548}}$,$^{10,3,5,6}$ 
\newauthor Yang~Chen$^{\orcidlink{0000-0002-3759-1487}}$,$^{11,12}$ 
Jacopo~Chevallard$^{\orcidlink{0000-0002-7636-0534}}$$^{13}$
and Marco~Dall'Amico$^{\orcidlink{0000-0003-0757-8334}}$$^{5,6}$
\\ 
\\
$^1$Sorbonne Universit\'e, CNRS, UMR 7095, Institut d'Astrophysique de Paris, 98 bis bd Arago, 75014 Paris, France\\
$^2$SISSA, via Bonomea 265, I-34136 Trieste, Italy\\
$^3$INAF, Osservatorio Astronomico di Padova, Vicolo dell’Osservatorio 5, I–35122, Padova, Italy\\
$^4$Instituto de Radioastronom{\'i}a y Astrof{\'i}sica, UNAM, Campus Morelia, Michoacan, M{\'e}xico, C.P. 58089, M{\'e}xico\\
$^5$Dipartimento di Fisica e Astronomia Galileo Galilei, Università di Padova, Vicolo dell’Osservatorio 3, I–35122 Padova, Italy\\
$^6$INFN-Padova, Via Marzolo 8, I–35131 Padova, Italy\\
$^7$Univ Lyon, Univ Lyon1, ENS de Lyon, CNRS, Centre de Recherche Astrophysique de Lyon UMR5574, F-69230 Saint-Genis-Laval, France\\
$^8$INAF, Osservatorio Astronomico di Roma, Via Frascati 33, I-00040, Monteporzio Catone, Italy\\
$^9$INFN-Trieste, Via Valerio 2, I-34127, Trieste, Italy\\
$^{10}$Universit\"at Heidelberg, Zentrum f\"ur Astronomie, Institut f\"ur Theoretische Astrophysik, Albert-Ueberle-Str. 2, D--69120 Heidelberg, Germany \\
$^{11}$Anhui University, Hefei 230601, China\\
$^{12}$National Astronomical Observatories, Chinese Academy of Sciences, Beijing 100101, China\\
$^{13}$Department of Physics, University of Oxford, Denys Wilkinson Building, Keble Road, Oxford OX1 3RH, UK\\
}

\date{Accepted XXX. Received YYY; in original form ZZZ}

\pubyear{2023}

\begin{document}
\label{firstpage}
\pagerange{\pageref{firstpage}--\pageref{lastpage}}
\maketitle

\begin{abstract}

We investigate the nebular emission produced by young stellar populations using the new \GLS model based on the combination of the \sevn population-synthesis code including binary-star processes and the \galaxev code for the spectral evolution of stellar populations. Photoionization calculations performed with the \cloudy code confirm that accounting for binary-star processes strongly influences the predicted emission-line properties of young galaxies. In particular, we find that our model naturally reproduces the strong \heiiopt/\hb ratios commonly observed at high \hb equivalent widths in metal-poor, actively star-forming galaxies, which have proven challenging to reproduce using previous models. Including bursty star formation histories broadens the agreement with observations, while the most extreme \heii equivalent widths can be reproduced by models dominated by massive stars. \GLS also enables us to compute, for the first time in a way physically consistent with stellar emission, the emission from accretion discs of X-ray binaries (XRBs) and radiative shocks driven by stellar winds and supernova explosions. We find that these contributions are unlikely to prominently affect the predicted \heiiopt/\hb ratio, and that previous claims of a significant contribution by XRBs to the luminosities of high-ionization lines are based on models predicting improbably high ratios of X-ray luminosity to star formation rate, inconsistent with the observed average luminosity function of XRBs in nearby galaxies. The results presented here provide a solid basis for a more comprehensive investigation of the physical properties of observed galaxies with \GLS using Bayesian inference.

\end{abstract}




\section{Introduction}
\label{sec:intro}

The \textit{James Webb Space Telescope} (\JWST) has opened a new window on the rest-frame ultraviolet and optical emission of young galaxies at the epoch of reionization. The emission-line signatures of such galaxies should contain valuable information on the sources that potentially contributed to reionize the Universe. Over the past several years, numerous studies have been conducted at various redshifts to characterize and analyse the ultraviolet and optical spectra of metal-poor, actively star-forming galaxies approaching the properties of these reionization-era galaxies \citep[e.g.,][]{stark2014, steidel2016, amorin2017, senchyna2017, nakajima2018, berg2022}. These observations have put a heavy strain on models designed to interpret the nebular emission from young galaxies, as some emission lines requiring highly energetic photons, such as \lheii-recombination lines, occasionally exhibit intensities so high that they cannot be reproduced by standard models fitting lower-ionization lines \citep[e.g.,][]{shirazi2012, nanayakkara2019, stanway2019, plat2019, schaerer2019, olivier2022}. 

Several production mechanisms have been suggested for these missing energetic photons. A most natural one is the hard radiation from hot, nearly pure-He stars produced through processes induced by binary interactions (such as envelope stripping, quasi-homogeneous evolution, and common-envelope ejection), often neglected in stellar population synthesis models \citep[e.g.,][]{eldridge2012, eldridge2017, goetberg2020}. This possibility is all the more compelling in that  $\sim70$ per cent of massive stars in nearby stellar populations are thought to be part of binary systems \citep[e.g.][]{sana2012, moe2017}. Yet, the \bpass models of \citet{stanway2018}, which incorporate the above processes, do not appear to produce enough energetic radiation to account for the observations \citep{stanway2019}. Other proposed origins of the surprisingly strong high-ionization lines in some metal-poor star-forming galaxies include fast radiative shocks, accretion on to compact objects and active galactic nuclei \citep[AGNs; e.g.,][]{izotov2012, nakajima2018, plat2019, schaerer2019, umeda2022, katz2023}. Such components are typically modelled independently of the stellar population and then scaled to reproduce the data under consideration. While insightful, this approach suffers from the weakness of not guaranteeing physical consistency between the different spectral components involved.

In this paper, we explore the emission-line properties of interstellar gas photoionized by young binary-star populations computed using a new spectral-synthesis model, built by combining the \sevn \citep{iorio2022} and \galaxev \citep{bruzual2003} population-synthesis codes, coupled with the \cloudy photoionization code \citep{ferland2017}. This spectral-synthesis model, called \GLS, enables us to compute, for the first time in a physically consistent way, the emission from stars, accretion discs of X-ray binaries (XRBs) and radiative shocks driven by stellar winds and supernova (SN) explosions. We do not consider here the emission from AGNs, which cannot be tightly related to the star formation properties of a galaxy, and whose effects have been studies in detail by \citet{plat2019}. We show how the \GLS model can better account for observations of ultraviolet and optical emission lines in metal-poor star-forming galaxies -- in particular strong \lheii emission -- than previous models of both single and binary stars. When combined with a Bayesian-inference tool such as \beagle \citep{chevallard2016}, the model presented here should provide valuable constraints on physical parameters of distant galaxies from observations of nebular emission.

The paper layout is as follows: in Section~\ref{sec:models}, we present the new \GLS model of spectral evolution of binary-star populations and our procedure to compute the associated nebular emission with \cloudy. In Section~\ref{sec:res}, we describe a reference observational sample and examine the ability of models with different stellar and nebular parameters to reproduce these observations. Then, in Section~\ref{sec:adds}, we consider additional effects, such as emission from XRB accretion discs and radiative shocks, potentially helpful to reproduce the most extreme emission-line properties of galaxies in the sample. We place our findings in the context of other recent studies in Section~\ref{sec:dsc}. Section~\ref{sec:ccl} summarizes our results.

\section{The models}
\label{sec:models}

In this section, we describe the models used to compute the nebular emission from young galaxies. We follow the approach of \citet[][see also \citealt{gutkin2016}]{charlot2001} and express the luminosity per unit frequency $\nu$ emitted at time $t$ by a star-forming galaxy as
\begin{equation}
L_{\nu}(t)=\int_0^t \mathrm{d}\tprime\, \psi(t-\tprime) \, S_{\nu}[\tprime,Z(t-\tprime)] \, T_{\nu}(t,\tprime)\,,
\label{eq:flux_gal}
\end{equation}
where $\psi(t-\tprime)$ is the star formation rate at time $t-\tprime$, $S_\nu[\tprime,Z(t-\tprime)]$ the luminosity produced per unit frequency per unit mass by a single generation of stars of age $\tprime$ and metallicity $Z(t-\tprime)$, and $T_\nu(t,\tprime)$ the transmission function of the interstellar medium (ISM). We compute $S_\nu$ for populations of single and binary stars using a combination of the \sevn \citep{iorio2022} and \galaxev \citep{bruzual2003} population-synthesis codes (Section~\ref{sec:mod_stellpop}) and $T_\nu$ using the \cloudy\ photoionization code \citep[][Section~\ref{sec:mod_photion}]{ferland2017}.

\subsection{Stellar population modelling}
\label{sec:mod_stellpop}

To describe the emission from populations of single and binary stars, we appeal to a combination of the \sevn population-synthesis code \citep{spera2015, spera2017, spera2019, mapelli2020, iorio2022} with the \galaxev code for the spectral evolution of stellar populations \citep{bruzual2003}. We briefly recall the main features of these codes in the following paragraph, referring the reader to the original studies for more details.

\sevn (Stellar EVolution for $N$-body) is an open-source binary population-synthesis code,\footnote{\url{https://gitlab.com/sevncodes/sevn}. The \sevn version used in this work is the release {\it Iorio22} (\url{https://gitlab.com/sevncodes/sevn/-/releases/iorio22}).} based on the interpolation of stellar properties (e.g., mass, radius, luminosity, core mass, core radius) from interchangeable libraries of pre-computed evolutionary tracks to describe the temporal evolution of stellar populations including binary-evolution processes. In this work, we adopt the \parsec (PAdova and TRieste Stellar Evolution Code) library of evolutionary tracks (including pre-main sequence evolution) for stars with initial masses between 2 and 600\,\Msun and metallicities in the range $10^{-11} \leq Z\leq 0.04$  \citep{bressan2012, chen2015, costa2019, costa2023, nguyen2022, santoliquido2023}. 

For stars ending their lives as supernovae (SNe), \sevn includes several prescriptions to compute the compact-remnant (neutron-star or black-hole) mass and natal kick depending on SN type, i.e., electron-capture \citep[][]{giacobbo2019}, core-collapse \citep{fryer2012} or pair-instability \citep{mapelli2020} SNe. We adopt here the delayed SN model, which predicts a smooth transition between maximum neutron-star mass and minimum black-hole mass. We generate natal kicks as described by \cite{giacobbo2020}, in agreement with the proper-motion distribution of young Galactic pulsars \citep{hobbs2005}, and with reduced kick magnitude for stripped and ultra-stripped SNe \citep{tauris2017}.

The binary-evolution processes incorporated in \sevn include, most notably, mass transfer driven by winds and Roche-lobe overflow, common envelope evolution, removal of stellar angular-momentum by magnetic braking, the effect of stellar tides on orbital motions, orbital decay through gravitational-wave emission and stellar mergers. These are described in depth in section~2.3 of \citet{iorio2022}. For convenience, we provide in Appendix~\ref{app:SEVN} a summary describing how SEVN handles binary-evolution products using the \parsec tracks adopted in this study. In the simulations presented in this paper, mass transfer is assumed to be always stable for donor stars on the main sequence and in the Hertzsprung gap. For other stars, mass transfer stability depends on the binary mass ratio and the evolutionary stage of the donor star \citep[following option `QCRS' in table~3 of][]{iorio2022}. Also, we set here the adjustable fraction of orbital energy converted into kinetic energy during common-envelope evolution to $\alpha_{\rm CE}=3$  \citep[section~2.3.3 of][see also \citealt{hurley2002}]{iorio2022}.

Finally, in this work, we adopt the formalism of quasi-homogeneous evolution introduced by \citet[][see section~2.3.2 of \citealt{iorio2022}]{eldridge2011}. According to this formalism, a low-metallicity, main-sequence star spun up by accretion of substantial material via stable Roche-lobe overflow mass transfer sees its core replenished with fresh hydrogen through rotational mixing and remains fully mixed until it burns all its hydrogen into helium (ending as a pure-He star), at nearly constant radius. Following \citet[][see also \citealt{eldridge2012}]{eldridge2011}, we include this evolution for stars with metallicity $Z \leq 0.004$ (which have weak-enough winds to prevent strong loss of angular momentum) accreting at least 5 per cent of their mass and with post-accretion masses greater than 10\,\Msun. Stars meeting these criteria brighten and see their temperature  increase as they burn hydrogen at nearly constant radius during quasi-homogeneous evolution \citep[e.g.,][]{eldridge2012}. Their inclusion augments the population of compact, hot luminous stars, which has important implications for the results presented in this paper.

We set the initial conditions as follows. For a given metallicity $Z$, we produce with \sevn\ a stochastic population of $10^6$ evolving binary pairs,\footnote{While we experimented with samples of up to $3\times10^6$ binary pairs, populations of $10^6$ pairs were found sufficient to provide stable evolution of the nebular properties of young galaxies.} with initial primary-star masses in the range $2\leq m\leq300\,\Msun$ drawn from a power-law IMF $\phi(m) \propto m^{-1.3}$ (where $\phi(m)dm$ is the number of stars created with masses between $m$ and $m+dm$). For each pair, we draw the ratio of initial secondary-star mass to initial primary-star mass ($q$), the orbital period ($P$) and the eccentricity ($e$) from the corresponding probability density functions (PDFs) adopted in \citet{sana2012}:  $\mathrm{PDF}(q) \propto q^{-0.1}$ with $q\in[0.1, 1.0]$; $\mathrm{PDF}(\mathcal{P}) \propto \mathcal{P}^{-0.55}$ with $\mathcal{P}=\log(P/\mathrm{day})\in[0.15,5.5]$;  and  $\mathrm{PDF}(e) \propto e^{-0.42}$ with $e\in[0,0.9]$.\footnote{Since the tables of \parsec tracks implemented in \sevn for this work do not include the evolution of stars less massive than $2\,\Msun$, we redraw a new pair if a secondary star is found below this limit.} The adopted primary-star IMF produces a larger proportion of massive stars than the standard Galactic-disc IMF, which has $\phi(m)\propto m^{-2.3}$ for $m\geq1\,\Msun$ \citep[e.g.,][]{kroupa2001, chabrier2003}. We make this choice to appropriately sample the high-mass end of the IMF among this base collection of one million pairs, from which we can then extract populations with different IMFs \citep[including][]{chabrier2003}.

To compute the spectral evolution of stellar populations obtained this way, we adopt an approach similar to that in the \galaxev code \citep{bruzual2003}. Specifically, we assign a spectrum to each \sevn star by selecting among the different stellar libraries listed in appendix~A of \citet[][]{sanchez2022}, supplemented by \wmbasic model atmospheres computed by \citet[][]{chen2015} for stars hotter than 50,000\,K. For a given metallicity, these models are interpolated at the mass-loss rate, effective temperature ($\log T_\mathrm{eff}$) and gravity ($\log g$) of the considered star. The spectra of stars hotter than the hottest \citet{chen2015} model (i.e., with $T_\mathrm{eff}>10^5$\,K) are approximated by black bodies.
We identify Wolf-Rayet (WR) stars of different types (WNL, WNE, WC and WO) in the populations generated with the \sevn code following the procedure outlined in section~3.2 of \citet{chen2015}, based on the H, C, N and O surface abundances of stars hotter than 25,000\,K. We appeal to the high-resolution version of the Potsdam-WR (\powr) models to describe the spectra of these stars \citep{hamman2004, sander2012, todt2015, hainich2019}. For each WR star, we select the \powr model with closest parameters in a way similar to that outlined in appendix~A of \citet{plat2019}, with the refinement that we also match the surface hydrogen fraction of WNL-type stars. For core-He burning stars less massive than 8\,\Msun and hotter than 20,000\,K, we adopt CMFGEN spectral models of stripped-envelope stars by \citet{goetberg2018}. All stellar spectra are re-sampled on a common wavelength scale ranging from 5.6\,\AA\ to 360\,$\mu$m in 16,902 steps using the \spectres tool \citep{carnall2017}.

In the course of this procedure, we record cases of neutron stars and black holes accreting mass from a companion, where the accretion disc can produce X-ray emission (Section~\ref{sec:add_XRBs}). Also, we record rates of pair-instability supernovae (PISNe), exploding \citep[i.e., non-failed;][]{spera2015} core-collapse SNe and type-Ia SNe, which can lead  to shock-driven nebular emission (Section~\ref{sec:add_shocks}).

In the remainder of this paper, we refer to the combined \sevn and \galaxev spectral evolution model as simply `\GLS'. More details on the coupling between \sevn and \galaxev will be provided elsewhere (Bruzual et al., in preparation), as well as predictions for the spectral evolution $S_{\nu}$ (Charlot et al., in preparation).

\subsection{Photoionization modelling}
\label{sec:mod_photion}

\begin{figure}
\centering
 \includegraphics[width=\columnwidth,trim= 0 15 0 0,clip]{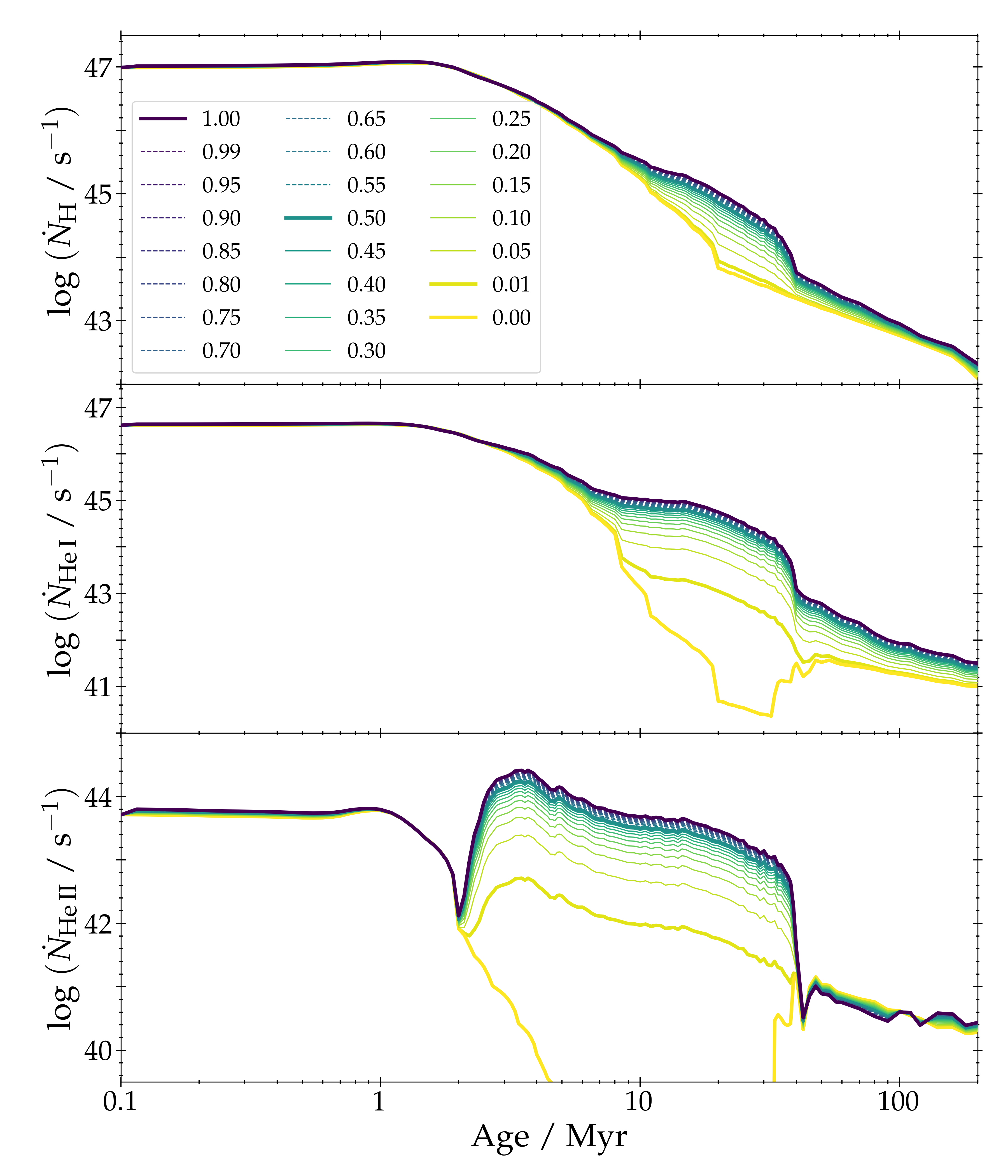}
 \caption{Rates of H- (top), \lhei- (middle) and \lheii- (bottom) ionizing photons as a function of age for `simple' (i.e., instantaneous-burst) stellar populations (SSPs) with different fractions of binary stars, \fbin (colour-coded as indicated), for a metallicity $Z=0.001$. The models assume a \citet{chabrier2003} IMF for primary stars and a \citet[][]{sana2012} distribution of secondary-to-primary mass ratios at age zero (single stars being modelled as non-interacting binaries) and are normalized to a total initial stellar mass of 1\,\Msun integrated over 0.1--300\,\Msun (see text for details).}
\label{fig:comp_partbin}
\end{figure}

The transmission function $T_{\nu}(t,\tprime)$ of the ISM in equation~\eqref{eq:flux_gal} incorporates absorption of stellar radiation by gas and dust in different phases of the ISM (ionized interiors and neutral envelopes of stellar-birth clouds, diffuse inter-cloud medium), as well as the production of nebular line-plus-continuum emission and dust emission \citep[e.g.,][]{charlot2001, dacunha2008, vidalgarcia2017}. Here, we are primarily interested in young star-forming galaxies with age close to the typical timescale for dissipation of giant molecular clouds \citep[$\sim10$\,Myr, ; e.g.,][]{murray2010, murray2011}. In this case, we can write $T_{\nu}(t,\tprime)\equiv T_{\nu}^+(\tprime)$, where $T_{\nu}^+(\tprime)$ is the transmission function of the ionized gas \citep[][]{gutkin2016}.

We compute $T_{\nu}^+(\tprime)$ using version~c17.00 of the \cloudy photoionization code \citep{ferland2017}. Following \citet{charlot2001}, we assume for simplicity that galaxies are ionization-bounded and that the transfer of radiation through ionized gas can be described galaxy-wide by means of `effective' parameters, the main ones being \citep[][see also section~2.1 of \citealt{plat2019}]{gutkin2016}:


\begin{itemize}
\setlength{\itemindent}{.2in}
    \item The hydrogen gas density, \nh.
    \item The gas-phase metallicity, which unless otherwise specified is taken to be equal to that of the ionizing stars, $Z_\mathrm{ISM}=Z$.
    \item The carbon-to-oxygen abundance ratio, \CO.
    \item The dust-to-metal mass ratio reflecting the depletion of heavy elements on to dust grains, \xid.
    \item The zero-age, volume-averaged ionization parameter, $\Uav\equiv\Uav(\tprime=0)$. In spherical geometry,  $ \Uav (\tprime)$ can be expressed in terms of the rate of ionizing photons produced by the evolving stellar population, $\Nhdot(\tprime)$, the volume-filling factor of the gas (i.e., the ratio of the volume-averaged hydrogen density to \nh), $\epsilon$, and the case-B hydrogen recombination coefficient, $\alpha_{\rm B}$, as \citep[e.g.,][]{panuzzo2003}:
    \begin{equation}
        \label{eq:defU}
        \Uav (\tprime)= \frac{3\alpha_B^{2/3}}{4c} \left[ \frac{3\Nhdot(\tprime)\epsilon^2\nh}{4 \pi} \right]^{1/3}\,.
    \end{equation}
\end{itemize}

We stop the photoionization calculations at the outer edge of the \hii region, when the electron density falls below 1 per cent of \nh (we adopt a default inner radius of 0.1\,pc). We adopt the prescription of \cite{gutkin2016} for the abundances and depletions of interstellar elements, as well as for scaling the total (primary+secondary) nitrogen abundance with that of oxygen. In this framework, the present-day solar (photospheric) metallicity is $\zsun=0.01524$, corresponding to a proto-solar metallicity $\zpsun=0.01774$ \citep{bressan2012}.

It is of interest to examine how $\Nhdot(\tprime)$ depends on the fraction of binary stars in a stellar population according to the models presented in Section~\ref{sec:mod_stellpop} \citep[see also section~2.4.1 of][]{eldridge2017}. This is shown in the top panel of Fig.~\ref{fig:comp_partbin} for an instantaneous-burst, `simple stellar population' [SSP; corresponding to $\psi(\tprime)=\delta(\tprime)$ in equation~\ref{eq:flux_gal}], for the typical metallicity $Z=0.001$ (i.e., about 6 per cent of solar) of galaxies in the sample of Section~\ref{sec:res_sample}. The middle and lower panels show the equivalent plots for \lhei- and \lheii-ionizing photons. In each panel, the different curves correspond to different zero-age binary fractions, $\fbin\equiv\fbin(\tprime=0)$, defined as $\fbin=\mathrm{B/(S+B)}$, where B is the number of binary pairs and S the number of single stars at $\tprime=0$. To isolate the effects of binary interactions, the IMF is taken to be exactly the same for binary and single stars. This is achieved by drawing single stars as binary systems whose member stars never interact, i.e., assuming a \citet{chabrier2003} IMF for primary stars and a \citet[][]{sana2012} distribution of secondary-to-primary mass ratios (see Section~\ref{sec:mod_stellpop}), with orbital parameters (large semi-major axes and circular orbits) such that primary and secondary stars never interact. For reference, the total IMF (including primary and secondary stars) drawn in this way has an effective near-\citet{salpeter1955} slope, i.e., roughly $\phi(m)\propto m^{-2.35}$, at $m>10\,\Msun$. All models are normalized to a total initial mass in stars of 1\,\Msun between 0.1 and 300\,\Msun.\footnote{This normalization is performed by drawing primary-star masses from a \citet{chabrier2003} IMF (and the associated secondary stars from a \citealt{sana2012} distribution) over the full range 0.1--300\,\Msun, even though only pairs with both primary- and secondary-star masses greater than 2\,\Msun are retained (the pairs drawn with both primary- and secondary-star masses greater than 2\,\Msun are matched to those in the base collection of Section~\ref{sec:mod_stellpop}). This has no influence on the predicted \Nhdot, \Nheidot and \Nheiidot, since main-sequence stars less massive than 2\,\Msun are cooler than $\sim14,000\,$K at any metallicity and do not produce significant ionizing radiation. Also, rejecting binaries with primary mass above 2\,\Msun and secondary mass below this limit has a negligible impact on ionizing fluxes: artificially including such draws by assigning them the light of pairs with same primary mass and a secondary mass of 2\,\Msun would increase \Nhdot, \Nheidot and \Nheiidot by less than one per cent at ages below 40\,Myr in Fig.~\ref{fig:comp_partbin}.}

Fig.~\ref{fig:comp_partbin} shows that, at ages younger than 1\,Myr, models with $\fbin>0$ produce slightly larger \Nhdot,  \Nheidot and \Nheiidot than single-star models, due to the presence of merged stars with masses up to $\sim600\,\Msun$ on the upper main sequence. Then, when the most massive stars leave the main sequence, all three quantities start to drop. The rate of \lheii-ionizing photons rises sharply again after 2\,Myr in models with $\fbin>0$. This is caused by the appearance of the hot ($>10^5\,$K), pure-He (WNE-like) products of the first massive stars to lose their H-rich envelope, either by dumping mass on their companion through Roche-lobe overflow, or by being driven to quasi-homogeneous evolution and burning all of their H to He after receiving mass from their companion, or through common-envelope ejection \citep[see, e.g.,][]{spera2019, iorio2022}. These stars have a comparatively weak influence on the rates of \lhei- and H-ionizing photons, which remain dominated by the much more numerous H-burning stars on the upper main sequence, at least until ages of several million years. Then, as the He-burning main sequence continues to develop, stellar mergers conspire to maintain the tip of the H-burning main sequence at a brighter and hotter point in binary-star models than in single-star models, leading to significantly larger \Nhdot,  \Nheidot and \Nheiidot. After a few tens of million years, the difference between models with $\fbin>0$ and $\fbin=0$ declines, especially for \Nheiidot after the disappearance of the brightest and hottest WNE-like stars (around 40\,Myr). The faintest of these stars, together with stellar mergers and stars undergoing quasi-homogeneous evolution, contribute to maintain \Nhdot and \Nheidot relatively strong in binary-star models relative to single-star models. The rise of \Nheidot and \Nheiidot at ages around 30\,Myr in single-star models is due to the development of the white-dwarf cooling sequence. From about 50 to 90\,Myr, the hot ionizing radiation from the most massive of these stars leads to slightly higher \Nheiidot than in binary-star models, where their population is reduced through binary-star interactions (their progenitors avoiding white-dwarf evolution through mass gain, or being driven to fainter evolution through mass loss).

A notable feature of Fig.~\ref{fig:comp_partbin} is that even small fractions of binary stars can have a major impact on the predicted rates of highly energetic photons. At an age of 3\,Myr, for example, the ratio of \lheii- to H-ionizing photons, \Nheiidot/\Nhdot, which is 2700 times larger in models with $\fbin=1$ than in those with $\fbin=0$, is already 54, 480 and 1300 times larger for $\fbin=0.01$, 0.1 and 0.3, respectively. It is 2200 times larger for $\fbin=0.7$, often adopted as typical of massive-star populations \citep[e.g.][]{sana2012}.

\begin{figure*}
\centering
 \includegraphics[width=\textwidth,trim= 0 20 0 0,clip]{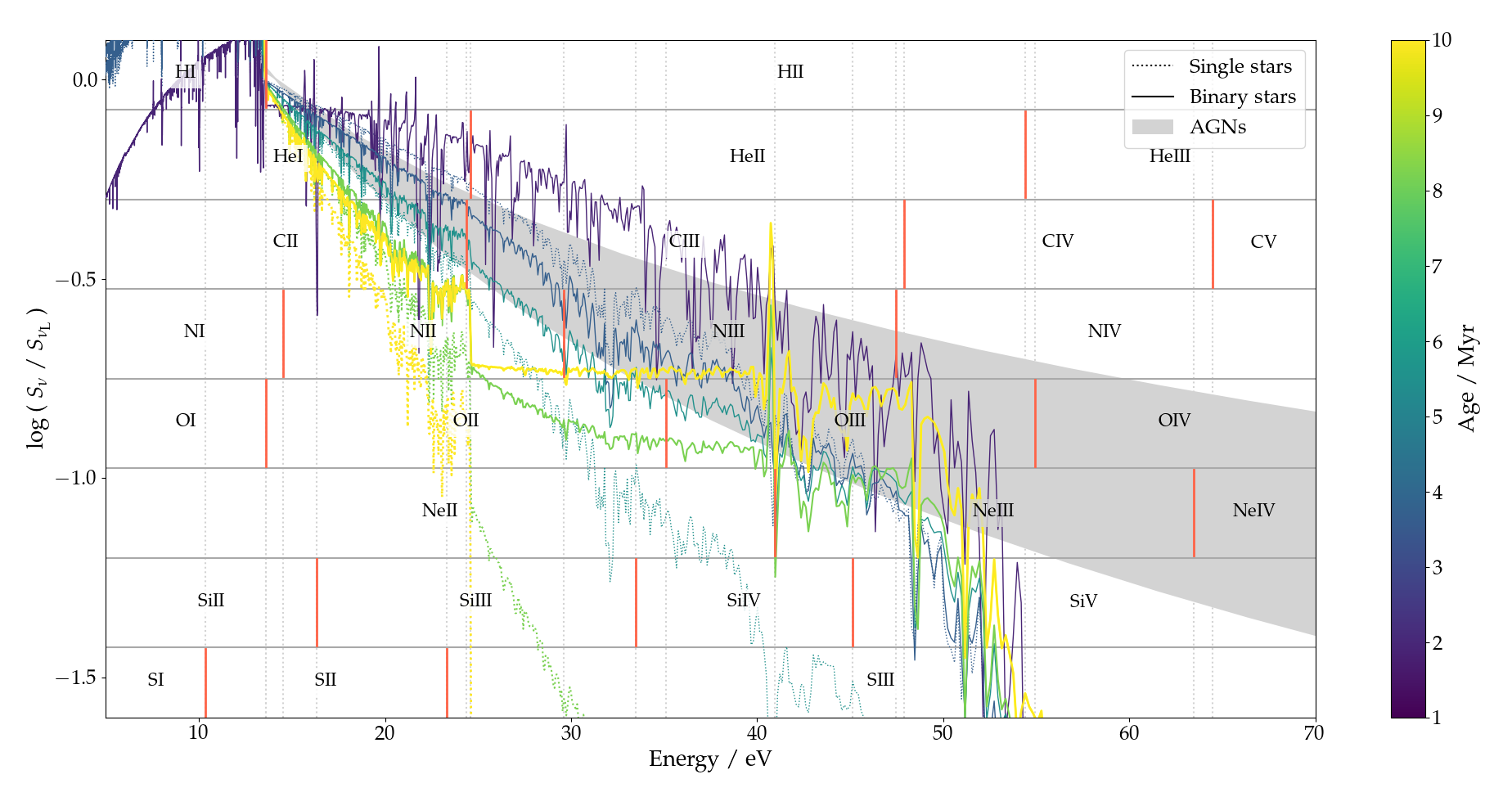}
 \caption{Spectral energy distribution $S_\nu$ entering equation~\eqref{eq:flux_gal} (in units of the luminosity per unit frequency at the Lyman limit) for the SSP models of Fig.~\ref{fig:comp_partbin} with pure binary stars ($\fbin=1$, solid lines) and pure single stars ($\fbin=0$, dotted lines) at the ages $\tprime=1$, 3, 5, 8 and 10\,Myr (colour-coded as indicated on the right-hand side). Also shown in grey is the area sampled by ionizing spectra of AGNs with power-law indices between $\alpha_\mathrm{AGN}=-2.0$ (bottom edge) and $-1.2$ (top edge; \citealt{feltre2016}). The ionization energies of some common species are indicated for reference.}
\label{fig:sed_z001}
\end{figure*}

In Fig.~\ref{fig:sed_z001}, we compare in more detail the ionizing spectra $S_\nu(\tprime)$ of SSPs with pure binary stars ($\fbin=1$, solid lines) and pure single stars ($\fbin=0$, dotted lines) at ages between 1 and 10\,Myr, for a metallicity $Z=0.001$. As expected from Fig.~\ref{fig:comp_partbin}, at an age of 1\,Myr, the spectra of both models nearly overlap, while at older ages, the model with binary stars exhibits an excess of energetic photons relative to that with single stars. Populations of binary stars will therefore boost the production of highly-ionized species in the gas they irradiate (we indicate for reference the ionization energies of some common species in Fig.~\ref{fig:sed_z001}). In fact, Fig.~\ref{fig:sed_z001} shows that at ages $\tprime\geq3\,$Myr in the binary-star model, the shape of $S_\nu(\tprime)$ at ionizing energies up to $\ga50$\,eV is similar to that of hard ionizing spectra of AGNs, $S_\nu\propto\nu^{\alpha_\mathrm{AGN}}$ with $-2.0\leq\alpha_\mathrm{AGN}\leq-1.2$ \citep[e.g.][grey shaded area]{feltre2016}.

\section{Comparison with observations}
\label{sec:res}

In this section, we compare the predictions of the models presented in Section~\ref{sec:models} with observations of ultraviolet and optical emission lines in metal-poor star-forming galaxies. We start by describing the observational sample (Section~\ref{sec:res_sample}), after which we examine the predictions of models with standard star and gas parameters (Section~\ref{sec:res_std}) and explore the influence of some key adjustable parameters on the results (Section~\ref{sec:res_param}).

\subsection{Observational sample}
\label{sec:res_sample}

We appeal to the compilation of emission-line measurements of star-forming galaxies by \citet[][based on the samples of \citealt{fosbury2003, erb2010, jaskot2013, stark2014, berg2016,berg2018,berg2019, izotov2016a,izotov2016b,izotov2017,izotov2018a,izotov2018b, nakajima2016, vanzella2016,vanzella2017, amorin2017, chisholm2017, schmidt2017, senchyna2017,senchyna2019, nanayakkara2019}]{plat2019}. From this data set, we select all metal-poor galaxies with confident oxygen abundances (i.e., determined using the direct-\Te or the strong-lines methods) less than $\logoh \approx8.2$, corresponding to $Z\la0.004$ \citep[see, e.g., table~2 of][]{gutkin2016}. With this selection, we retain 102 galaxies, of which 20 are classified as Lyman-continuum (LyC) leakers, with fractions of escaping LyC photons up to $\sim0.6$. The galaxies in this sample span wide ranges in redshift, $0\la z\la7$, ionization parameter, $-3.6\la\logU\la-1.0$, C/O abundance ratio, $0.1 \la (\CO)/\COsol \la1.2$ and specific star formation rate, $1\la \psi/M\la 1,000\,\mathrm{Gyr}^{-1}$ (where $M$ denotes the galaxy mass). For reference, we also consider the emission-line properties of 84 AGNs and candidate AGNs from \citet{diaz1988,kraemer1994,dors2014,stark2015,nakajima2018}. We refer the reader to section~3 and tables~1 and 2 of \citet{plat2019} for more details about this sample.

We complement this data set with recent emission-line measurements for a few more local analogues to primeval galaxies, including the extremely metal-poor ($Z < 0.1\,\zsun$) starburst-dwarf galaxies SBS~0335-052E \citep{wofford2021},  J1631+4426, J104457 and I~Zw~18~NW \citep{umeda2022}, 7 extremely metal-poor galaxy candidates from \citet{kojima2021} and 6 \lciv emitters from \citet{senchyna2022}.

In this paper, we compare  the predictions of our model with the observed intensity ratios and equivalent widths (EWs) of selected ultraviolet (\civ, \heii, \oiii and \ciii) and optical (\oiiiopt, \oiiopt, \oiopt, \heiiopt, \hb, \ha and \niiopt) emission lines commonly observed in metal-poor star-forming galaxies. We will pay particular attention to the \lheii lines, which require ionization energies above 54.4\,eV and represent a challenge for current models.

\subsection{Standard models}
\label{sec:res_std}

We start by exploring how models with `standard' parameters compare with the observations described above. As in Section~\ref{sec:mod_photion}, we consider stellar populations with metallicity $Z=0.001$, typical of the observational sample. For the photoionized gas, we adopt the same metallicity ($\zism=0.001$) and a density ($\nh=100\,\mathrm{cm}^{-3}$), ionization parameter ($\logU=-2.0$) and dust-to-metal mass ratio ($\xid=0.3$) typical of metal-poor star-forming galaxies \citep[e.g.,][]{berg2016, gutkin2016, mingozzi2022}. These choices of \zism\ and \xid\ imply a gas-phase oxygen abundance $\logoh\approx7.53$ \citep[table~2 of][]{gutkin2016}, which according to fig.~6 of \citet{berg2016} corresponds to a typical carbon-to-oxygen abundance ratio $\CO\approx0.17$, i.e., about 40 per cent of the solar ratio $\COsol = 0.44$ \citep[][]{gutkin2016}. This is close to the typical value $\CO\approx0.18$ reported by \citet{izotov2023} in low-redshift LyC-leaking galaxies with $\logoh\la8.1$. Yet, given that galaxies in the observational sample of Section~\ref{sec:res_sample} exhibit \CO values up to 120 per cent of solar, we expect some of these galaxies to show stronger carbon lines than predicted by our standard models. We report the above standard nebular parameters in Table~\ref{tab:cloudy_params}. We perform the photoionization calculations at stellar-population ages up to 10\,Myr, i.e., roughly the timescale of dissipation of giant molecular clouds in star-forming galaxies \citep{murray2011,ma2015}.

\begin{table}
\caption{Nebular parameters of the `standard' models of Section~\ref{sec:res_std}.}
\centering
\begin{tabular}{lll}
\hline
 Parameter & Value & Description\\
 \hline
\zism & 0.001 & Interstellar metallicity \\
\nh & $100\ \mathrm{cm}^{-3}$ & Hydrogen density \\
\CO & 0.17  & Carbon-to-oxygen abundance ratio \\
\logU & $-2.0$ & Volume-averaged ionization\\
&&parameter at age $\tprime=0$\\
\xid & 0.3 & Dust-to-metal mass ratio \\
 \hline
\end{tabular}
\label{tab:cloudy_params}
\end{table}

\begin{figure*}
\centering
 \includegraphics[width=\textwidth]{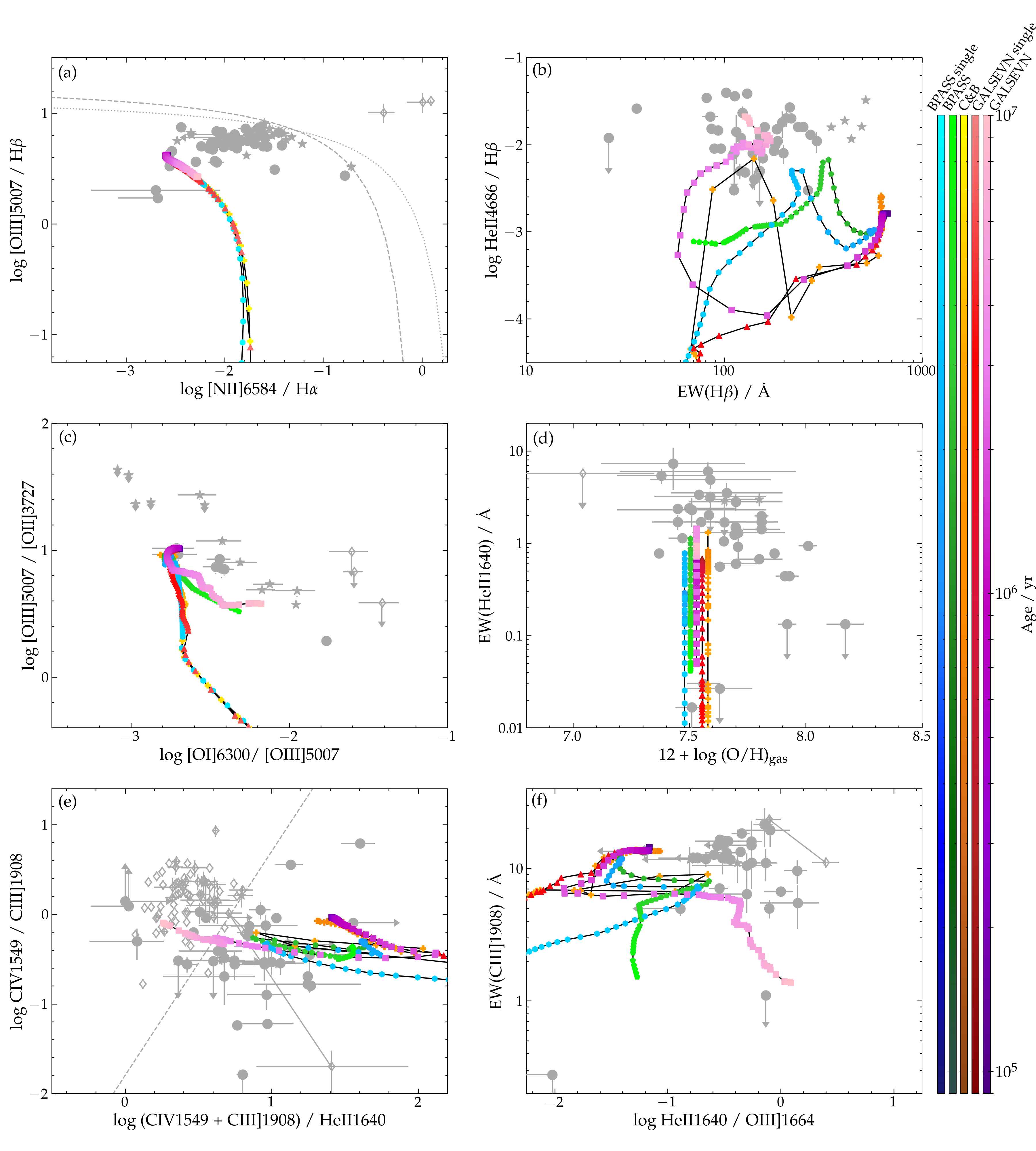}
 \caption{Optical and ultraviolet emission-line properties of the reference observational sample of star-forming galaxies (filled grey circles: non LyC leakers; filled grey stars: LyC leakers) and AGNs (open grey diamonds) described in Section~\ref{sec:res_sample}. The diagrams show different combinations of equivalent widths and ratios of the \civ, \heii, \oiii and \ciii, \oiiiopt, \oiiopt, \oiopt, \heiiopt, \hb, \ha and \niiopt nebular emission lines, and the gas-phase oxygen abundance in (d). All line fluxes are corrected for attenuation by dust, as prescribed in the original studies (arrows show 1$\sigma$ upper limits). In (a), the dotted and dashed lines show the criteria of \citet{kewley2001} and \citet{kauffmann2003}, respectively, to separate AGN-dominated from star-forming galaxies, while in (e), the dashed line shows the equivalent criterion proposed by \citet{nakajima2018}. The curves interspersed with filled squares, triangles, pentagons, hexagons and crosses show the predictions of the \protect\GLS binary- and single-star models, the \protect\bpass binary- and single-star models and the \protect\CB single-star model, respectively, colour-coded with age as indicated on the right (the \protect\bpass models, available only at ages above 1\,Myr, do not sample the youngest ages in the colour bars). In (d), the models are shown for clarity at slightly different absciss\ae, although all have $\logoh=7.53$. In all models, the ionizing stellar population is an SSP with metallicity $Z=0.001$ and the same zero-age \citet{chabrier2003} IMF as in Fig.~\ref{fig:comp_partbin}, while the parameters of the photoionized gas are those listed in Table~\ref{tab:cloudy_params}.}
\label{fig:diags_oldmodels}
\end{figure*}


In Fig.~\ref{fig:diags_oldmodels}, we show the intensity ratios and equivalent widths of emission lines predicted by the SSP models of Fig.~\ref{fig:comp_partbin} with pure binary stars ($\fbin=1$, labelled `\GLS') and pure single stars ($\fbin=0$, `\GLS single') at ages up to 10\,Myr (colour-coded as indicated on the right). It is instructive to start by examining the behaviour of these models in the \heiiopt/\hb-versus-EW(\hb) diagram (Fig.~\ref{fig:diags_oldmodels}b), where observations are traditionally challenging to reproduce with photoionization models powered by young stellar populations \citep[e.g.,][]{plat2019, schaerer2019}. All models start at high EW(\hb) on the right-hand side of the diagram. As massive stars leave the main sequence and the first red super-giant stars appear in the single-star \protect\GLS model, \Nhdot and \Nheiidot decrease while the continuum at \hb increases, causing both EW(\hb) and \heiiopt/\hb to drop without ever reaching the area of the diagram occupied by the observations. 

Remarkably, in the binary-star \protect\GLS model, the rise in \Nheiidot triggered by the appearance of hot WNE-like stars with blue spectra around 3\,Myr (Fig.~\ref{fig:comp_partbin}) makes both \heiiopt/\hb and EW(\hb) rise in such a way that the models spend all the time thereafter up to 10\,Myr in the same location as the observations. It is worth recalling that, as shown by Fig.~\ref{fig:comp_partbin}, similar results would be obtained with models containing only a minor fraction of binary stars ($\fbin\ga0.3$).

\begin{figure}
\centering
 \includegraphics[width=\columnwidth,trim= 0 10 0 0,clip]{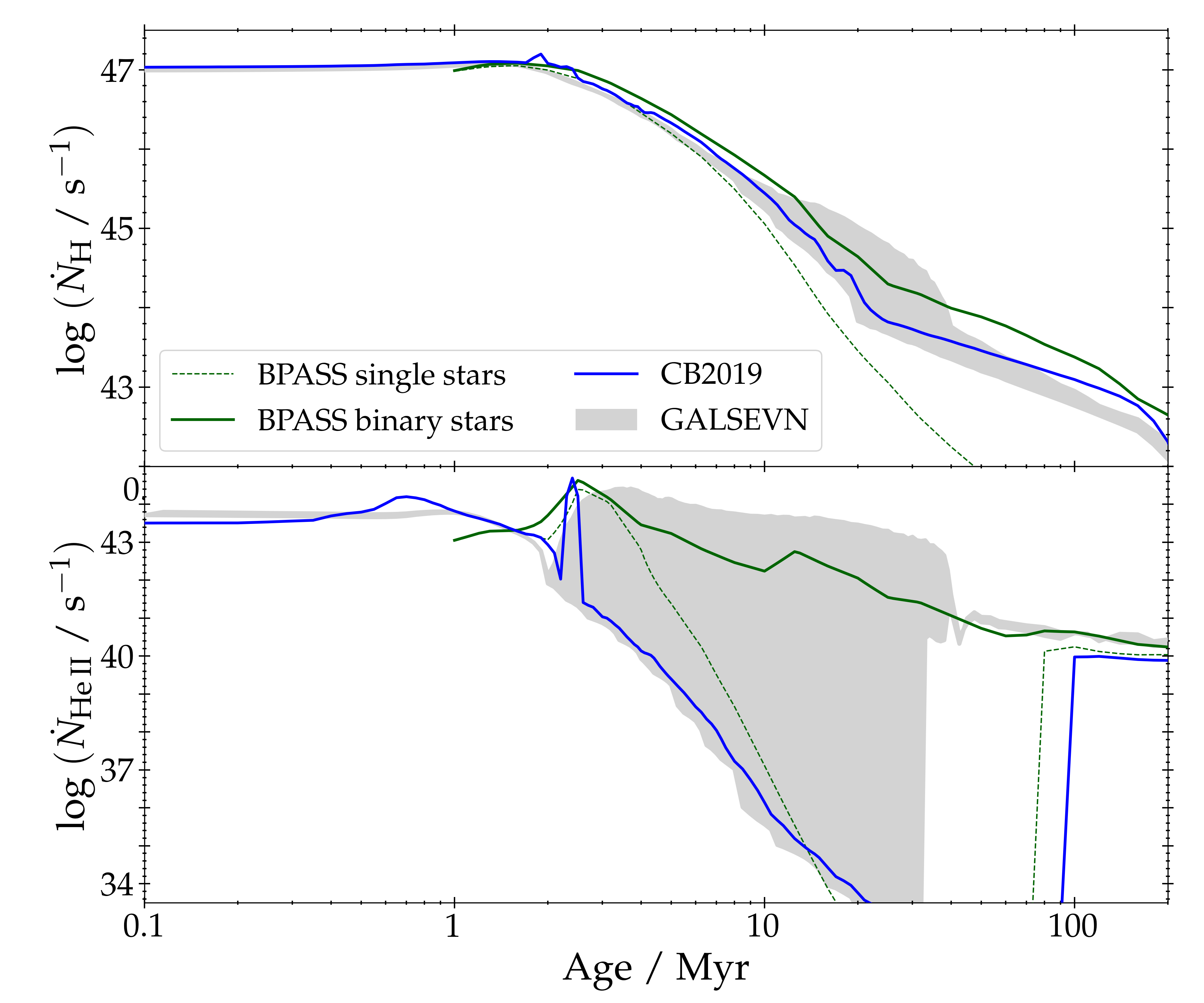}
 \caption{Rates of H- (top) and \lheii- (bottom) ionizing photons as a function of age for SSPs of metallicity $Z=0.001$ computed with different models, as indicated (the \protect\bpass models are available only at ages above 1\,Myr). In each panel, the grey shaded area shows the area covered by the \protect\GLS models with different fractions of binary stars from Fig.~\ref{fig:comp_partbin}. All models have the same zero-age \citet{chabrier2003} IMF normalized to a total initial stellar mass of 1\,\Msun integrated over 0.1--300\,\Msun as in Fig.~\ref{fig:comp_partbin}.}
\label{fig:comp_nphot_oldmodels}
\end{figure}

Also shown for comparison in Fig.~\ref{fig:diags_oldmodels}(b) are photoionization calculations performed using the \bpass v2.2.1 models of \citet{stanway2018} for single and binary stars, as well as the \CB single-star model used by \citet[][based on an updated version of the \citealt{bruzual2003} \galaxev code]{plat2019}, for a metallicity $Z=0.001$ and the same nebular parameters as in Table~\ref{tab:cloudy_params}. The \cloudy calculations for these models were conducted similarly to those for the \protect\GLS models as part of the present study. We note the striking difference between the \CB and \protect\GLS single-star models at ages around 2\,Myr, when the brief ($\sim0.2\,$Myr) WR phase that makes \heiiopt/\hb rise up to the observed range in the \CB model is absent from the \protect\GLS model. This can also be seen in the evolution of \Nhdot and \Nheiidot shown in Fig.~\ref{fig:comp_nphot_oldmodels}. Both models being based on the same library of stellar spectra (described in Section~\ref{sec:mod_stellpop}), this difference originates from recent updates (in particular in the opacities) of the \parsec-track library of \citet{bressan2012} and \citet{chen2015}, leading to reduced mass loss, and hence fewer WR stars at low metallicity \citep{costa2021, nguyen2022}.

The brief WR phase at early ages is also present (although to a lesser extent) in the \bpass models for both single and binary stars, which are based on the Cambridge \stars code \citep[][see also Fig.~\ref{fig:comp_nphot_oldmodels}]{eggleton1971, eldridge2008}. After this phase and the disappearance of the most massive stars, EW(\hb) and \heiiopt/\hb both drop rapidly in the single-star model, as in the \protect\GLS and \CB single-star models. Instead, in the \bpass binary-star model, \heiiopt/\hb is maintained at a fairly high level by the appearance of envelope-stripped and spun-up stars, still without reaching the region of the diagram probed by the observations. Given that in \GLS, \Nheiidot is dominated by WNE stars after a few Myr, the discrepancy between the \GLS and \bpass binary-star models in Fig.~\ref{fig:diags_oldmodels}(b) can likely be attributed to a distinct population of these stars in \bpass, particularly as both models use similar spectra for them (from the PoWR library).

The behaviour of the models in the other diagrams of Fig.~\ref{fig:diags_oldmodels} follows from the same arguments. In the classical \citet[][hereafter BTP]{baldwin1981} diagram to distinguish AGNs from star-forming galaxies (Fig.~\ref{fig:diags_oldmodels}a), at early ages, all models lie at the high-\oiiiopt/\hb, low-\niiopt/\ha end characteristic of low metallicity and high ionization parameter \citep[see, e.g., fig.~2 of][]{gutkin2016}. The \protect\GLS and \bpass binary-star models stay in this area at ages up to 10\,Myr, while the \oiiiopt/\hb ratio of single-star models rapidly drops after the disappearance of photons more energetic than $\sim35.1\,$eV capable of ionizing O$^{2+}$. The implied drop in \oiiiopt luminosity is also the reason for the difference between binary- and single-star models in the \oiiiopt/\oiiopt-versus-\oiopt/\oiiiopt diagram (Fig.~\ref{fig:diags_oldmodels}c). In Fig.~\ref{fig:diags_oldmodels}(d), none of the models (plotted for clarity at slightly different absciss\ae\ even though all have $\logoh=7.53$) reaches the highest observed $\mathrm{EW}(\heii)\sim2$--8\,\AA. In this diagram, only the \protect\GLS binary-star models maintains $\mathrm{EW}(\heii)>0.5\,$\AA\ at ages $\tprime\ga3\,$Myr, while the \bpass binary-star model declines to $\mathrm{EW}(\heii)\approx0.03\,$\AA\ and all single-star models to $\mathrm{EW}(\heii)\approx0$.

The difference between the \protect\GLS binary-star model and all other models is more striking in Fig.~\ref{fig:diags_oldmodels}(e), where \GLS is the only model able to account for the properties of star-forming galaxies similar to those of AGNs at low (\civ+\ciii)/\heii and high \civ/\ciii. Again, albeit without certainty, the difference between the \GLS and \bpass binary-star models likely arises from distinct populations of WNE stars, which dominate the production of \lheii-ionizing photons. Similarly, in Fig.~\ref{fig:diags_oldmodels}(f), this model at ages $\tprime\ga2\,$Myr is closest to the observational sequence at $\heii/\oiii>0.3$, although it does not appear to reach sufficiently high EW(\ciii) at a given \heii/\oiii.

On the whole, Fig.~\ref{fig:diags_oldmodels} shows the need to include binary stars and their inherent physical processes in modelling the emission-line properties of metal-poor, star-forming galaxies. It also validates reasonable agreement between \protect\GLS binary-star models and (non-extreme) observations of such galaxies in our sample, although discrepancies between some observed and modelled spectral features remain to be explored.

\subsection{Influence of adjustable model parameters}
\label{sec:res_param}

So far, we have considered only models with metallicity $Z=0.001$ and nebular parameters fixed at the standard values of Table~\ref{tab:cloudy_params}. Given the diversity of objects in the observational sample of Section~\ref{sec:res_sample}, it is relevant to explore the effect of changing these parameters. Here, we focus in turn on the influence of metallicity, volume-averaged ionization parameter and carbon-to-oxygen abundance ratio. We do not explore changes in \nh and \xid, which have been shown by \cite{plat2019} to have a minimal impact on the emission-line ratios presented in Fig.~\ref{fig:diags_oldmodels}. We adopt the same \citet{chabrier2003} IMF as in the standard models.

\subsubsection{Metallicity}
\label{sec:res_param_met}

\begin{figure*}
 \centering
 \includegraphics[width=\textwidth]{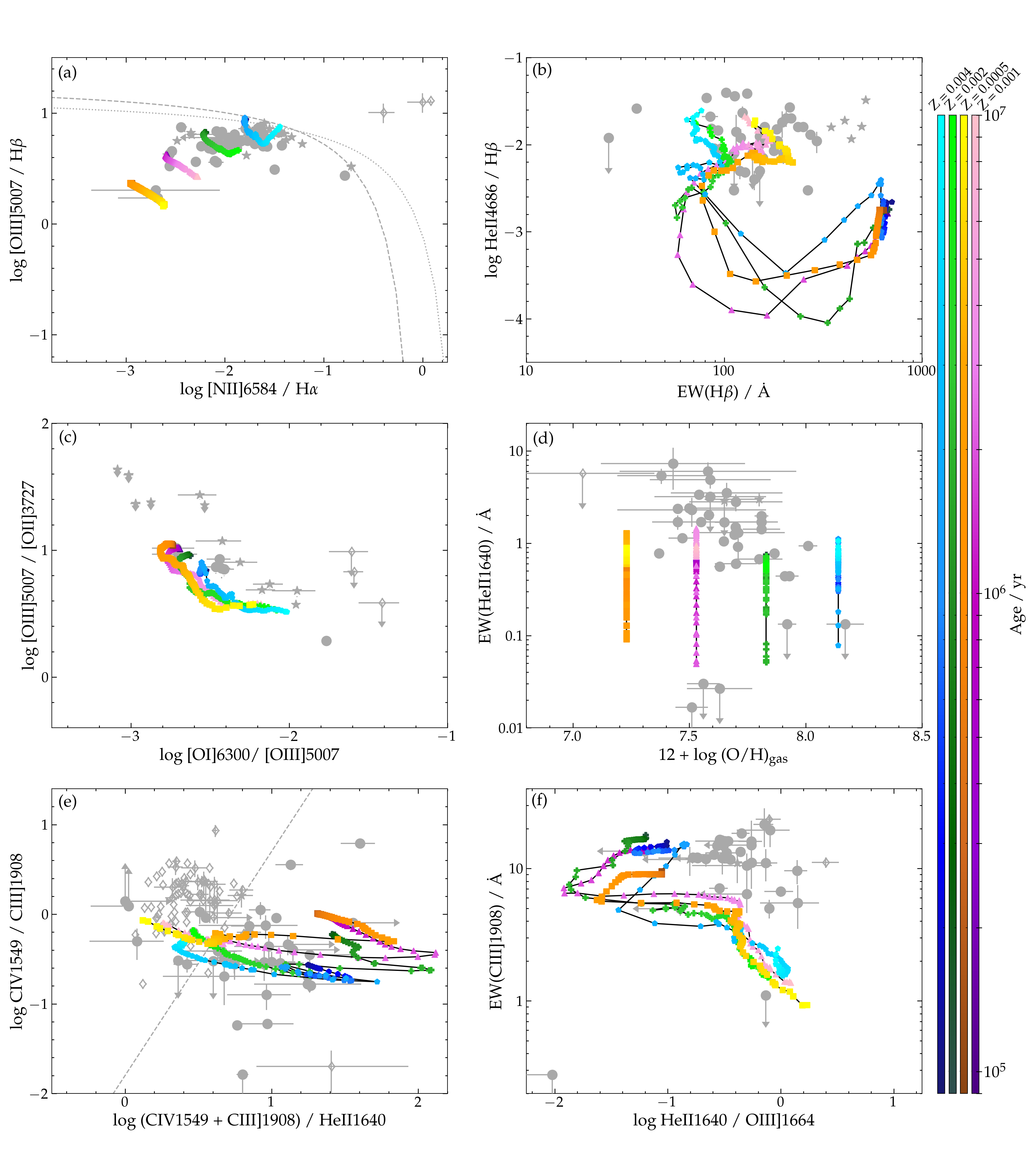}
 \caption{Same as Fig.~\ref{fig:diags_oldmodels}, but for \protect\GLS binary-star models with different metallicities, as indicated at the tip of the colour bars. The curves interspersed with filled squares, triangles, crosses and pentagons refer to the metallicities $Z=0.0005$, 0.001, 0.002 and 0.004, respectively.}
\label{fig:diags_met}
\end{figure*}

Fig.~\ref{fig:diags_met} shows the evolution of \protect\GLS binary-star models with metallicities $Z=\zism=0.0005$, 0.001, 0.002 and 0.004 in the same diagrams as in Fig.~\ref{fig:diags_oldmodels}, with all other parameters fixed at their values in Table~\ref{tab:cloudy_params}.
An increase in metallicity makes cooling through collisionally-excited metal transitions more efficient, which reduces the electronic temperature \Te. This causes the luminosity ratios of metal lines to H and He lines to rise and the ratios of high- to low-ionization lines to drop (Figs~\ref{fig:diags_met}a, c, e and f). The increase of \niiopt/\ha with $Z$ in Fig.~\ref{fig:diags_met}(a) is further enhanced by the inclusion of secondary nitrogen enrichment in the nebular-emission models (Section~\ref{sec:mod_photion}). At $Z=0.004$, we note the emergence of classical WR stars at early ages following the increased mass loss of evolved massive stars with metal-rich envelopes, which causes a brief excursion of the model at high \heiiopt/\hb and \heii/\oiii in Figs.~\ref{fig:diags_met}(b) and (f).

As Fig.~\ref{fig:diags_met}(d) shows, while metallicities in the range $0.0005\leq Z\leq0.004$ can fully describe the range of \logoh probed by the observational sample, metallicity has little influence on the maximum EW(\heii) achievable with the \protect\GLS binary-star model.

\subsubsection{Ionization parameter and C/O abundance ratio}
\label{sec:res_param_neb}

\begin{figure*}
 \centering
 \includegraphics[width=\textwidth]{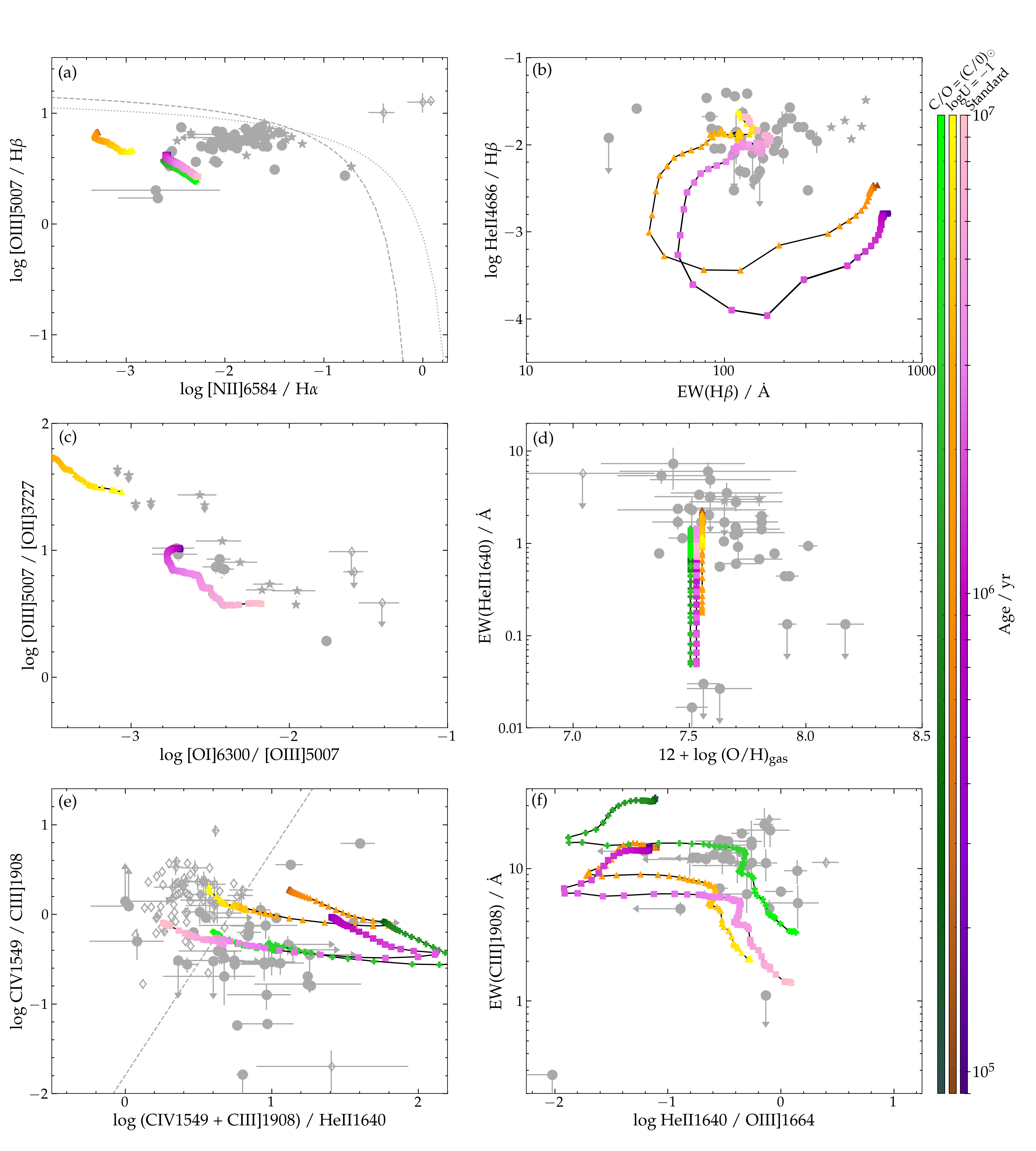}
 \caption{Same as Fig.~\ref{fig:diags_oldmodels}, for three \protect\GLS binary-star models with metallicity $Z=0.001$: a model with the standard parameters of Table~\ref{tab:cloudy_params} ($\logU = -2$, $\CO = 0.17$; filled squares), one with enhanced ionization parameter ($\logU = -1$, $\CO = 0.17$; filled triangles), and one with enhanced \CO abundance ratio ($\logU = -2$, $\CO = \COsol=0.44$; nearly overlapping with the standard model in (b) and (c); filled crosses), as indicated on the right.}
\label{fig:diags_neb}
\end{figure*}

In Fig.~\ref{fig:diags_neb}, we compare the standard \protect\GLS binary-star model of Fig.~\ref{fig:diags_oldmodels} with two \protect\GLS binary-star models: one with ionization parameter $\logU=-1$ instead of $-2$; and one with $\CO=\COsol=0.44$ instead of 0.17. All models have metallicity $Z=0.001$.

The effect of increasing \logU is to increase the probability of multiple ionizations, and hence to strengthen high-ionization lines (see equivalent widths of \heii and \ciii in Figs~\ref{fig:diags_neb}d and f) and increase ratios of high-to-low ionization lines (e.g., \lciv/\lciii and \loiii/\loiiopt; Figs~\ref{fig:diags_neb}c and e). In Fig.~\ref{fig:diags_neb}(a), the drop in \lniiopt/\ha as \logU increases results from the conversion of N$^{+}$ into N$^{2+}$. In Fig.~\ref{fig:diags_neb}(b), the drop in EW(\hb) is caused by the increase in H-column density, and hence in the absorption of H-ionizing photons by dust at fixed $Z$ and \xid, as the Str\"omgren radius increases \citep[e.g.,][]{plat2019}. Overall, Fig.~\ref{fig:diags_neb} shows that enhancing the ionization parameter does not significantly improve the agreement between \protect\GLS binary-star models and the observations considered here. We note that, while ionization parameters as high $\logU = -1$ have been favoured for only a few objects in this observational sample \citep[see table~1 of][]{plat2019}, such parameters may be more common in young star-forming galaxies near the epoch of reionization \citep[e.g.,][]{cameron2023}. 

Increasing the \CO abundance ratio at fixed metallicity means raising the abundance of carbon while reducing that of all other metals \citep[e.g.,][]{gutkin2016}. This results in a strengthening of C lines and a slight weakening of O lines, with negligible effect on H and He lines, as Fig.~\ref{fig:diags_neb} shows. Interestingly, the adoption of solar \CO in place of the standard value in Table~\ref{tab:cloudy_params} \citep[favoured by the trend with \logoh in fig.~6 of ][]{berg2016} appears to provide better agreement with the data in Fig.~\ref{fig:diags_neb}(f). Such high \CO ratio at low metallicity is in agreement with the measured properties of some metal-poor galaxies in our sample, particularly those in the subsamples from \citet{berg2016,berg2019}, \citet{amorin2017} and \citet{stark2014}. Recently, \citet{deugenio2023} also reported the existence of a metal-poor star-forming galaxy with strong \lciii emission corresponding to slightly super-solar \CO ratio at redshift $z\sim12.5$.

\subsubsection{O/Fe abundance ratio}
\label{sec:res_zism_zstar}

\begin{figure*}
 \centering
 \includegraphics[width=\textwidth]{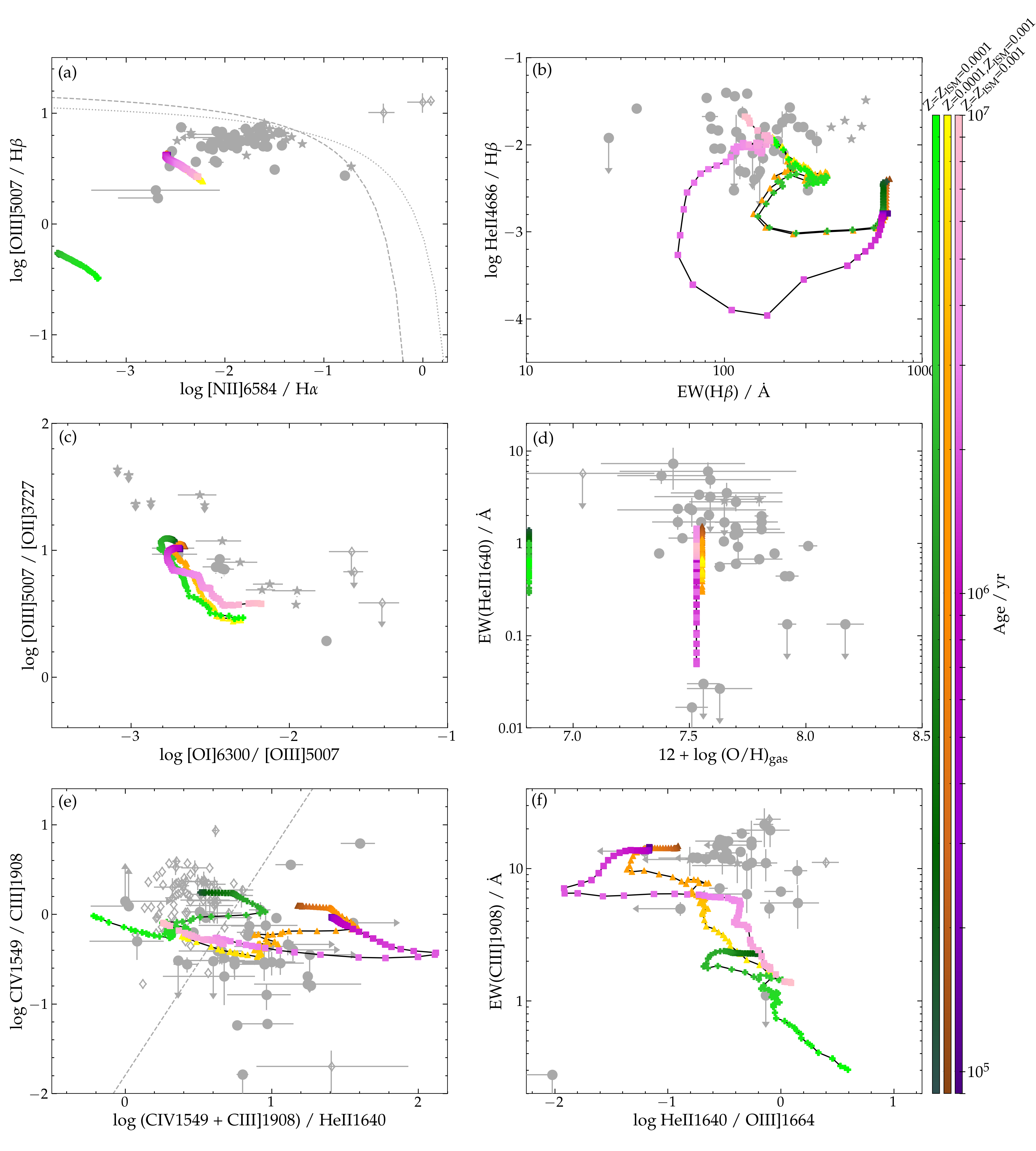}
 \caption{Same as Fig.~\ref{fig:diags_oldmodels}, for three \protect\GLS binary-star models: a model with the standard parameters of Table~\ref{tab:cloudy_params} (with $Z=\zism=0.001$; filled squares), one exploring the influence of super-solar O/Fe (with $Z=0.0001$ and $\zism=0.001$ ; filled triangles), and one investigating the impact of a global reduction of metallicity (with $Z=\zism=0.0001$ ; filled crosses). See main text for details. In (d), the model with $\zism = 0.0001$, which corresponds to $\logoh=6.53$, is shown at the plot limit of $\logoh=6.80$ to retain the same scale as in Fig.~\ref{fig:diags_oldmodels}.}
\label{fig:diags_gsmet}
\end{figure*}

Several recent studies have found evidence for super-solar ratios of $\alpha$ to iron-peak elements (e.g., O/Fe) in metal-poor star-forming galaxies at various redshifts, as expected for chemically young objects enriched by core-collapse SNe \citep[e.g.,][]{steidel2016, strom2018, topping2020, cullen2021, senchyna2022}. Since Fe dominates the opacity of stellar ultraviolet spectra (which is relatively insensitive to O/H), while O, the most abundant heavy element in \hii regions, dominates gas-phase metallicity (interstellar iron being almost entirely depleted on to dust grains), line emission of gas with super-solar O/Fe photoionized by metal-poor stars may be explored using models with fixed O/Fe, but adopting a lower total metallicity for the stars than for the gas \citep[e.g.,][]{steidel2016}.

In Fig.~\ref{fig:diags_gsmet}, we compare the standard \protect\GLS binary-star model of Fig.~\ref{fig:diags_oldmodels} (with $Z=\zism=0.001$) with two new \protect\GLS binary-star models: one exploring the influence of super-solar O/Fe by setting the stellar metallicity to 10 per cent of the ISM metallicity (i.e., $Z=0.0001$ versus $\zism=0.001$) and the other investigating the impact of a global reduction of metallicity, with $Z=\zism=0.0001$.

At fixed $\zism=0.001$, lowering the stellar metallicity from $Z=0.001$ to 0.0001 implies bluer stellar-population spectra and harder ionizing radiation, since stars evolve at higher effective temperature  \citep[e.g.,][]{bressan2012, plat2019}. The \hb equivalent width increases significantly in Fig.~\ref{fig:diags_gsmet}(b) (due to both higher \Nhdot and fainter 4861-\AA\ continuum), as does the \heiiopt/\hb ratio at ages until $\sim3\,$Myr, when the ionizing radiation becomes dominated by pure-He, WNE-type stars at both metallicities. A similar effect is seen at early ages in the (\civ+\ciii)/\heii (Fig.~\ref{fig:diags_gsmet}e) and \heii/\oiii ratios (Fig.~\ref{fig:diags_gsmet}f).

The model with $Z=\zism=0.0001$ in Fig.~\ref{fig:diags_gsmet} produces a luminosity ratio and equivalent widths of \hb and \heii comparable to those in the model with the same $Z=0.0001$ but higher $\zism=0.001$ (Figs~\ref{fig:diags_gsmet}b and d). This is because these emission lines are primarily sensitive to the shape of the stellar-population spectrum. Instead, as expected from Fig.~\ref{fig:diags_met}, the ratios of metal-to-H and He lines decrease significantly when \zism is reduced from 0.0001 to 0.001 (Figs~\ref{fig:diags_gsmet}a, e and f).

On the whole, Fig.~\ref{fig:diags_gsmet} shows that adopting super-solar O/Fe significantly affects the predicted emission-line properties of metal-poor star-forming galaxies, although we note that the impact on the equivalent width of \heii is relatively modest in Fig.~\ref{fig:diags_gsmet}(d).

\section{Inclusion of additional effects}
\label{sec:adds}

In the previous section, we have explored the effects of stellar and several ISM parameters (ionization parameter, \CO and O/Fe abundance ratios) on the predicted emission-line properties of binary-star SSPs computed with the \protect\GLS code. We showed that, while these models produce significantly more energetic radiation than previous single- and binary-star models, they do not appear to reach the largest observed equivalent widths of lines with high ionization energies, such as \heii and \ciii. In the following paragraphs, we consider additional effects associated with the evolution of stellar populations, which  could potentially influence these observables. Unless otherwise specified, in all models in this section, we consider binary-star populations with $Z=0.001$ and the standard nebular parameters listed in Table~\ref{tab:cloudy_params}.

\subsection{Multiple bursts of star formation}
\label{sec:add_SFH}
 
\begin{figure*}
 \centering
 \includegraphics[width=\textwidth]{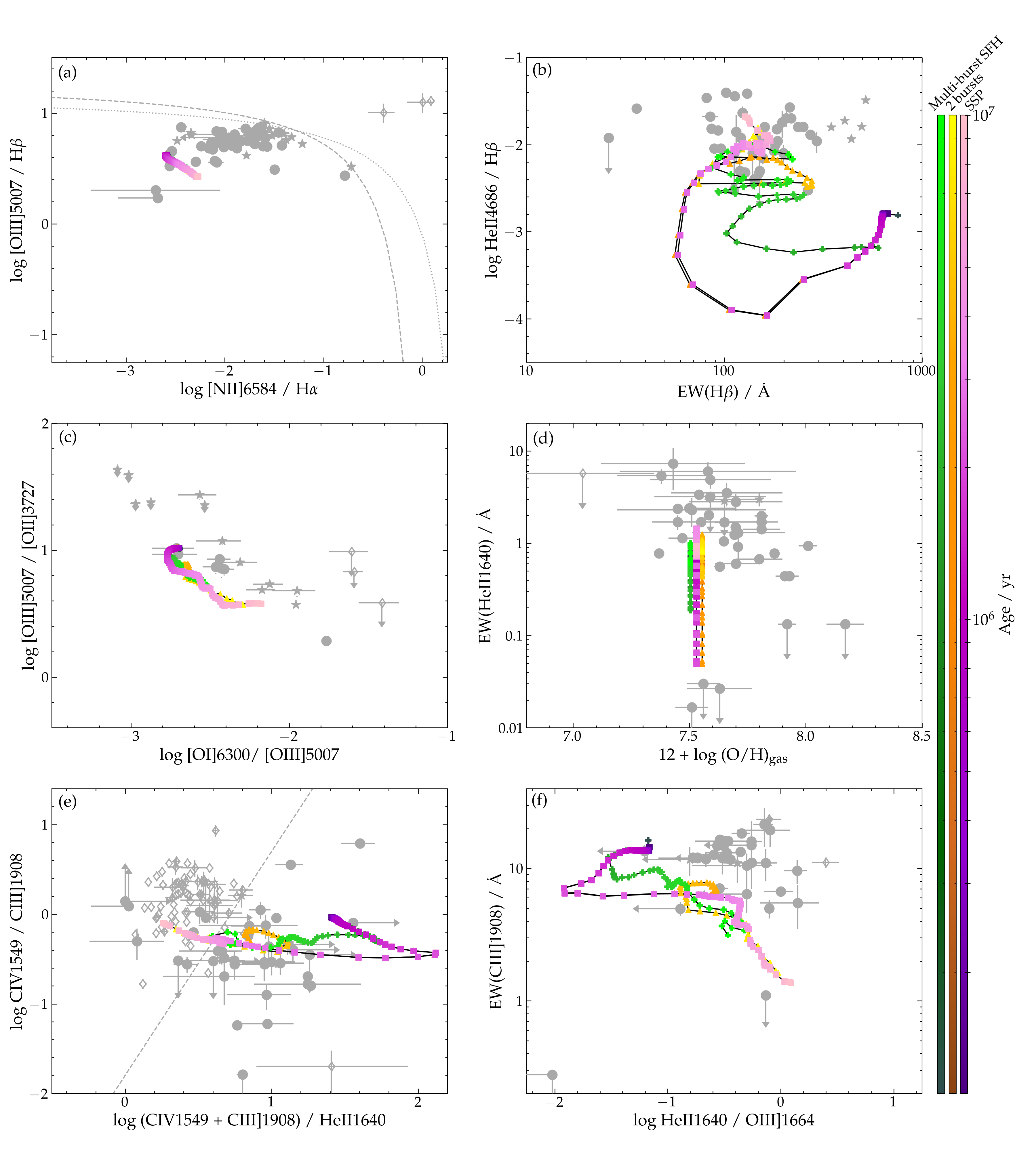}
 \caption{Same as Fig.~\ref{fig:diags_oldmodels}, for three \protect\GLS binary-star models with metallicity $Z=0.001$ and the standard nebular parameters of Table~\ref{tab:cloudy_params}: a plain SSP model (in green, with filled squares); one with a second burst of star formation of same amplitude as the first one added at the age of 3\,Myr (in orange, with filled triangles); and one with the star formation history of a typical galaxy experiencing several bursts of different amplitudes at ages up to 10\,Myr in the \protect\sphinx cosmological simulations of \citet[][in green, with filled crosses; see text for details]{rosdahl2022}.}
\label{fig:diags_sfh}
\end{figure*}

So far, we have considered the emission-line properties of SSPs only, while even the most metal-poor nearby star-forming galaxies show evidence of several episodes of star formation \citep[e.g.,][]{izotov2004, tolstoy2009, senchyna2019}. In Fig.~\ref{fig:diags_sfh}, we show the effect of adding, at the age of 3\,Myr, a second burst of star formation of the same amplitude as the first one to the evolution of an SSP. The sudden input of massive main-sequence stars boosts the rate of ionizing photons at that age, leading to a temporary rise in not only EW(\hb), but also \civ/\ciii and EW(\ciii) (see the difference between the orange and purple curves in Figs~\ref{fig:diags_sfh}b, e and f). As a result, the model probes a different region of the observational space than sampled by a pure SSP, the effect being less noticeable in the other diagrams of Fig.~\ref{fig:diags_sfh}.

We can explore more realistic histories of star formation for metal-poor galaxies by appealing to the \sphinx suite of cosmological (radiation-hydrodynamical) simulations of young galaxies in the epoch of reionization \citep{rosdahl2022}. The green curve in Fig.~\ref{fig:diags_sfh} shows the emission-line properties obtained for a typical galaxy experiencing several bursts of different amplitudes at ages up to 10\,Myr in these simulations. The effects of three major bursts following the onset of star formation are clearly distinguishable through the associated boosts in EW(\hb), \civ/\ciii and EW(\ciii). This enables the model to probe new regions of the observational space not sampled by the SSP and two-burst models.

\begin{figure*}
 \centering
 \includegraphics[width=\textwidth]{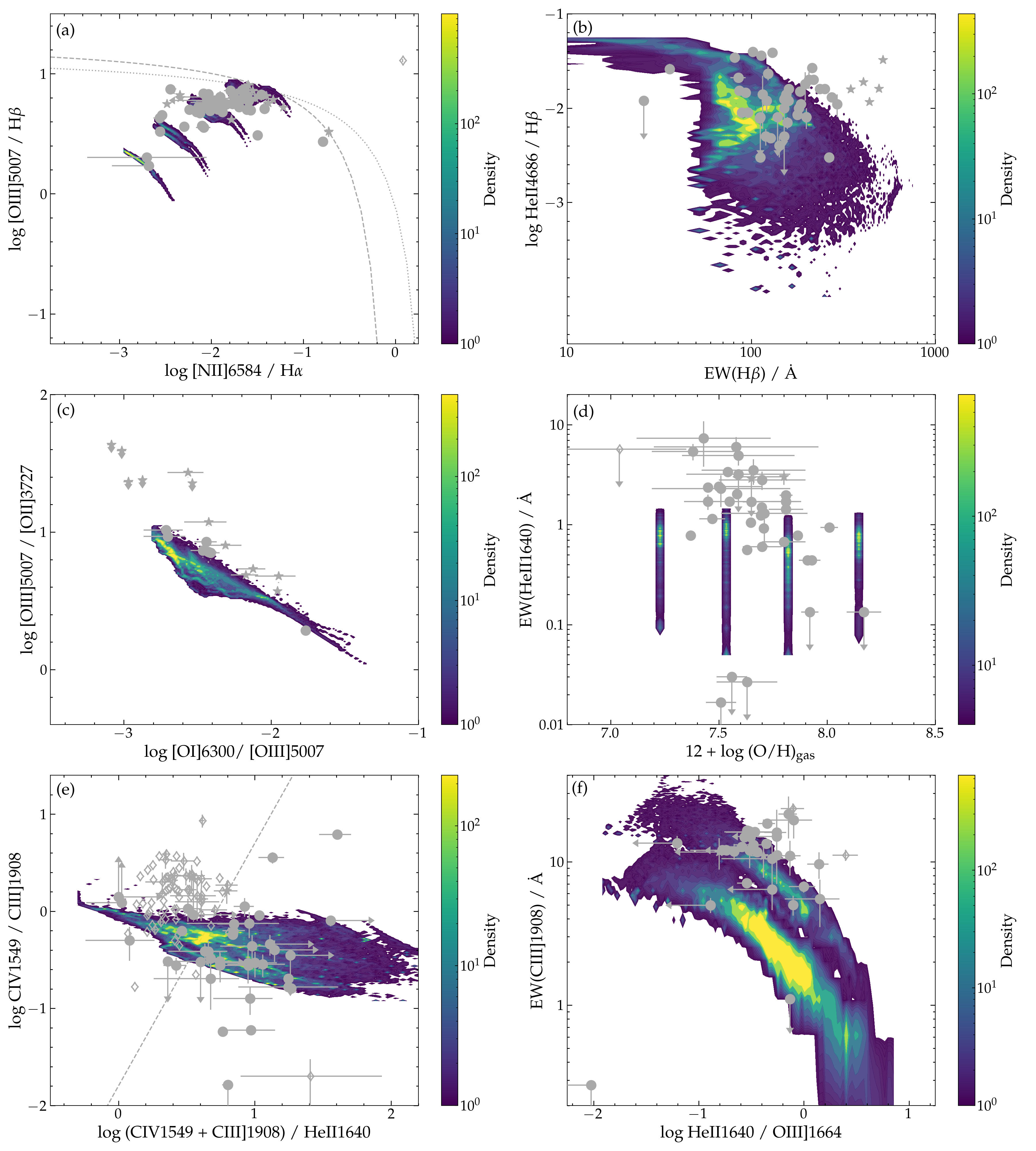}
 \caption{Same as Fig.~\ref{fig:diags_oldmodels}, but for 800 \protect\GLS models computed using the star formation histories of 100 \protect\sphinx galaxies from \citet{rosdahl2022}, combined alternatively with four metallicities ($Z=\zism=0.0005$, 0.001, 0.002 and 0.004) and two \CO abundance ratios (\CO=0.17 and 0.44), at ages between 2 and 10\,Myr (younger models are not shown to avoid the artificial peak in model properties at early ages). The 2-dimensional histograms of models are colour-coded according to the density scale shown on the right.}
\label{fig:contours_blz}
\end{figure*}

To illustrate the observational space sampled by models spanning a wider range of parameters, we show in Fig.~\ref{fig:contours_blz} the emission-line properties of 800 models computed using the star formation histories or 100 \sphinx galaxies, combined alternatively with four metallicities ($Z=\zism=0.0005$, 0.001, 0.002 and 0.004) and two \CO abundance ratios (\CO=0.17 and 0.44), at ages between 2 and 10\,Myr. The aim here is not to fully nor uniformly explore the space of adjustable parameters, but rather to provide global insight into the influence of star formation history, metallicity and \CO ratio on the observables of interest to us. The two (yellow) peaks of the model density distributions appearing in almost all diagrams of Fig.~\ref{fig:contours_blz} correspond to the two values of \CO sampled (see Fig.~\ref{fig:diags_neb}). 

Fig.~\ref{fig:contours_blz} shows that the restricted set of models considered here samples quite widely the observational space. In particular, it is worth emphasizing that the area of the \heiiopt/\hb-versus-EW(\hb) diagram with the highest density of models coincides with that where the bulk of observations lie, which was out of reach of previous models (Fig.~\ref{fig:diags_oldmodels}). Yet, some areas remained out of reach of these models, most notably those populated by galaxies with both largest \heiiopt/\hb and EW(\hb)  (the most extreme of which being classified as LyC leakers; Fig.~\ref{fig:contours_blz}b), \heii equivalent widths in excess of $\sim1.5$\,\AA\ (Fig.~\ref{fig:contours_blz}d), largest \civ/\ciii (Fig.~\ref{fig:contours_blz}e) and both largest EW(\ciii) and \heii/\oiii (Fig.~\ref{fig:contours_blz}f).

\subsection{Top-heavy initial mass function}
\label{sec:add_IMF}

Since the ionizing radiation of a young galaxy is controlled by massive stars, the upper part of the IMF could have a significant impact on the predicted emission-line properties of the models. So far, we have adopted in all models the standard \citet{chabrier2003} IMF truncated at 0.1 and 300\,\Msun, which can lead in binary-star populations to the presence of merged stars with masses up to $\sim600\,\Msun$  (Section~\ref{sec:mod_photion}). To explore the influence of a top-heavy IMF, we adopt the generalised Rosin-Rammler distribution function \citep[e.g.,][]{chabrier2003, wise2012, goswami2022},
\begin{equation}
    \label{eq:imf_wise}
    \phi(m) \propto m^{-2.3} \exp\left[- \left( \frac{\mchar}{m} \right) ^{1.6}\right]\,,
\end{equation}
which approaches the standard \citet{chabrier2003} IMF at large $m$ but is exponentially cut off at $m<\mchar$. \citet{wise2012} used this function with $\mchar=40\,\Msun$ to explore the formation of Population~III stars in early dwarf galaxies, while \citet{goswami2022} adopted $200\leq \mchar\leq300\,\Msun$ to explore chemical enrichment by PISNe in extremely metal-poor galaxies. Here, we consider values of \mchar in the range 50--100\,\Msun to explore the emission-line signatures of stellar populations strongly dominated by massive stars. 

\begin{figure}
 \centering
 \includegraphics[width=0.95\columnwidth,trim= 10 5 0 0,clip]{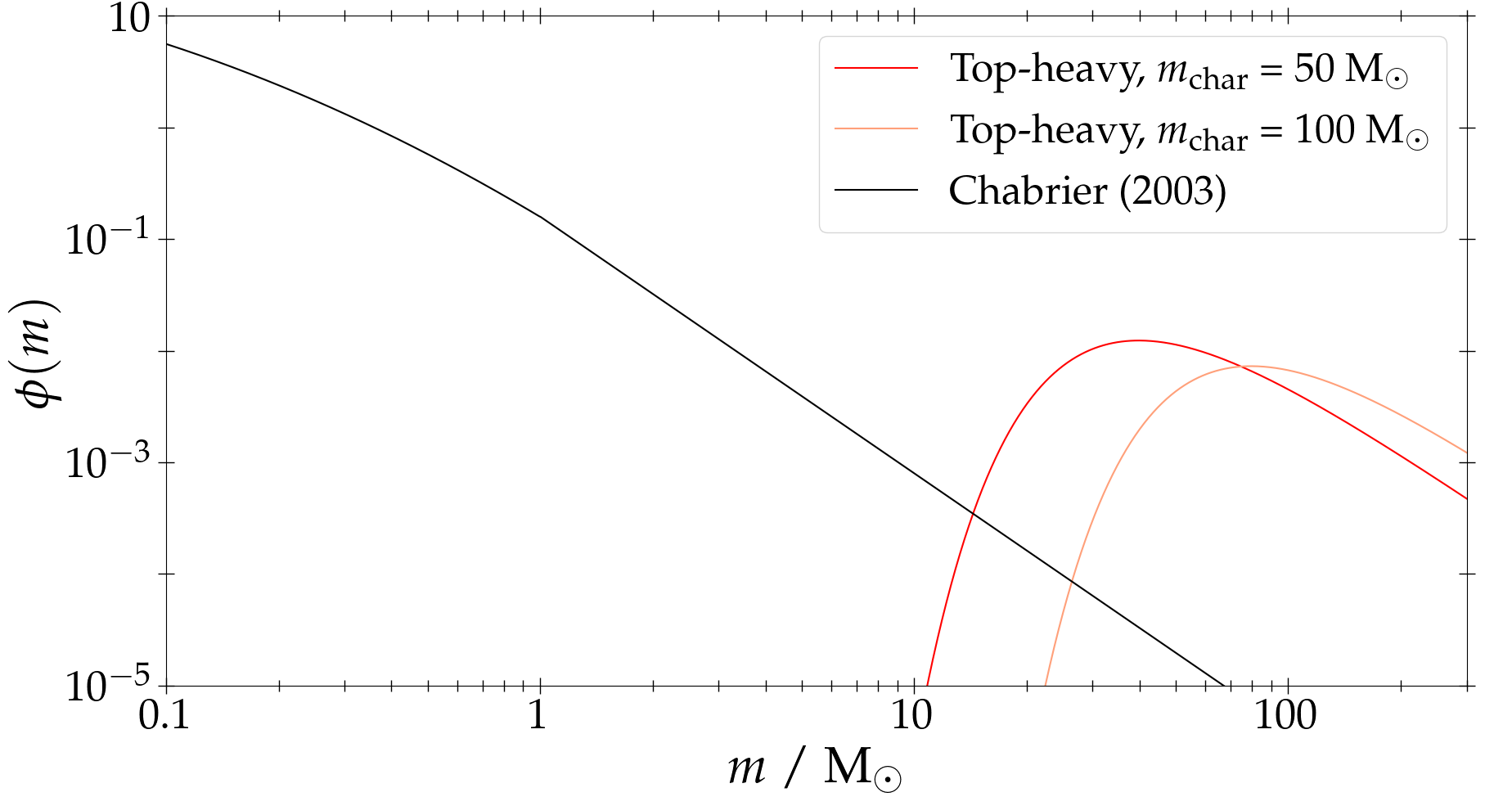}
 \\\includegraphics[width=0.95\columnwidth,trim= 10 5 0 0,clip]{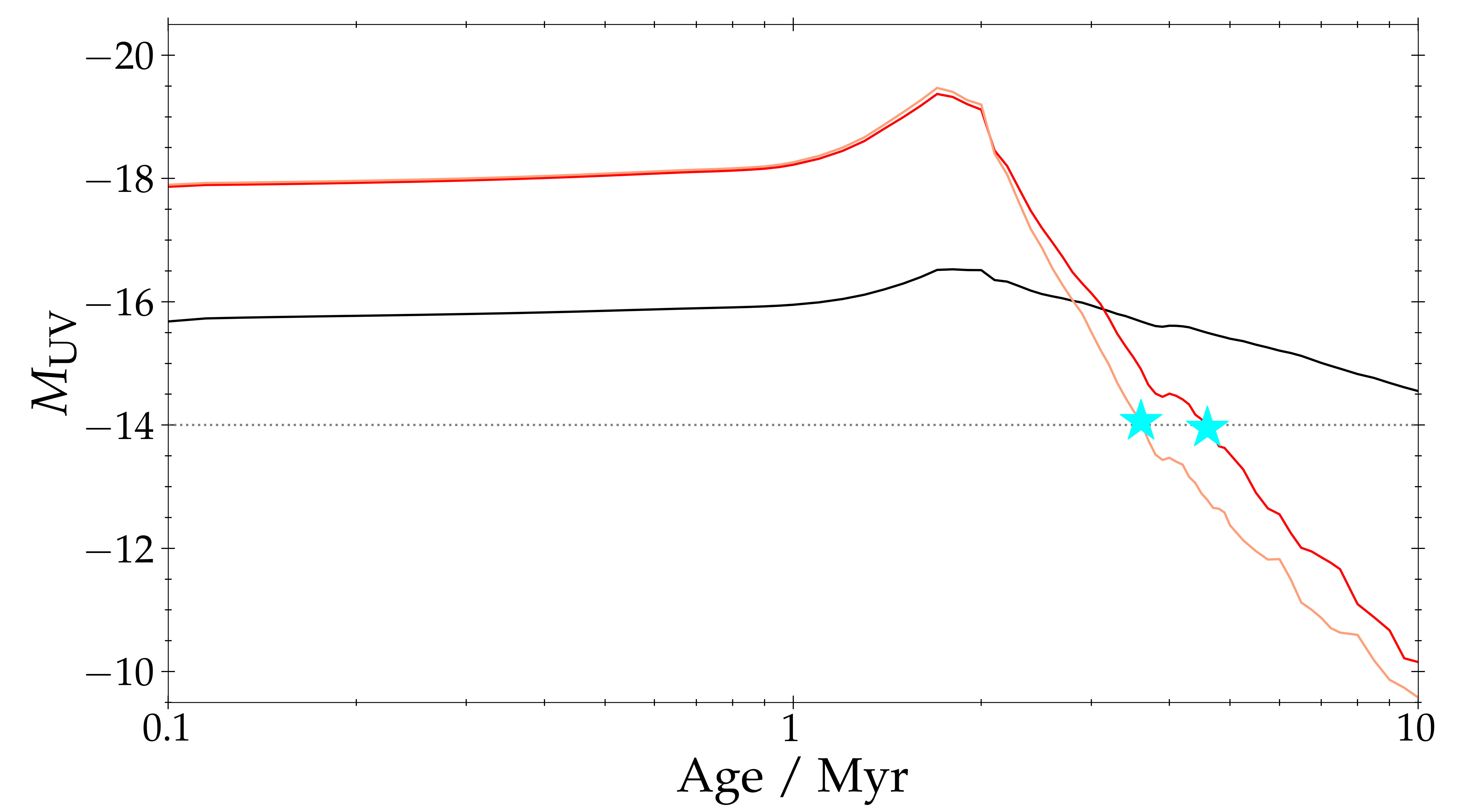}
 \caption{Top: top-heavy IMFs computed using equation~\eqref{eq:imf_wise} for two values of \mchar (red: $50\,\Msun$; salmon: 100\,\Msun), compared to the standard \citet{chabrier2003} IMF (in black). All IMFs are normalized to a total mass in stars of 1\,\Msun. Bottom: absolute ultraviolet AB magnitude at 1500\,\AA, \Muv, plotted against age for three \protect\GLS binary-star models computed with the three IMFs shown in the top panel, assuming a total initial mass in stars of $10^6\,\Msun$, metallicity $Z=0.001$ and the standard nebular parameters of Table~\ref{tab:cloudy_params}. For reference, the dotted horizontal line marks the magnitude $\Muv=-14\,$mag typical of nearby, extremely metal-poor star-forming galaxies (reached by the top-heavy-IMF models at the ages marked by the cyan stars).}
\label{fig:imf_wise}
\end{figure}

The upper panel of Fig.~\ref{fig:imf_wise} shows two versions of the top-heavy IMF of equation~\eqref{eq:imf_wise}, for $\mchar=50\,\Msun$ and 100\,\Msun, compared to the standard \citet{chabrier2003} IMF. In the lower panel, we show the evolution of the absolute AB magnitude at 1500\,\AA, \Muv, of \protect\GLS binary-star models computed with these three IMFs, assuming a total initial mass in stars of $10^6\,\Msun$, metallicity $Z=0.001$ and the standard nebular parameters of Table~\ref{tab:cloudy_params}. Since most of the mass in the models with top-heavy IMFs is concentrated in massive stars with low mass-to-light ratios, at fixed total initial stellar mass, these models are about 2\,mag brighter in \Muv than that with a standard \citet{chabrier2003} IMF at early ages. The brightening in all models at ages before 2\,Myr arises in part from stellar evolution on the main sequence, and in part from stellar mergers. Then, once the bulk of stars in the top-heavy IMF models leave the main sequence and complete their evolution, the population fades rapidly. Instead, the ultraviolet luminosity decreases more gradually in the model with standard \citet{chabrier2003} IMF, as stars of lower and lower mass progressively leave the main sequence. 

\begin{figure*}
 \centering
 \includegraphics[width=\textwidth]{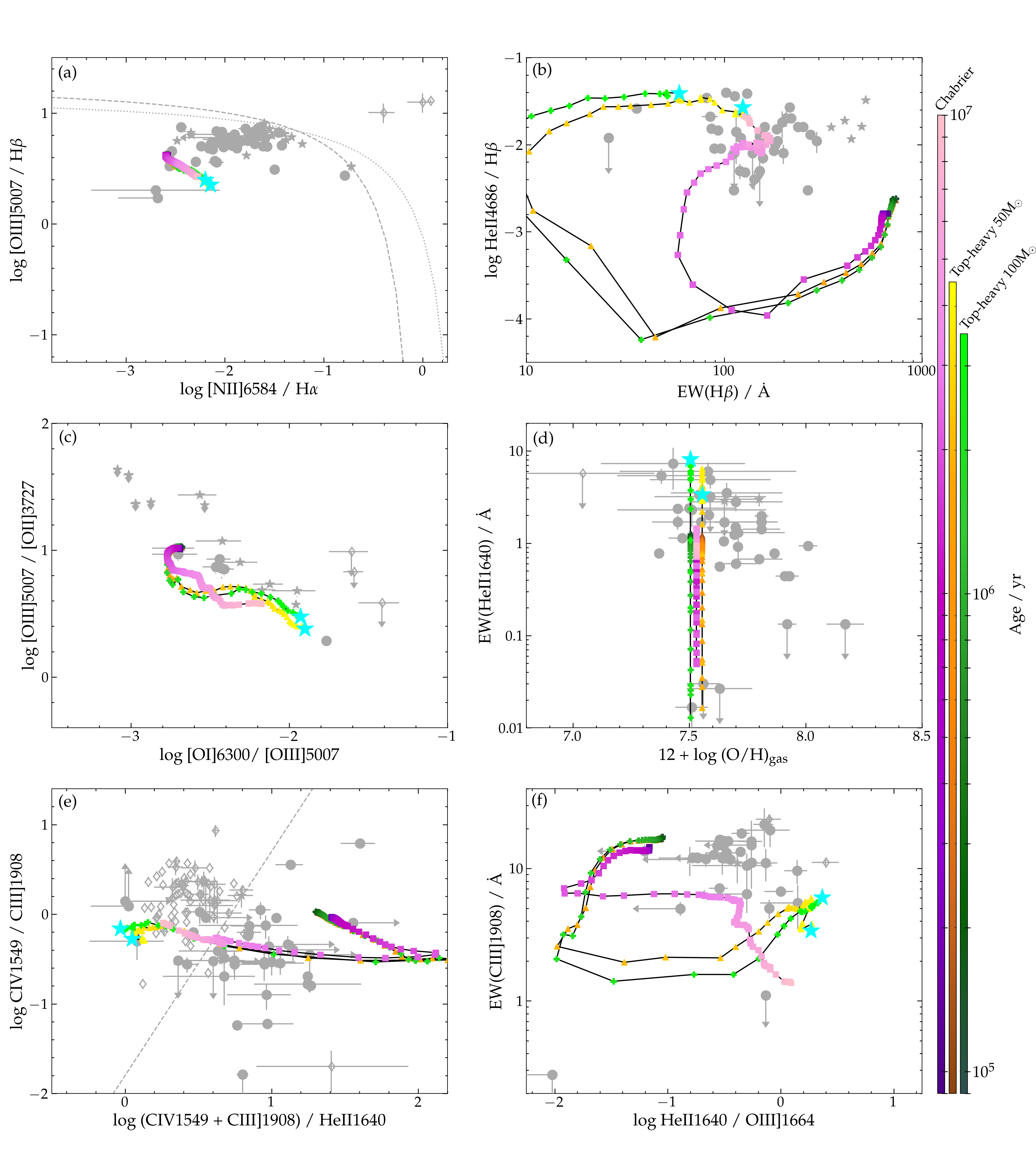}
 \caption{Same as Fig.~\ref{fig:diags_oldmodels}, for the three \protect\GLS binary-star models with different IMFs shown in the lower panel of Fig.~\ref{fig:imf_wise} (colour-coded as indicated on the right). The curves interspersed with filled squares, triangles and crosses refer to the models with a \citet{chabrier2003} IMF and two top-heavy IMFs (equation~\ref{eq:imf_wise}) with characteristic masses of $\mchar=50$ and 100\,\Msun, respectively. Cyan stars mark the points beyond which the models with top-heavy IMFs become fainter than $\Muv=-14\,$mag and are not shown (see text and Fig.~\ref{fig:imf_wise} for details).}
\label{fig:diags_wise}
\end{figure*}


In Fig.~\ref{fig:diags_wise}, we present the emission-line properties of the three models shown in the lower panel of Fig.~\ref{fig:imf_wise}. To avoid cluttering the diagrams, we do not show the properties of models fainter than $\Muv=-14\,$mag (limit marked by a cyan star), corresponding roughly to the typical luminosity of nearby, extremely metal-poor star-forming galaxies \citep[see, e.g., table~1 of][]{senchyna2019}. At early ages, when massive stars on the upper main sequence dominate the emission, the three models overlap in all diagrams. Once the bulk of stars in the top-heavy IMF models leave the main sequence to become red super-giant stars, at ages around 2\,Myr, the implied strong drop in ionizing-photon rate and rise in 4861-\AA\ continuum emission make the \hb equivalent width decline more abruptly than in the standard model (Fig.~\ref{fig:diags_wise}b). Meanwhile, after reaching a minimum, the \heiiopt/\hb ratio (Fig.~\ref{fig:diags_wise}b) and \heii equivalent width (Fig.~\ref{fig:diags_wise}d) promptly rise again due to the emergence of hot, pure-He products of massive stars (Section~\ref{sec:mod_photion}). In Fig.~\ref{fig:diags_wise}(f), the \ciii equivalent width shows similar behaviour to EW(\heii) relative to the standard model. EW(\hb) takes longer than EW(\heii) and EW(\ciii) to increase again (Fig.~\ref{fig:diags_wise}b), as red super-giant stars have stronger continuum emission at 4861\,\AA\ than at ultraviolet wavelengths. 

Remarkably, Fig.~\ref{fig:diags_wise} shows that, in diagrams involving ultraviolet lines, the two \protect\GLS binary-star SSP models with top-heavy IMFs, fixed metallicity $Z=0.001$ and $\CO=0.17$ investigated here reach regions of the observational space not sampled by the more complete grid of models with standard \citet{chabrier2003} IMF considered in Fig.~\ref{fig:contours_blz} (up to \heii equivalent withs of $\sim10\,$\AA; Fig.~\ref{fig:diags_wise}b). Although these SSP models spend less than a few Myr with such extreme emission-line properties, they can potentially be combined with more complex star formation histories. In fact, they could represent a transient phase in the evolution of a galaxy, when a cluster of massive stars happens to form and dominates the light for a while.

\subsection{Emission from accretion discs of X-Ray binaries}
\label{sec:add_XRBs}

Several recent studies have suggested that the unusually strong \heii emission observed in the spectra of many metal-poor star-forming galaxies may originate from accretion discs of X-ray binaries, where a compact object (neutron star or stellar-mass black hole) accretes material from a companion \citep[e.g.,][]{schaerer2019,umeda2022,katz2023}. Emission from the hot accretion discs of these systems peaks at X-ray energies \citep[e.g.,][]{mitsuda1984, fragos2013, mirocha2014, senchyna2019}. In detail, the accretion rate depends on the evolutionary stage and chemical evolution of the donor (companion) star, the mass ratio between the donor and the compact object and their orbital separation.


In this context, the \protect\GLS model, which follows the evolution of the physical properties and the emission from stars in binary systems, provides an ideal framework to compute in a self-consistent way the contribution by accretion discs of XRBs to the integrated light of stellar populations. In fact, in Appendix~\ref{app:XRBs_calc}, we present a simple model to compute the spectral energy distribution of XRB accretion discs in \protect\GLS binary-star populations, based on a library of multicolour-disc spectra including potential Compton upward scattering of soft disc photons by energetic coronal electrons  \citep{mirocha2014, steiner2009}. This model, which also accounts for X-ray variability \citep[e.g.,][]{tanaka1996, chen1997}, provides good agreement with the observed X-ray luminosities of various samples of nearby, metal-poor star-forming galaxies \citep{douna2015, brorby2016, lehmer2019}, as well as with the average X-ray luminosity function of XRBs in the five most metal-poor star-forming galaxies observed by \citet{lehmer2019}. 

\begin{figure}
 \includegraphics[width=\columnwidth]{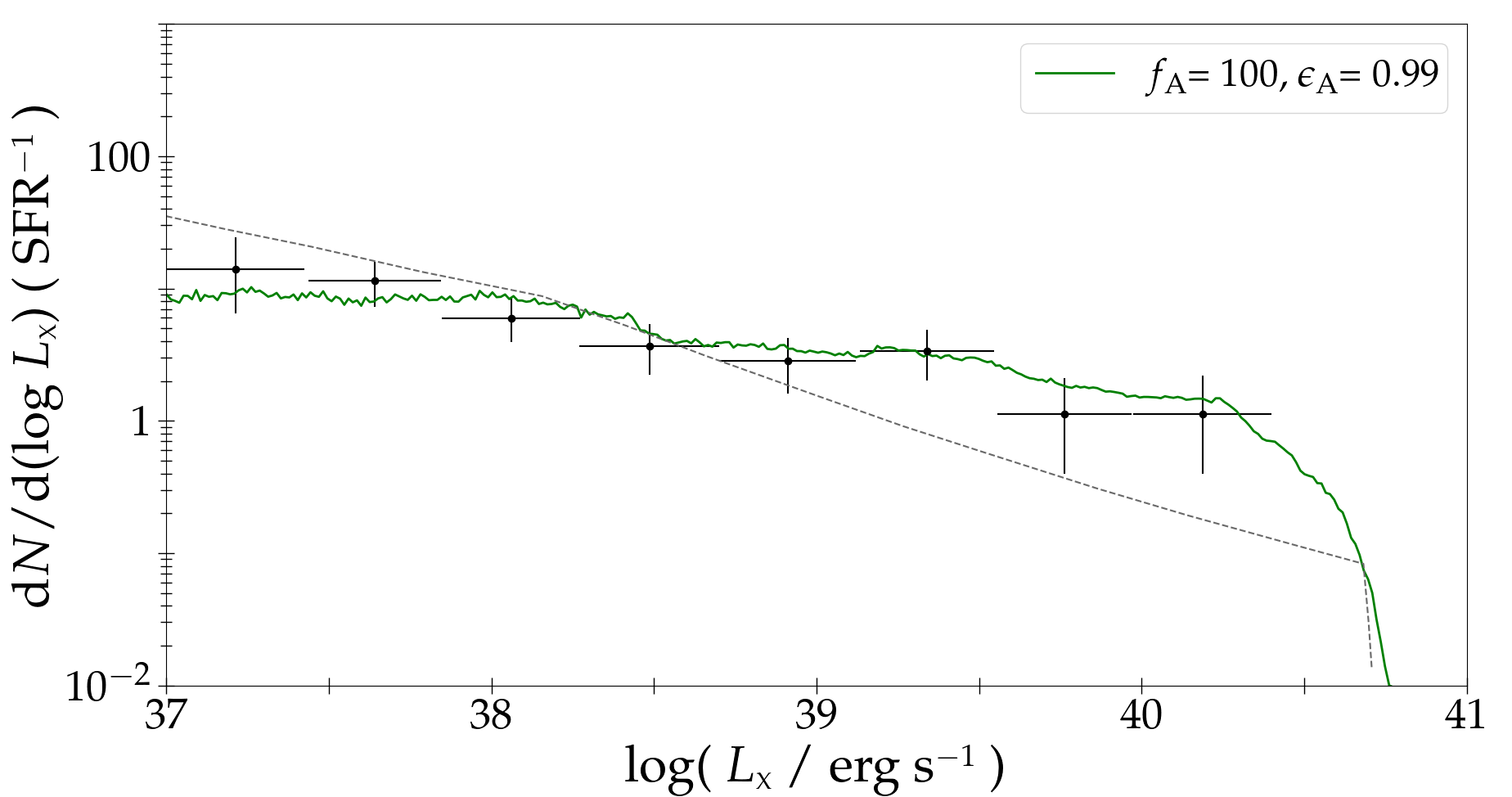}
 \caption{X-ray luminosity function (in the 0.5--8\,keV energy band) of XRBs for a \protect\GLS binary-star population of metallicity $Z=0.008$ with binary fraction $\fbin=0.7$, assuming 100\,Myr of constant star formation at the rate $\psi=1\Msun\,\mathrm{yr}^{-1}$. Accretion is assumed to proceed according to the two-phase model described by equation~\eqref{eq:twophase_ap}, with $\fa=100$ and $\ea=0.99$. Also shown for comparison is the average luminosity function derived from {\it Chandra} observations of the five most metal-poor star-forming galaxies in the sample studied by  \citet[][data points with error bars]{lehmer2019}. The black dotted line shows an empirical model proposed by these authors (see Section~\ref{sec:add_XRBs} for details).}
  \label{fig:LF_opt}
\end{figure}

As shown in Appendix~\ref{app:XRBs_calc}, such agreement with observations can be obtained with models in which most of the mass is accreted during episodes of X-ray outbursts, at a much higher rate than during the quiescent phase, the actual strength and duration of outbursts being degenerate. Fig.~\ref{fig:LF_opt} shows the X-ray luminosity function of XRBs computed assuming that 99 per cent of the mass is accreted by the compact object during outbursts at a rate 100 greater than in the quiescent phase, for a \protect\GLS binary-star population corresponding to 100\,Myr of constant star formation at the rate $\psi=1\Msun\,\mathrm{yr}^{-1}$. The model has a binary-star fraction $\fbin=0.7$ \citep[typical of massive-star populations; e.g.,][]{sana2012} and the metallicity $Z=0.008$, close to the typical metallicity of the five metal-poor galaxies for which \citet{lehmer2019} derived the average luminosity function shown by the black data points with error bars, which is well reproduced by the model. In Appendix~\ref{app:XRBs_calc}, we further show that metallicity has only a minor influence on the predicted X-ray luminosity function (Fig.~\ref{fig:LF_met}).

\begin{figure}
\centering
 \includegraphics[width=\columnwidth,trim= 0 0 0 0,clip]{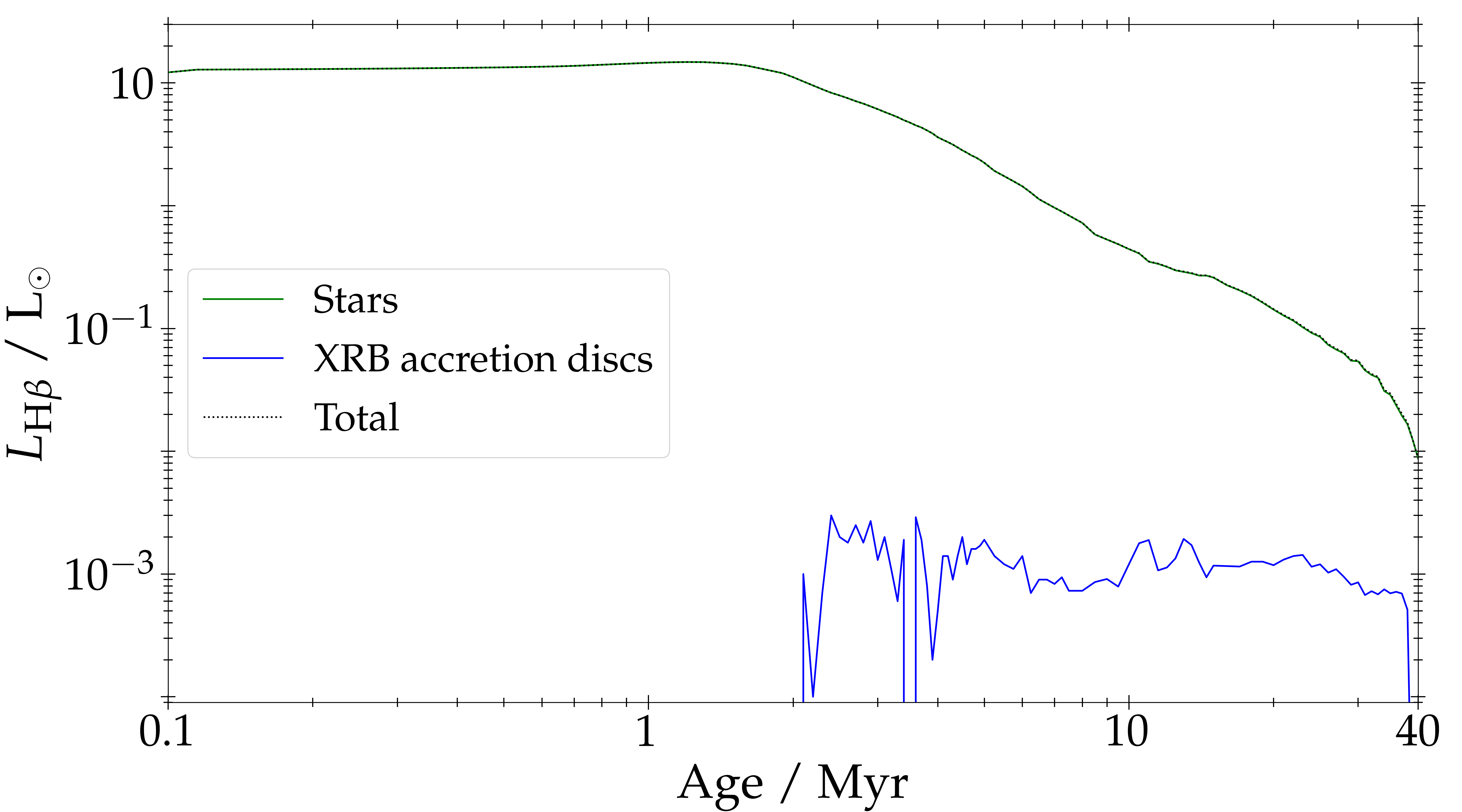}
 \\\includegraphics[width=\columnwidth,trim= 0 0 0 0,clip]{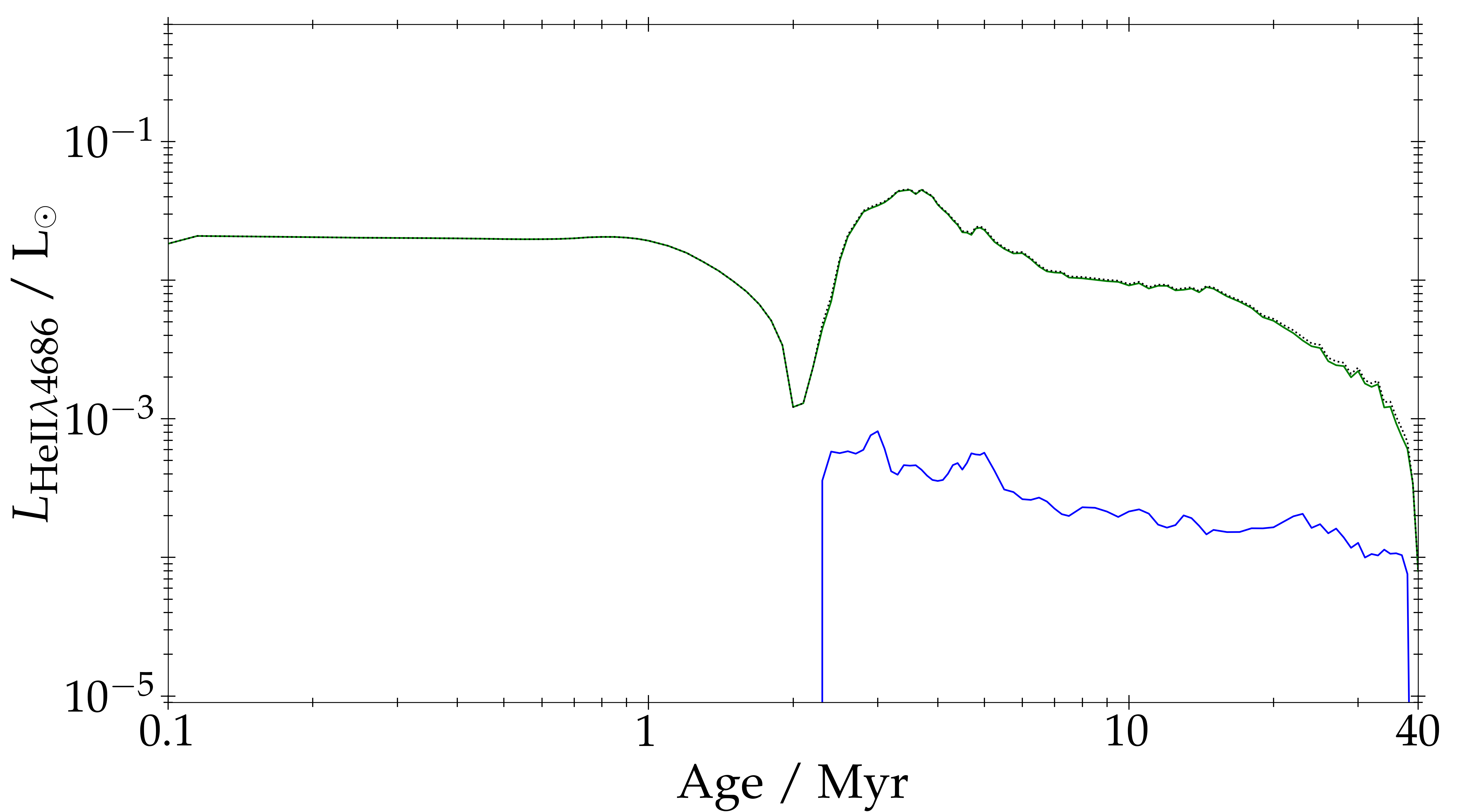}
 \caption{Contribution by XRB accretion discs (in blue), compared to the stellar contribution (in green), to the total luminosities (in black) of \hb (top panel) and \heiiopt (bottom panel) plotted against age, for a \protect\GLS binary-star SSP with metallicity $Z=0.001$. The model has the same zero-age \citet{chabrier2003} IMF normalized to a total initial stellar mass of 1\,\Msun integrated over 0.1--300\,\Msun as in Fig.~\ref{fig:comp_partbin}.}
\label{fig:lum_XRBs}
\end{figure}

In Fig.~\ref{fig:lum_XRBs}, we show the contribution by XRB accretion discs to the total \hb and \heiiopt luminosities predicted by such a model for a \protect\GLS binary-star population of standard metallicity $Z=0.001$. The population of XRBs grows with time, as black holes and neutron stars form. Yet, the contribution by ionizing radiation from XRB accretion discs to the total line luminosities remains negligible, reaching at most a few percent at ages greater than about 30\,Myr. We conclude that, based on the XRB luminosity function observed in nearby galaxies, XRB accretion discs are unlikely to contribute significantly to the strong \lheii emission from metal-poor star-forming galaxies in the sample of Section~\ref{sec:res_sample}.

\subsection{Emission from radiative shocks}
\label{sec:add_shocks}

Radiative shocks, such as those expected from massive-star winds and SN blast waves, have spectral signatures characterized by strong high-ionization lines \citep[e.g.,][]{dopita1996, alarie2019}, similar to those observed in some metal-poor star-forming galaxies \citep{izotov2012, plat2019}. As noted by \cite{plat2019}, the high gas densities  ($\nh\gtrsim10^4$\,cm$^{-3}$) measured from the \ciiid\ doublet in some metal-poor star-forming galaxies could be suggestive of the presence of radiative shocks from SNe. So far, however, no predictive phenomenological model has been proposed to link shocks to other galaxy properties. 

In Appendix~\ref{app:rad_shocks}, we propose a simple prescription to compute the emission from radiative shocks generated by stellar winds and SN blast waves consistently with the emission from stars and photoionized \hii regions in young star-forming galaxies. This prescription allows line luminosities to be computed from knowledge of the rates of energy injection into the ISM by stellar winds and SN explosions, based on line fluxes tabulated as a function of shock velocity ($\Vs$) and preshock density (\nh) in grids of radiative-shock models \citep[e.g.,][]{dopita1996, alarie2019}. We estimate the rate at which stellar winds inject energy into the ISM by summing the individual contributions of all stars (ignoring for simplicity the influence of binary-star processes on this rate; see Appendix~\ref{app:rad_shocks}). For energy injection by SNe, we consider the rate of SN explosions predicted by \sevn, including pair-instability, type-II and type-Ia SNe. We ignore non-exploding, `failed' SNe leading to the formation of black holes \citep{spera2015, spera2017}. For reference, the rates of energy injection computed in this way are roughly consistent with those reported by \citet[][see their fig.~6]{leitherer1992} under slightly different assumptions.

\begin{figure}
\centering
 \includegraphics[width=\columnwidth,trim= 0 0 0 0,clip]{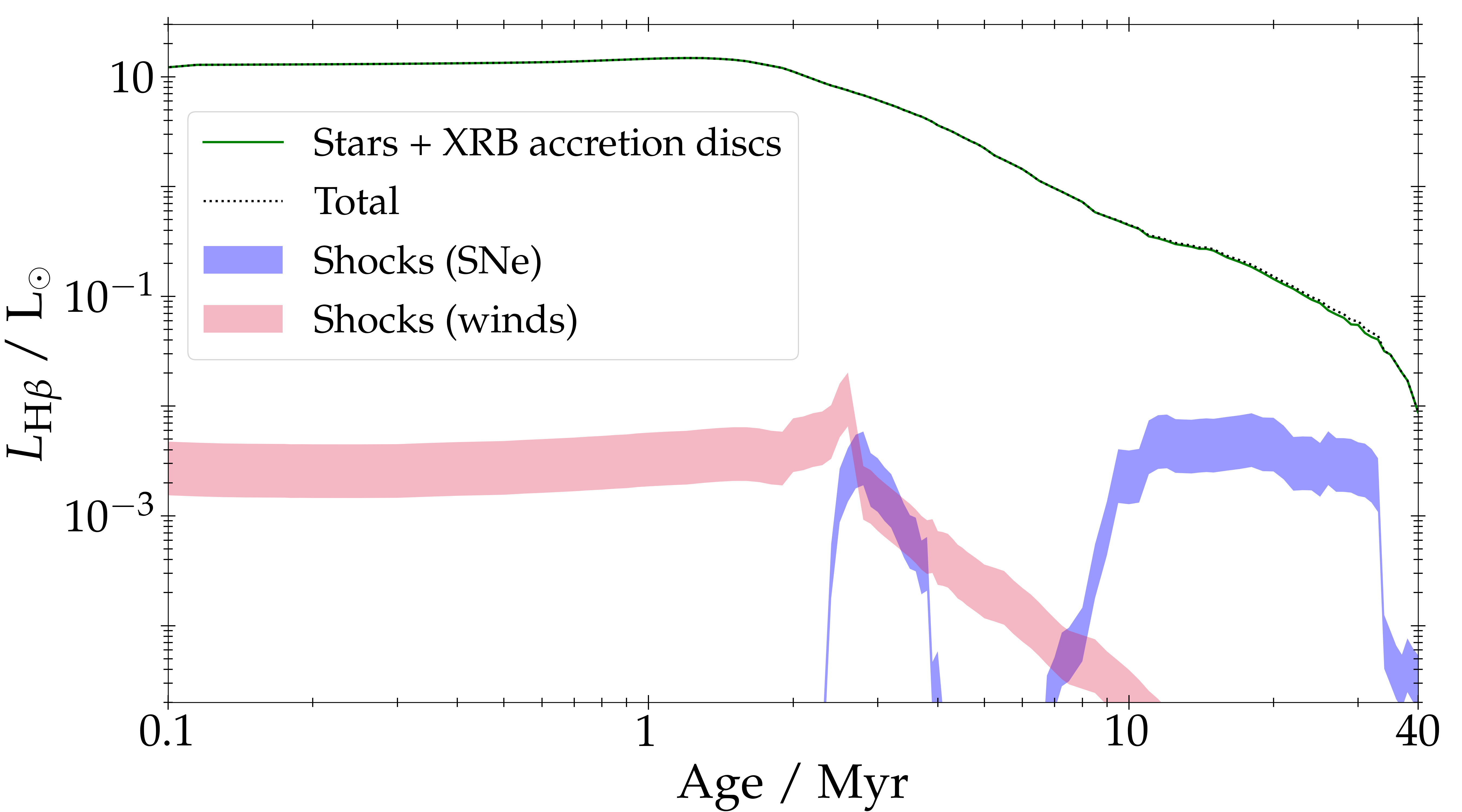}
 \\\includegraphics[width=\columnwidth,trim= 0 0 0 0,clip]{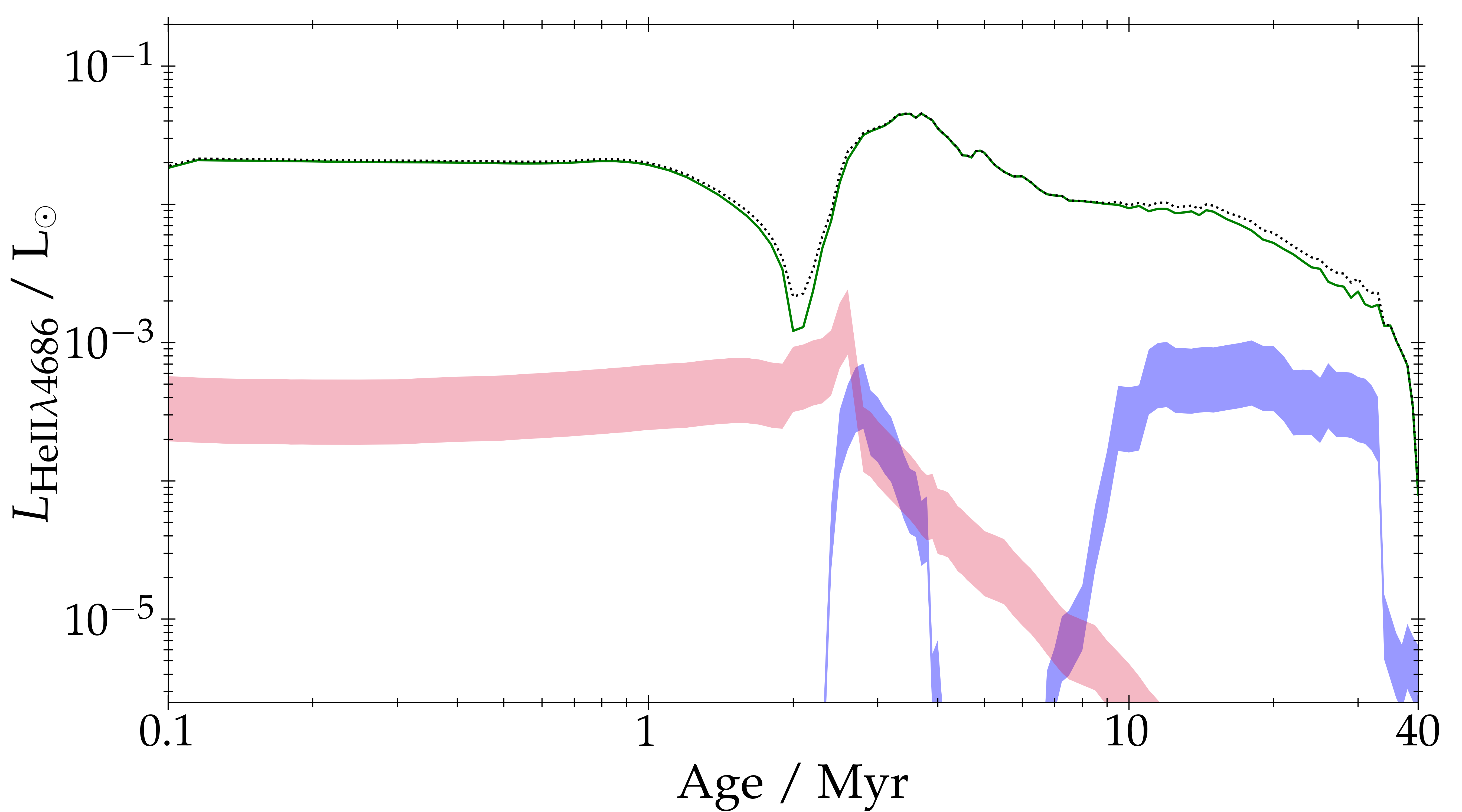}
 \\\includegraphics[width=\columnwidth,trim= 0 0 0 0,clip]{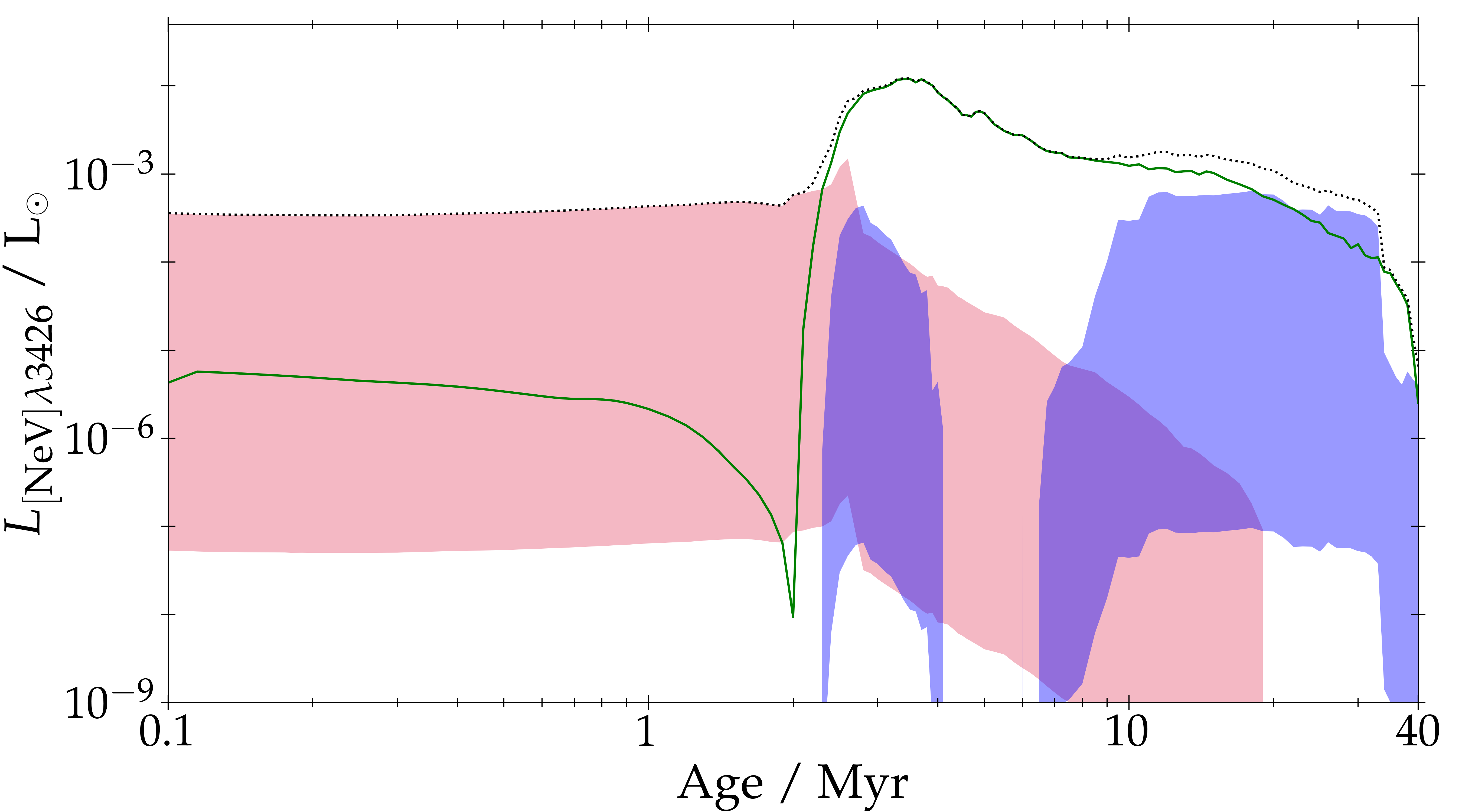}
 \caption{Contribution from shocks driven by stellar winds (in red) and SNe (in blue) to the total \hb (top panel), \heiiopt (middle panel) and \nevopt (bottom panel) luminosities of a \protect\GLS binary-star SSP with metallicity $Z=0.001$, plotted against age (see text Section~\ref{sec:add_shocks} for details). The contributions are shown for the range of shock velocities $100 \leq \Vs \leq 1000\,\mathrm{km\,s}^{-1}$ and the density $\nh=100\,\mathrm{cm}^{-3}$. In each panel, the black dotted line shows the total line emission obtained by adding the maximum possible shock contribution (independently for stellar winds and SNe) to the emission from stars and XRB accretion discs (green solid line). The model has the same zero-age \citet{chabrier2003} IMF normalized to a total initial stellar mass of 1\,\Msun integrated over 0.1--300\,\Msun as in Fig.~\ref{fig:comp_partbin}.}
\label{fig:lum_shocks}
\end{figure}

The top-two panels of Fig.~\ref{fig:lum_shocks} show the resulting contributions of radiative shocks driven by stellar winds and SN explosions to the total \hb and \heiiopt luminosities of a \GLS binary-star population with standard metallicity $Z=0.001$. These contributions, obtained by inserting line fluxes from the recent radiative-shock calculations of \citet[][based on \mappings]{alarie2019} into equations~\eqref{eq:lhb_winds}--\eqref{eq:lxalarie}, are shown for the full available range of shock velocities, $100 \leq \Vs \leq 1000\,\mathrm{km\,s}^{-1}$, and $\nh=100\,\mathrm{cm}^{-3}$ (Table~\ref{tab:cloudy_params}). In each panel, the black dotted line shows the total line emission obtained by adding the maximum possible shock contribution (independently for stellar winds and SNe) to the emission from stars and XRB accretion discs (green solid line).

Fig.~\ref{fig:lum_shocks} shows that radiative shocks in this model account at most for only a few per cent of the total \hb and \heiiopt line emission at ages below 10\,Myr, with no significant impact on the  global \heiiopt/\hb ratio, except for an extremely brief period around 2\,Myr, when shocks from stellar winds can produce around 40 per cent of the total \heii emission. At ages beyond 10\,Myr, the shock contribution increases as the rates of ionizing photons produced by the stars begin to decrease, but it remains with no significant impact on the line luminosities and emission-line ratios. The two peaks in SN-shock contribution originate from PISNe at ages around 4\,Myr, and type-II SNe at ages around 10\,Myr, type-Ia SNe appearing only after about 30\,Myr and in small number. We note that adopting energies greater than $10^{51}\,\mathrm{erg}$ for the most massive PISNe (footnote~\ref{foot:pisn}) would increase the magnitude of the first peak, but only for a brief spike at the earliest ages. 

It is instructive to also examine the predictions of our model for the luminosity of the \nevopt emission line, which has been suggested as a potential signature of radiative shocks in blue compact-dwarf galaxies \citep[e.g.,][]{izotov2012}. The bottom panel of Fig.~\ref{fig:lum_shocks} shows that, indeed, radiative shocks driven by stellar winds and SN explosions can significantly contribute to, and even dominate, \nevopt emission in metal-poor star-forming galaxies.

Overall, for what concerns the direct focus of the present study, we conclude from Fig.~\ref{fig:lum_shocks} that radiative shocks driven by stellar winds and SN explosions are unlikely to contribute prominently to the strong \lheii emission observed in metal-poor star-forming galaxies in the sample of Section~\ref{sec:res_sample}.

\section{Discussion}
\label{sec:dsc}

In Sections~\ref{sec:res} and \ref{sec:adds} above, we have seen that the new \GLS spectral-evolution model of binary-star populations presented in Section~\ref{sec:models} predicts significantly harder ionizing radiation than previous models of single- and binary-star populations, making it possible to reproduce, in particular, the high \heiiopt/\hb ratios and large \hb equivalent widths commonly observed in metal-poor star-forming galaxies. The most extreme \heii emitters can be accounted for by invoking populations entirely dominated by massive stars (Section~\ref{sec:add_IMF}). The model also allows us to compute, in a physically consistent way, the contribution by accretion discs of X-ray binaries to the ionizing radiation of a star-forming galaxy (Section~\ref{sec:add_XRBs}). Interestingly, we find that reproducing the observed luminosity function of X-ray binaries in nearby, metal-poor star-forming galaxies implies that XRB accretion discs must have, on average, a negligible impact on the predicted luminosities of high-ionization lines. While consistent with the results from several previous studies \citep[e.g.,][]{jaskot2013, senchyna2017, senchyna2020, saxena2020}, this finding contrasts with some recent claims \citep{schaerer2019, umeda2022, katz2023}, which we now examine in more detail. 

\citet{schaerer2019} used an observed anti-correlation between X-ray luminosity per unit star formation rate (\Ltotx/SFR) and metallicity \citep{douna2015, brorby2016}, combined with an estimate of the rate of \lheii- ionizing photons (\Nheiidot) per unit  \Ltotx in the metal-poor star-forming galaxy I\,Zw\,18, to conclude that XRBs are likely to be the main source of \lheii emission in metal-poor star-forming galaxies. The founding assumption of this reasoning, that \lheii emission in I\,Zw\,18 is produced by XRBs, has since been ruled out by \citet{kehrig2021}, who find that the high-mass binary dominating the X-ray emission lies about 200\,pc away from the \lheii-emission peak. Also, the prediction by \citet{schaerer2019} that the \heiiopt/\hb line-luminosity ratio should correlate with \Ltotx/SFR does not appear to be confirmed observationally by the sample of 11 metal-poor star-forming galaxies with high-quality constraints on both X-ray and \lheii emission studied by \citet[][see their fig.~9]{senchyna2020}. 

In fact, by arbitrarily scaling \ares multicolour-disc models for a wide range of black-hole masses \citep{mirocha2014} added to \bpass ionizing spectra of binary-star populations with constant star formation, \cite{senchyna2020} find that \Ltotx/SFR in excess of $10^{42}$\,erg\,s$^{-1}/\Msun\,$yr$^{-1}$ is required for XRBs to significantly boost \lheii emission. This is 10--100 times larger than observed in the 11 metal-poor star-forming galaxies of their sample, suggesting that XRBs are not primarily responsible for \lheii emission in these galaxies. Similarly, \citet{saxena2020} find no significant difference in the \Ltotx/SFR ratios of galaxies with and without \lheii emission at redshift $z\sim3$ in the Chandra Deep Field South. These results are consistent with our finding that accretion discs of XRBs negligibly affect \lheii emission (Fig.~\ref{fig:comp_partbin}).


\citet{katz2023} also explore the emission-line properties of gas photoionized by sources with different \Ltotx/SFR ratio by 
arbitrarily scaling an \ares multicolour-disc model, for a fixed black-hole mass of 25\,\Msun, added to the \bpass ionizing spectrum of a 5\,Myr-old binary-star SSP.\footnote{To compute \Ltotx/SFR, \citet{katz2023} define an `effective' star formation rate based on the rate of H-ionizing photons emitted by the 5\,Myr-old \bpass SSP (H.~Katz, private communication).} They find that reproducing the surprisingly high \oiiioptc/\oiiiopt ratio of 0.048 (0.055 when dust-corrected) in the \JWST/NIRSpec spectrum of the metal-poor star-forming galaxy S04590 at redshift $z=8.5$ requires  $L_{\mathrm{X,2-10 keV}}^\mathrm{tot}/\mathrm{SFR}>10^{41}$\,erg\,s$^{-1}/\Msun\,$yr$^{-1}$, which translates into $\Ltotx/\mathrm{SFR}>3\times10^{41}$\,erg\,s$^{-1}/\Msun\,$yr$^{-1}$ when adopting the energy band 0.5--8\,keV used by \citet{senchyna2020} and in Fig.~\ref{fig:brorby_met} of the present paper. Again, this is much larger than observed in extremely metal-poor star-forming galaxies \citep{senchyna2017, senchyna2020}, and than predicted by our physically consistent model of the emission from stars and XRB accretion discs, anchored on the average observed XRB luminosity function of Fig.~\ref{fig:LF_opt}. 

It is important to bear in mind that the simple prescription presented in Section~\ref{sec:add_XRBs} pertains to the average $\langle\Ltotx/\mathrm{SFR}\rangle\approx2.1\times10^{40}$\,erg\,s$^{-1}/\Msun\,$yr$^{-1}$ produced by XRBs in the five metal-poor star-forming galaxies considered by \citet{lehmer2019} to build the X-ray luminosity function of Fig.~\ref{fig:LF_opt}, and that the associated standard deviation of $1.4\times10^{40}$\,erg\,s$^{-1}/\Msun\,$yr$^{-1}$ implies that this ratio can vary from galaxy to galaxy. The 10 star-forming galaxies with $\logoh<8.0$ in the sample of \citet[][see Fig.~\ref{fig:brorby_met} below]{douna2015} exhibit a very similar mean $\langle\Ltotx/\mathrm{SFR}\rangle\approx2.7\times10^{40}$\,erg\,s$^{-1}/\Msun\,$yr$^{-1}$ with a standard deviation of $2.4\times10^{40}$\,erg\,s$^{-1}/\Msun\,$yr$^{-1}$. Yet, the X-ray luminosity per unit star formation rate required by \citet{katz2023} to significantly impact the \oiiioptc/\oiiiopt ratio of S04590 appears improbably high, as it is, respectively, $20\,\sigma$ and $11\,\sigma$ larger than the mean observed values for metal-poor galaxies in the samples of \citet{lehmer2019} and \citet{douna2015}.

Using a slightly different approach, \citet{umeda2022} appeal to Markov Chain Monte Carlo techniques to select the combination of a blackbody and power-law ionizing spectra and gas parameters providing the best fits to the emission-line properties of three extremely metal-poor galaxies with strong \heiiopt emission (included in the sample of Section~\ref{sec:res_sample}): J1631+4426, J104457 and I\,Zw\,18\,NW. \citet{umeda2022} then show that, for each galaxy, the best-fitting combination of blackbody+power-law ionizing spectra can be approached by a combination of the accretion-disc model spectrum of \citet{gierlinski2009} and the spectrum of a \bpass single-star SSP with an age between about 2 and 8\,Myr (we note that the emission-line properties actually produced by these spectra are not compared to the observations in their paper). The \Ltotx/SFR ratios of these models are $5.0\times10^{40}$\,erg\,s$^{-1}/\Msun\,$yr$^{-1}$ (J1631+4426), $1.0\times10^{41}$\,erg\,s$^{-1}/\Msun\,$yr$^{-1}$ (J104457) and $3\times10^{41}$\,erg\,s$^{-1}/\Msun\,$yr$^{-1}$ (I\,Zw\,18\,NW), reaching again up to $11\,\sigma$ above the mean observed value in the \citet{douna2015} sample of metal-poor star-forming galaxies.\footnote{\label{foot:umedasfr}To compute \Ltotx/SFR, \citet{umeda2022} define an `effective' star formation rate based on the 1500\,\AA\ luminosity produced by the \bpass single-star SSP of their best-fitting model \citep[using the conversion in][]{kennicutt1998}.}


It is of interest to check how the emission-line and X-ray properties of the collection of models shown in Fig.~\ref{fig:contours_blz}, which include self-consistent scaling of accretion-disc with stellar emission, compare with those of J1631+4426, J104457 and I\,Zw\,18\,NW. These 800 models, built using the star formation histories of 100 \sphinx galaxies \citep{rosdahl2022}, are far from sampling the whole available space of stellar and gas parameters, as they have fixed $\nh=100\,\mathrm{cm}^{-3}$, $\logU=-2$ and $\xid=0.3$, only two \CO abundance ratios (\CO=0.17 and 0.44) and four metallicities ($Z=\zism=0.0005$, 0.001, 0.002 and 0.004), for ages between 2 and 10\,Myr. Therefore, the use of sophisticated fitting algorithms \cite[e.g., \beagle,][]{chevallard2016} is not appropriate in this case, and we simply explore the properties of the models that, from this limited library, are best able to approximate the observed emission-line properties of the three galaxies considered by \citet{umeda2022}. After some experimentation and to better account for the physical and chemical conditions in these galaxies, it appeared opportune to probe higher ionization parameters, carbon-to-oxygen and dust-to-gas ratios, so we extended the library to include a few more values: $\logU=-1.7$ and $-1.4$, \CO=0.09 and \xid = 0.1. 

\begin{figure}
 \centering
 \includegraphics[width=\columnwidth]{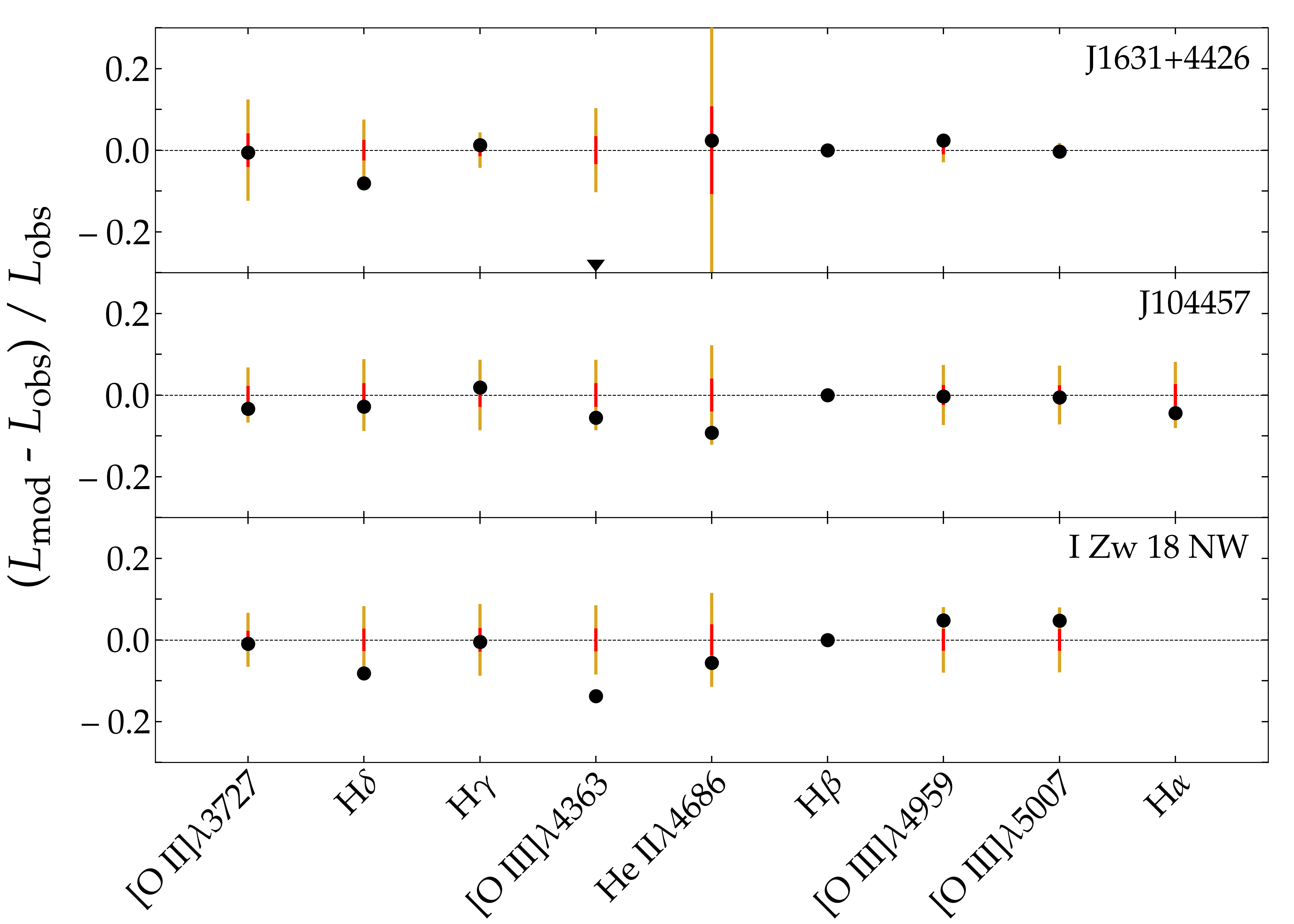}
 \caption{Relative offset between line luminosities of the best-fitting (i.e., minimum-$\chi^2$) \protect\GLS model among the limited library described in Section~\ref{sec:dsc}, and observed line luminosities in the three metal-poor galaxies J1631+4426, J104457 and I\,Zw\,18\,NW (from top to bottom) studied by \citet{umeda2022}. All luminosities are normalized to that of \hb. The 1-$\sigma$ (3-$\sigma$) observational errors are shown in red (gold). For J1631+4426, the \oiiioptc point falls outside the plot, at $(L_\mathrm{mod}-L_\mathrm{obs})/L_\mathrm{obs}\approx-0.50$, as indicated by the down-pointing black triangle.}
\label{fig:umeda_fit_lines}
\end{figure}

We present the results of this exercise in Fig.~\ref{fig:umeda_fit_lines}, which shows the relative offset between luminosity of the best-fitting (i.e., minimum-$\chi^2$) model and observed luminosity, for the lines in common between the two sets. Remarkably, among the restricted library of models at hand, those shown in Fig.~\ref{fig:umeda_fit_lines} appear to provide reasonable fits to all but one emission line. In particular, the models reproduce the strong observed \heiiopt luminosities of these galaxies, corresponding to \heiiopt/\hb ratios of 0.023 (J1631+4426), 0.018 (J104457) and 0.034 (I\,Zw\,18\,NW), i.e., totally out of reach of previous stellar population synthesis models (Fig.~\ref{fig:diags_oldmodels}b). 

The only line that the limited set of models considered here does not seem to naturally reproduce is \oiiioptc, whose observed luminosities in two of the three galaxies are higher than predicted by the best-fitting models, especially in the case of J1631+4426. The observed \oiiioptc/\oiiiopt ratios are 0.048 (J1631+4426), 0.032 (J104457) and 0.036 (I\,Zw\,18\,NW). Intriguingly, the value for J1631+4426 is the same as that found in the high-redshift galaxy S04590 studied by \citet{katz2023}, for which no simple explanation, other than an improbably high contribution by XRBs (see above) or a high cosmic-ray background (not applicable to nearby galaxies), could be found. Exploring other potential causes of strong \oiiioptc emission, such as exceptional heating perhaps related to temperature inhomogeneities \citep[e.g.,][]{cameron+katz2023}, is beyond the scope of the simple exercise considered here.

For reference, the \Lx/SFR ratios of the best-fitting models in Fig.~\ref{fig:umeda_fit_lines}, computed by estimating the SFR from the 1500-\AA\ luminosity by analogy with \citet[][see footnote~\ref{foot:umedasfr}]{umeda2022}, span the range 3.6--7.2$\times10^{40}$\,erg\,s$^{-1}/\Msun\,$yr$^{-1}$, compatible with the observations of \citet{lehmer2019} and \citet{douna2015}. The best-fitting metallicities and volume-averaged ionization parameters are $(0.0005, -2.0)$, $(0.001, -1.4)$ and $(0.0005, -1.7)$ for J1631+4426, J104457 and I\,Zw\,18\,NW, respectively, close to the values $(0.0004, -1.6)$, $(0.001, -2.0)$ and $(0.0005, -1.7)$ derived by \citet{umeda2022} using blackbody+power-law ionizing spectra. For reference, the best-fitting \CO and \xid values are $(0.09,0.1)$, $(0.44,0.3)$ and $(0.09,0.1)$, respectively.


We therefore conclude that the prediction from the \GLS model presented in this paper, that XRB accretion discs have little influence on the high-ionization line properties of metal-poor star-forming galaxies, appears robust when confronted with observations of both the emission-line and X-ray properties of such galaxies.

\section{Summary}
\label{sec:ccl}

We have explored the emission-line properties of interstellar gas photoionized by young binary-star populations computed with the new \GLS model built by combining the \sevn \citep{iorio2022} and \galaxev \citep{bruzual2003} population-synthesis codes. This model allows us to compute in a physically consistent way, for the first time, the emission from stars, accretion discs of X-ray binaries and supernova-driven radiative shocks. We compared the predictions of this model with observed UV and optical emission-line properties of a sample of about 120 metal-poor star-forming galaxies in a wide redshift range. We can summarize our main results as follows.

\begin{itemize}
\setlength{\itemindent}{.2in}

\item We confirm that, as well known from previous work \citep[e.g.,][]{eldridge2012, eldridge2017, goetberg2020}, the inclusion of binary-star processes (such as envelope stripping, quasi-homogeneous evolution, common-envelope ejection and star-star mergers) boosts and hardens the ionizing emission from young stellar populations. Such effect is achieved assuming even a relatively low fraction of binary stars, the rates of H-, \lhei- and \lheii-ionizing photons changing only little for stellar populations with $0.3\la\fbin\le1.0$ (Fig.~\ref{fig:comp_partbin}).

\item We find that the \heiiopt/\hb ratios and \hb equivalent widths of \GLS SSP models with metallicity $Z\sim0.001$ reproduce remarkably well the observations of galaxies in our sample at all ages between 3 and 10\,Myr, providing better agreement with the data than achievable with previous single-star and binary-star models (Fig.~\ref{fig:diags_oldmodels}b).

\item The inclusion of bursty star formation histories, such as those predicted by the \sphinx simulations of young galaxies in the epoch of reionization \citep{rosdahl2022}, allows broader sampling of the observational space than achievable with simpler models, and hence, should be encouraged when interpreting observations (Figs~\ref{fig:diags_sfh}--\ref{fig:contours_blz}).

\item We find that the most extreme \heii equivalent widths (and other high-ionization signatures) of galaxies in our sample can be reproduced by models with a top-heavy IMF with characteristic mass $\sim50\,\Msun$, which could represent a transient phase in the evolution of a galaxy (Fig.~\ref{fig:diags_wise}).

\item Reproducing the average luminosity function of X-ray binaries per unit star formation rate in the most metal-poor star-forming galaxies observed by \citet{lehmer2019} requires that the bulk of mass in such systems be accreted during episodes of X-ray outbursts, at a much higher rate than during the quiescent phase, the actual strength and duration of these outbursts being degenerate (Appendix~\ref{app:XRBs_calc}). This implies that XRBs contribute negligibly to the H- and \lheii-line emission from young star-forming galaxies, in agreement with expectations from several previous studies \citep[e.g.,][see Fig.~\ref{fig:lum_XRBs}]{jaskot2013, senchyna2017, senchyna2020, saxena2020}. We find that the claims in some recent studies that XRB accretion discs might contribute significantly to the emission from high-ionization UV and optical lines \citep{schaerer2019, umeda2022, katz2023} are based on models predicting improbably high ratios of X-ray luminosity to star formation rate (Section~\ref{sec:dsc}).

\item Finally, by combining the rates of energy injection into the ISM by stellar winds and (pair-instability, type-II and type-Ia) SN explosions predicted by the \GLS model with the radiative-shock models of \citet[][see also \citealt{dopita1996}]{alarie2019}, we find that shocks are unlikely to contribute significantly to the H- and \lheii-line emission from young star-forming galaxies (Fig.~\ref{fig:lum_shocks}).
    
\end{itemize}

The work presented in this paper provides only first examples of the properties of the new \GLS model of the emission from binary-star populations. The application of this model to interpret the spectral properties of young star-forming galaxies using Bayesian inference (e.g., with the \beagle tool) requires building a comprehensive library spanning full ranges of stellar and nebular parameters and the exploration of a comprehensive set of emission lines. This will be the subject of a forthcoming paper.


\section*{Acknowledgements}

We are grateful to M. Celeste Artale, Jeremy Blaizot, Andrea Lapi, Jean-Pierre Lasota, Jordan Mirocha, Christophe Morisset and Ad\`ele Plat for helpful discussions. 
AB acknowledges financial support from PRIN-MIUR 2017.
GB acknowledges financial support from the National Autonomous University of M\'exico (UNAM) through grants DGAPA/PAPIIT IG100319 and BG100622.
MM, GC, and GI acknowledge financial support from the European Research Council (ERC) through the ERC Consolidator Grant DEMOBLACK, under contract No.~770017. MM acknowledges financial support from the German Excellence Strategy via the Heidelberg Cluster of Excellence (EXC 2181 - 390900948) `STRUCTURES'. 
GC acknowledges support from the Agence Nationale de la Recherche grant POPSYCLE No.~ANR-19-CE31-0022.
YC acknowledges support from the National Natural Science Foundation of China (NSFC) No.~12003001. 
JC acknowledges financial support from the ERC through the ERC Advanced Grant `FirstGalaxies', under contract No. 789056.
MD acknowledges financial support from the Cariparo Foundation under grant 55440.


\section*{Data Availability}

The data underlying this article will be shared on reasonable request to the corresponding author. The code \sevn is publicly available via the following link: \url{https://gitlab.com/sevncodes/sevn.git}.


\bibliographystyle{mnras}
\bibliography{paper.bib} 


\appendix

\section{Handling of binary-evolution products in SEVN}
\label{app:SEVN}

The \sevn code evolves stars in a binary system by interpolating from a set of precomputed single-star tracks (in this work, we use the \parsec track library; see Section~\ref{sec:mod_stellpop}). The evolution of a star involves interpolation from four tracks bracketing its mass and metallicity (see \citealt{iorio2022} for details). However, binary processes, such as wind mass transfer, Roche-Lobe overflow and stellar mergers, can significantly alter the properties of a star. In such cases, \sevn searches for new tracks to interpolate from to better match the current stellar properties. This is achieved through different strategies tailored to specific situations, as follows.

\begin{itemize}
\item{Mass loss/accretion}: for main-sequence stars, \sevn checks whether the net cumulative mass variations due to binary processes exceed 1~per cent of the current stellar mass. In such cases, \sevn searches for new interpolating tracks providing a better match to the total mass of the star at the same percentage of stellar life.
The evolution of stars with decoupled He and CO cores is driven by core properties \citep{hurley2002}. For this reason, we do not allow stars outside the main sequence to change stellar tracks unless the core mass has changed (e.g., during stellar mergers) or the star has completely lost its envelope.
\item{Envelope stripping}: if a star completely loses its envelope, \sevn switches to pure-He stellar-evolution tables, selecting the pure-He interpolating tracks that best match the mass of the bare He core at the same percentage of stellar life. In this work, we use the \parsec pure-He stellar-evolution tables described in \cite{iorio2022}. When a pure-He star loses its He envelope, \sevn models the bare CO core as a remnant: if the core is massive enough to produce a black hole or neutron star, its properties remain constant until remnant formation, otherwise it immediately becomes a white dwarf.
\item{Stellar merger}: depending on the properties of the merging stars, \sevn employs different strategies to select the interpolating tracks for the merger product. If two main-sequence stars merge, \sevn changes stellar evolution tracks similarly to cases of mass loss or accretion. Instead, if at least one of the merging stars is  evolved, \sevn searches for new interpolating tracks by matching the core mass at the same percentage of stellar life. For mergers between a `standard' star and a stripped, pure-He star, \sevn searches for new interpolating  tracks in the H-rich tables matching the new core mass. We assume that mergers between stars and remnants (white dwarfs, neutron stars, black holes and bare-CO cores) completely destroy the star, leaving only the remnant, and no mass is accreted on to the remnant.
\end{itemize}

Additional details about the strategy for selecting interpolating tracks of binary-evolution products in \sevn can be found in \cite{iorio2022}.

\section{Emission from XRB accretion discs}
\label{app:XRBs_calc}

\subsection{Context}
 \label{sec:context_ap}

We identify X-ray binaries among a population of evolving binary pairs generated with the \sevn code (Section~\ref{sec:mod_stellpop}) by searching for black holes and neutron stars whose masses increase with time, i.e., with positive mass accretion rate across a dynamical time step, $\Mdotacc>0$. For spherical infall, the maximum luminosity that can theoretically be produced by an accretion disc in hydrostatic equilibrium is the Eddington luminosity corresponding to the balance between radiation pressure and gravity  \citep[][see also \citealt{hurley2002}]{cameron1967}. For accretion of fully-ionized H+He material, with hydrogen mass fraction X, on to a compact object of mass \Mc, this limit can be expressed as
\begin{equation}\label{eq:ledd_ap}
\Ledd \approx 1.26\times10^{38} \left(\frac{2}{1+\mathrm{X}}\right)\left(\frac{M_c}{M_{\odot}}\right) \ \mathrm{erg\ s}^{-1}\,.
\end{equation}
The Eddington luminosity is related to an Eddington accretion rate, \Mdotedd, through the formula \citep[e.g.,][]{shakura1973}
\begin{equation}
\Ledd =\eta\Mdotedd c^2\,,
\end{equation}
where $\eta$ is the radiative efficiency (i.e. the fraction of binding energy radiated away by the accretion flow) and $c$ the speed of light.

Observationally, accretion on to compact objects is known to occasionally produce luminosities in excess of the theoretical Eddington limit, perhaps as a result of radiation-driven inhomogeneities \citep[e.g.,][]{begelman2002}. Examples include tidal disruption events \citep[e.g.,][]{rees1988} and ultra-luminous X-ray sources \citep[e.g.,][]{bachetti2014, pinto2016, rodriguezcastillo2020}, while some evidence suggests that most low-mass XRBs may have undergone a phase of super-Eddington accretion \citep{kalogera1998}. For super-Eddington accretion, corresponding to $\Mdotacc>\Mdotedd$, the total accretion luminosity, noted \Lacc, can exceed the Eddington luminosity by a logarithmic factor $\sim\ln(\Mdotacc/\Mdotedd)$ \citep[see][]{shakura1973, begelman2006}. We adopt here the convenient approximation \citep[][see also \citealt{lapi2014}]{watarai2006}
\begin{equation}\label{eq:lacc_ap}
\Lacc \approx  2 \ln \left(1 + \frac{\Mdotacc}{2 \Mdotedd}\right)\Ledd\,,
\end{equation} 
which, for sub-Eddington accretion ($\Mdotacc\la\Mdotedd$), reduces to
\begin{equation}
\Lacc \approx\eta\Mdotacc c^2\,.
\end{equation}
For super-Eddington accretion, the weaker-than-linear increase of \Lacc with \Mdotacc in equation~\eqref{eq:lacc_ap} \citep[illustrated by fig.~7 of][]{watarai2006} reflects how the accreting gas becomes optically too thick to radiate away all the dissipated energy, causing some radiation to be trapped and advected inward. For simplicity, we fix here the radiative efficiency to the standard value $\eta\approx0.1$ corresponding to the fraction of energy liberated by a gas particle falling from far away on to the innermost stable circular orbit of a stationary (non-spinning) black hole \citep[e.g.,][]{shakura1973}.

In \sevn, the maximum rate at which a compact object can accrete mass is $\dot{M}_\mathrm{acc}^\mathrm{max}=\fedd\Mdotedd$, where \fedd is an adjustable factor \citep[][]{spera2019, iorio2022}.\footnote{The quantity \fedd is noted $\eta_\mathrm{Edd}$ in \citet{iorio2022}.} This factor can be set to a value greater than unity to account for super-Eddington accretion. In practice, we compute the mass accretion rate as $\dot{M}_{\mathrm{c}}/(1-\eta)$, where $\dot{M}_{\mathrm{c}}$ is the mass-growth rate of the compact object recorded in \sevn. In the setup of the code used here, the mass lost by the donor that is not accreted by the compact object (if $\Mdotacc>\dot{M}_\mathrm{acc}^\mathrm{max}$) is assumed to be lost from the system \citep[other options are described in][]{iorio2022}.
 
 \subsection{Spectral modelling}
 \label{sec:specmod_ap}
 
Observations of X-ray binaries show that these stars commonly undergo X-ray outbursts, during which their luminosity can exceed that in the quiescent state by up to several orders of magnitude \citep[e.g.,][]{tanaka1996, chen1997}. While this transience phenomenon is often discussed in the context of low-mass XRBs, where the donor star is typically less massive than a few \Msun \citep[e.g.,][]{yan2015}, it is also observed in high-mass XRBs \citep[e.g.,][]{martin2014, vandeneijnden2022}. Even high-mass XRBs classified as `persistent' (i.e., non-transient sources emitting X-rays continuously over long periods) exhibit variability accompanied by (sometimes giant) X-ray flares \citep[e.g.,][]{hertz1992, pottschmidt2003, fuerst2010}. Such X-ray outbursts appear to be correlated with variability at all wavelengths from ultraviolet to infrared \citep[e.g.,][]{hynes2003, sonbas2019, lopeznavas2020, yang2022}.

The physical origin of X-ray variability is unclear. For several decades, the so-called disc-instability model has been put forward, according to which, as material falls on to the disc, a thermal-viscous instability develops, leading to recurrent burst-accretion events \citep[e.g.,][see the reviews by \citealt{lasota2001, hameury2020b}]{meyer1981, faulkner1983, vanparadijs1996, hameury2020a}. While the realistic incorporation of thermodynamics and the ability to reproduce fairly well observations of X-ray outbursts have made this model popular, it does suffer from several weaknesses and shortcomings, notably with regard to angular-momentum transport and the inclusion of several poorly-constrained parameters (see \citealt{hameury2020b} for details). Hence, any attempt to implement the disc-instability model in \sevn would be highly uncertain. 

In this context, we prefer to adopt here a more empirical approach to account for variability when modelling the spectral properties of XRB accretion discs. We assume that any accretion event spreads over two phases: a quiescent phase during which accretion proceeds at the rate provided by \sevn, $\dot{M}_{\mathrm{c}}/(1-\eta)$, and an outburst phase during which it is boosted by a factor \fa. We assume that a fraction $\ea$ of the mass is accreted during the outburst phase. For the accreted mass $\delta\Mc$ to be conserved over the time interval $\delta \tprime$, we thus write
\begin{equation}
\Mdotacc(\tprime) =
\left \{
\begin{array}{ll}
\frac{1}{1-\eta}\frac{\delta\Mc}{\delta \tprime} & \mbox{during} \quad  (1-\ea)\delta\tprime\,,\\ \\
\frac{\fa}{1-\eta}\frac{\delta\Mc}{\delta \tprime} & \mbox{during} \quad \frac{\ea}{\fa}\delta \tprime\,,
\end {array}
\right.
\label{eq:twophase_ap}
\end{equation}
where $\delta\Mc$ is the mass growth of the compact object in a time interval $\delta \tprime$ of the evolution of an SSP (adopting the same notation as in equation~\ref{eq:flux_gal} of Section~\ref{sec:models}). For reference, $\delta\Mc$ represents at most a few percent of \Mc. 

We describe the spectral distribution of the power \Lacc(\tprime) produced by the hot accretion disc at time \tprime (equation~\ref{eq:lacc_ap}) by means of a multicolour-disc model \citep{mitsuda1984}, i.e., a modified blackbody spectrum reflecting the gas-temperature distribution in the disc. To this end, we use the \ares\footnote{\url{https://ares.readthedocs.io/en/latest/}} code of \citet{mirocha2014} and compute a comprehensive library of multicolour disc models for compact objects accreting at the Eddington limit, parametrized in terms of the accreting-object mass, \Mares, and the maximum radius of the accretion disc, \rmax. Following \citet{senchyna2020}, we fix \rmax to $10^4$ gravitational radii (we find that decreasing or increasing \rmax by an order of magnitude would, respectively, lower by about 20~per cent or increase negligibly the X-ray luminosities and rates of \lheii-ionizing photons predicted by our models). By design, each spectrum is normalized to the Eddington luminosity given by equation~\eqref{eq:ledd_ap} for $\Mc=\Mares$, for a pure-H gas ($\mathrm{X}=1$).

XRB spectra also often present a high-energy tail component, whose strength can vary over time, and which is thought to arise from Compton upward scattering of soft disc photons by energetic coronal electrons \citep[e.g.,][]{remillard2006, steiner2009}. Based on the analysis of 20 XRBs with multi-epoch X-ray luminosities in the range $\sim0.5$--$50\times10^{39}\,\mathrm{erg\,s}^{-1}$, \citet{sutton2013} conclude that the spectra of sources with sub-Eddington accretion rates may be well represented by multi-colour discs, while those of sources with super-Eddington accretion rates tend to exhibit `hard-luminous' and `soft-luminous' components (in roughly equal proportions among the sample), presumably depending on inclination. For each \Mares, we therefore compute both a spectrum including Comptonization of disc photons \citep[implemented in \ares using the \simpl model by][]{steiner2009} and a spectrum not including it. If accretion is sub-Eddington, i.e. for $\Lacc(\tprime)\leq\Ledd$, we adopt the pure multi-colour-disc spectrum. Otherwise, we randomly draw between the pure multi-colour-disc and the Comptonized spectra.\footnote{Since the \ares spectral library pertains to objects accreting at the Eddington rate, in practice, for a given \Lacc, we adopt the spectrum corresponding to the \ares model for a compact object accreting at the Eddington luminosity $\Ledd=\Lacc$ given by equation~\eqref{eq:ledd_ap} for a pure-H gas, i.e., $\Mares= \Lacc/(1.26\times10^{38}\mathrm{erg\ s}^{-1})\,\Msun$.} We note that the results presented in this paper about the X-ray luminosities and \lheii-ionizing photon rates of star-forming galaxies do not depend sensitively on this refinement of spectral modelling at the highest energies.

\begin{figure}
 \includegraphics[width=\columnwidth]{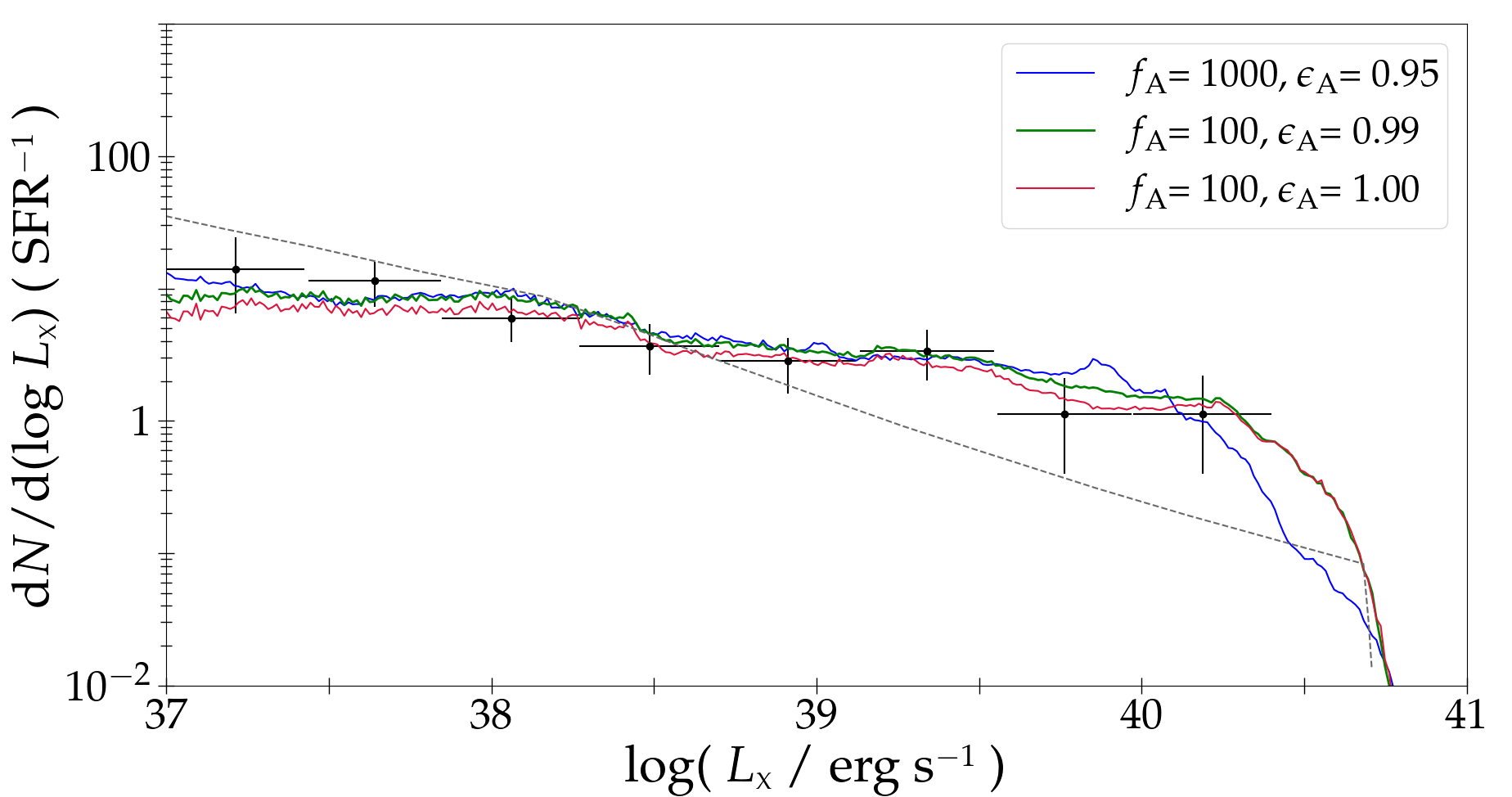}
 \caption{X-ray luminosity functions of XRBs for \protect\GLS populations with binary fraction $\fbin=0.7$, assuming 100\,Myr of constant star formation at the rate $\psi=1\Msun\,\mathrm{yr}^{-1}$. Accretion is taken to proceed according to the two-phase model of equation~\eqref{eq:twophase_ap}, for different choices of the fraction $\ea$ of mass accreted while the rate is boosted by a factor \fa, as indicated. All models have the metallicity $Z=0.008$, corresponding to the metallicity of the five most metal-poor star-forming galaxies observed with {\it Chandra} for which \citet{lehmer2019} derived the average, completeness-corrected luminosity function shown by the black data points with error bars. The black dotted line shows an empirical model proposed by these authors (see text for details).}
\label{fig:LF_models}
\end{figure}

The above approach allows us to compute the spectrum of every hot accretion disc of XRB in a \protect\GLS binary-star SSP of age \tprime, from which we extract the disc luminosity in the 0.5--8\,keV energy band, noted \Lx. For reference, the quantity \Lx amounts to typically between 30 and 90 per cent of \Lacc. Fig.~\ref{fig:LF_models} shows X-ray luminosity functions (XLFs) of XRBs obtained in this way for populations with binary-star fraction $\fbin=0.7$ \citep[typical of massive-star populations; e.g.,][]{sana2012}, assuming 100\,Myr of constant star formation, for different choices of the parameters \fa and \ea in equation~\eqref{eq:twophase_ap}. All models have the metallicity $Z=0.008$, corresponding to the metallicity of the five most metal-poor star-forming galaxies observed with {\it Chandra} for which \citet{lehmer2019} derived the average, completeness-corrected XLF of XRBs shown by the black data points with error bars (normalized per unit star formation rate in the same way as the models). The black dotted line shows an empirical model, incorporating also potential contamination by background point sources from the cosmic X-ray background, whose contribution to the XLF is however estimated to be minor, decreasing from about 20 per cent at $\Lx=10^{37}\,\mathrm{erg\,s}^{-1}$ to about 3 per cent at $\Lx=10^{40}\,\mathrm{erg\,s}^{-1}$ \citep[for details, see][]{lehmer2019}. We therefore ignore this component in our model. Furthermore, we adopt here $\fedd=100$ to set the maximum allowed accretion rate in \sevn (Section~\ref{sec:context_ap}), as significantly lower values do not produce enough high-luminosity sources, while adopting higher values negligibly affects the predicted XLFs, indicating that only a few systems can potentially have $\Mdotacc>100\Mdotedd$. 

We find that the predicted XLF depends in a somewhat degenerate way on the parameters \fa and \ea of the two-phase model, the main requirement for reproducing the observations being that the bulk of mass be accreted during outbursts. This is illustrated by the three models shown in Fig.~\ref{fig:LF_models}, accreting, during transient outbursts,  fractions $\ea=1.00$, 0.99 and 0.95 of the mass at $\fa=100$, 100 and 1000 times the quiescent rate, respectively. In the first model, the entire mass is accreted during outbursts, while in the second model, outburst- and quiescent-accretion phases have similar durations, and in the third model, outburst accretion lasts less than 1/50$^\mathrm{th}$ of quiescent accretion. All three models provide roughly similar agreement with the observed XLF of XRBs compiled by \citet{lehmer2019} in Fig.~\ref{fig:LF_models}. The maximum effective accretion rate for these models is $\fa\fedd=10^4$--$10^5$ times the Eddington rate \Mdotedd, corresponding to a maximum accretion luminosity $\Lacc\approx17$--$22\Ledd$ (equation~\ref{eq:lacc_ap}), compatible with the 10--$100\Ledd$ obtainable from radiation-driven inhomogeneities in accretion discs \citep[e.g.,][]{begelman2002}. We note that about a third of XRBs exhibit super-Eddington accretion rates during the phase of standard accretion, compared to three quarters during the phase of enhanced accretion (for $\fa=100$).

The model with a single phase of enhanced accretion is the simplest conceptually. The one with $\ea=0.99$ and $\fa=100$ provides formally the best fit (mininum $\chi^2$) to the data, while the one with quisecent accretion $\sim50$ times longer than enhanced accretion is the closest conceptually to that proposed by, e.g., \citet{li2012} for the growth of supermassive black holes in quasars. Since all three models reproduce the observed XLF of XRBs measured by \citet{lehmer2019} in nearby metal-poor galaxies, they predict similar contributions of XRB accretion discs to the ionizing-radiation budget of a star-forming galaxy modelled with the \protect\GLS code. Here, we adopt the model with $\ea=0.99$ and $\fa=100$, which provides the best fit to the observations in Fig.~\ref{fig:LF_models}. Any of the other two models would provide similar results on the emission from XRB accretion discs presented in this paper.

\begin{figure}
 \includegraphics[width=\columnwidth]{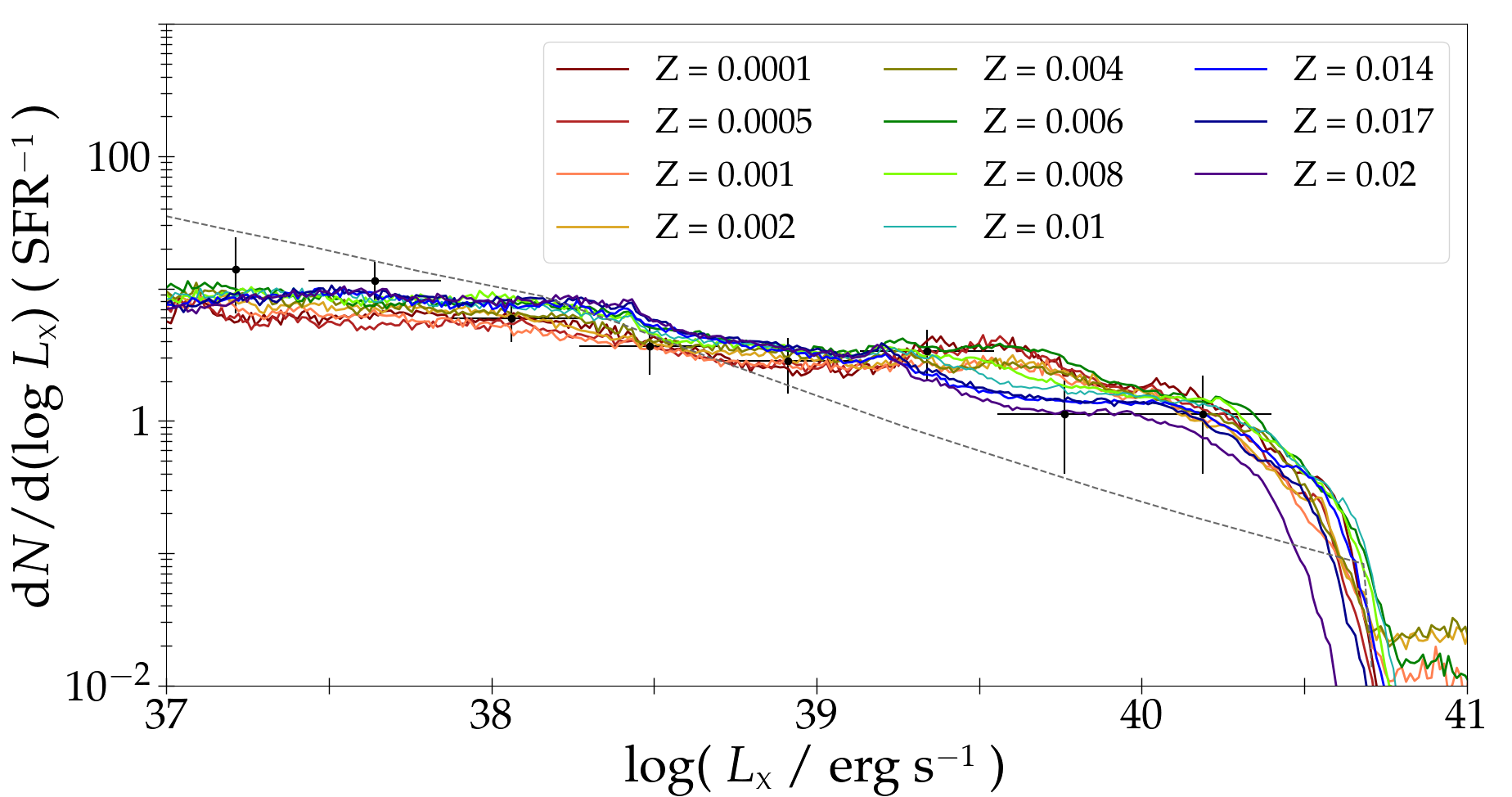}
 \caption{Same as Fig.~\ref{fig:LF_models}, for \protect\GLS binary-star populations of different metallicities in the range $0.0001\leq Z\leq 0.02$, as indicated. In all models, accretion is taken to proceed according to the two-phase model described by equation~\eqref{eq:twophase_ap}, with $\fa=100$ and $\ea=0.99$.}
\label{fig:LF_met}
\end{figure}

In Fig.~ \ref{fig:LF_met}, we show the XLFs obtained at fixed $\fa=100$ and $\ea=0.99$, for \protect\GLS binary-star populations of different metallicities in the range $0.0001\leq Z\leq 0.02$. Remarkably, all XLFs remain compatible with the observational determination of \citet{lehmer2019}. This is because the XLF is largely dominated by systems composed of low-mass compact objects with relatively low-mass companions: nearly 80~per cent of overall accretion occurs in systems composed of a compact object lighter than 20\,\Msun \citep[with progenitor main-sequence mass typically smaller than 30\,\Msun; e.g.,][]{spera2015} and a companion lighter than 25\,\Msun. Since the evolution of stars lighter than 30\,\Msun is not significantly affected by line-driven stellar winds, we do not expect a strong dependence of XRB properties on metallicity. The main metallicity-dependent effect comes from the reduced population of massive compact objects at high metallicity \citep[e.g.,][]{dray2006, mapelli2013}, apparent in the slight decline in the XLF at the highest X-ray luminosities in Fig.~ \ref{fig:LF_met}, while agreement with the observed luminosity function is maintained. 

\begin{figure}
  \includegraphics[width=\columnwidth]{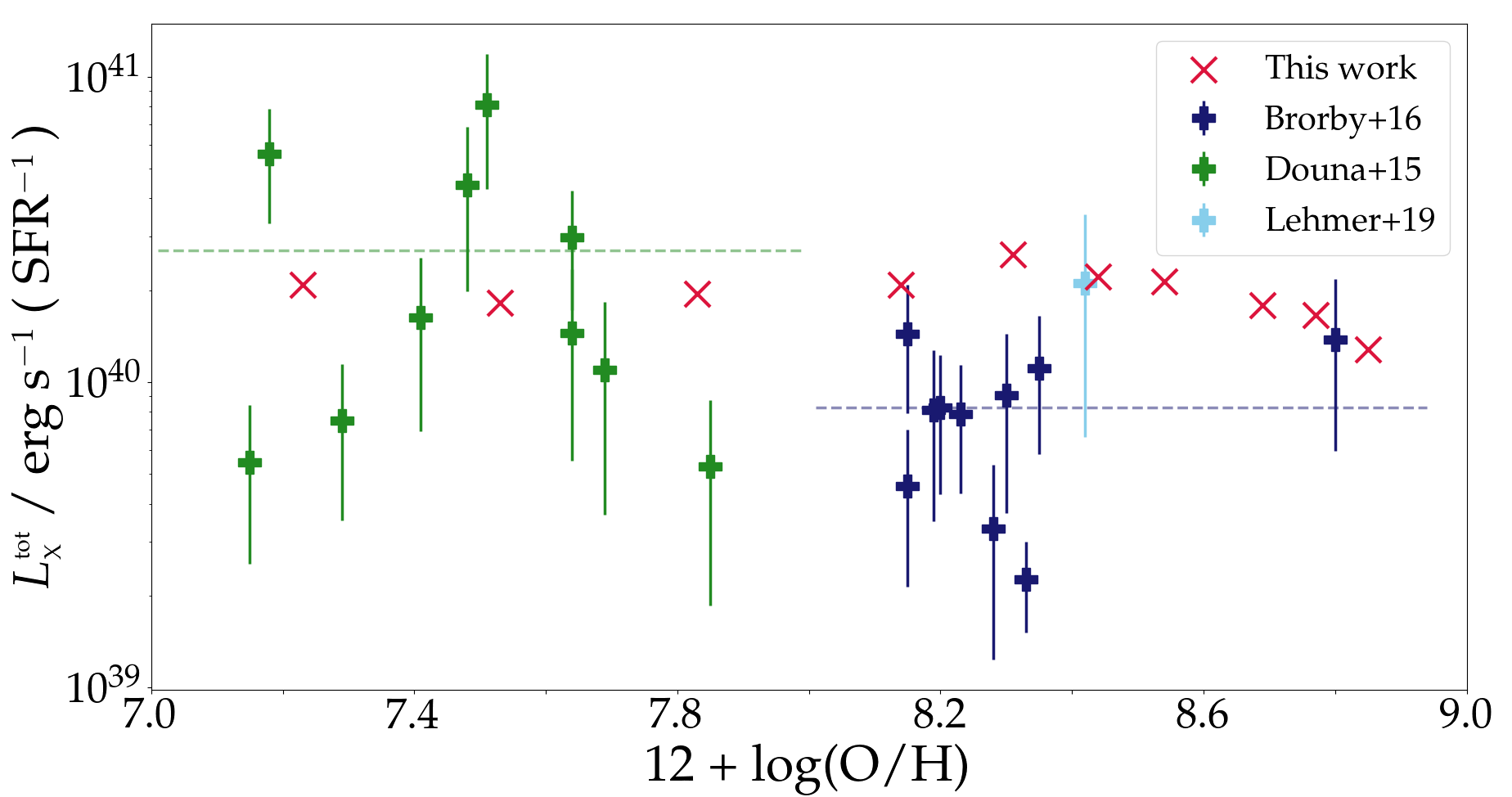}
  \caption{Total X-ray luminosity \Ltotx (per unit star formation rate) obtained by integrating the luminosity functions of Fig.~\ref{fig:LF_met}, for metallicities in the range $0.0005\leq Z\leq 0.02$ (red crosses). The data are from {\it Chandra} observations of a sample of 10 nearby star-forming galaxies with $\logoh < 8.0$ compiled by \citet[][green points ; see also \citealt{brorby2014}]{douna2015} and a sample of nearby analogues of distant Lyman-break galaxies compiled by \citet[][dark blue points]{brorby2016}. The light-blue point (and error bars) at $\logoh\approx8.4$ shows the mean \Ltotx/SFR (and associated standard deviation) of the five metal-poor star-forming galaxies used by \citet{lehmer2019} to derive the average XLF of Fig.~\ref{fig:LF_models}. Dotted horizontal green and blue lines represent the mean \Ltotx/SFR of galaxies in the \citet{douna2015} and \citet{brorby2016} samples, respectively. \citet{brorby2016} estimate the typical uncertainty on \logoh at 0.14.}
  \label{fig:brorby_met}
\end{figure}

In Fig.~\ref{fig:brorby_met}, we compare the total X-ray luminosities per unit star formation rate, \Ltotx/SFR, obtained by integrating the XLFs of the binary-star models with different metallicities (converted to gas-phase O abundances using table~2 of \citealt{gutkin2016} for $\xid=0.3$) of Fig.~\ref{fig:LF_met} with those of the sample of 10 nearby star-forming galaxies with $\logoh < 8.0$ compiled by \citet[][green points ; see also \citealt{brorby2014}]{douna2015} and the sample of nearby analogues of distant Lyman-break galaxies compiled by \citet{brorby2016}. Also shown is the mean \Ltotx/SFR (and associated standard deviation) of the five metal-poor star-forming galaxies used by \citet{lehmer2019} to derive the average XLF in Fig.~\ref{fig:LF_models}.\footnote{For each galaxy, \Ltotx/SFR was obtained by integrating the observed individual luminosity function in fig.~3 of \citet{lehmer2019} and then dividing by the star formation rate in their table~1.} As expected from Fig.~\ref{fig:LF_models}, the model with $Z=0.008$ (corresponding to $\logoh\approx8.4$) lies near this mean value. 
More generally, Fig.~\ref{fig:brorby_met} shows that the total X-ray luminosity per unit star formation rate predicted by our models is well within the ranges found by \citet[][green points]{douna2015} and \citet[][light blue points]{lehmer2019} and near the upper end of that reported by \citet[][dark blue points]{brorby2016}. 

We consider the reasonable agreement between \protect\GLS models of binary-star populations and X-ray observations of nearby, metal-poor star-forming galaxies and analogues of distant Lyman-break galaxies in Figs~\ref{fig:LF_met}--\ref{fig:brorby_met} as support for our simple approach to model the emission from accretion discs of X-ray binaries consistently with the spectral modelling of stars.

\section{Line emission from shocks driven by stellar winds and supernova explosions in star-forming galaxies}
\label{app:rad_shocks}

The energy deposited into the ISM by stellar winds and supernova explosions \citep[e.g.,][]{leitherer1992,leitherer1995} can drive fast shocks contributing to line emission in star-forming galaxies \citep[e.g.,][]{shull1979, izotov2012}. We express the rate of total energy deposition into the ISM at SSP age \tprime (equation~\ref{eq:flux_gal}) as
\begin{equation}
\label{eq:edot}
\dot{E}_\mathrm{tot}(\tprime)=\effw \,\dot{E}_\mathrm{W}(\tprime)+\effsn\,\rsn(\tprime) \,{}E_\mathrm{SN}\,,
\end{equation}
where $\dot{E}_\mathrm{W}$ is the rate of energy input by stellar winds, $E_\mathrm{SN}\approx10^{51}\,\mathrm{erg}$ is the typical energy released by a SN explosion \citep[e.g.,][]{arnett1996},\footnote{\label{foot:pisn}This is generally the case for type-II, type-Ia and low-mass pair-instability SNe, but the most massive PISNe can reach $\sim100$ times this value \citep[see][]{heger2002}. This has little effect on our conclusions (see Section~\ref{sec:add_shocks}).} $\rsn(\tprime)$ is the rate of SN explosions at time \tprime, and \effw and \effsn are thermalization efficiency factors. This energy injection will generate shocks, whose radiative properties we wish to estimate. 

Emission-line predictions for fast radiative shocks have been computed using the {\footnotesize MAPPINGS} photoionization code \citep[e.g.,][]{dopita1996,allen2008,alarie2019}. These usually pertain to high-velocity, steady-flow, fully radiative, plane-parallel magneto-hydrodynamic shocks and are traditionally expressed in flux per unit area (erg\,cm$^{-2}$\,s$^{-1}$). We need to scale these predictions to the total energy deposition rate from equation~\eqref{eq:edot} to obtain estimates of emission-line luminosities (in erg\,s$^{-1}$). This can be achieved by considering the total radiative flux of a shock. Assuming that the entire flux of mechanical energy through the shock is radiated, \citet{dopita1996} express this quantity as (see their equation~3.3)
\begin{equation}
\label{eq:ftot_shock}
F_\mathrm{tot}^\mathrm{shock}=2.28\times10^{-3}\left(\frac{\Vs}{100\,\mathrm{km\,s}^{-1}}\right)^{3.0}\left(\frac{\nh}{\mathrm{cm}^{-3}}\right)\,\mathrm{erg\,cm}^{-2}\,\mathrm{s}^{-1}\,,
\end{equation}
where $\Vs$ denotes the shock velocity and \nh the hydrogen density. The fraction of this flux radiated in the \hb recombination line in the plasma column behind the shock can be written \citep[equation~3.4 of][]{dopita1996}
\begin{equation}
\label{eq:fhb_shock}
F_{\mathrm{H}\beta}^\mathrm{shock}=7.44\times10^{-6}\left(\frac{\Vs}{100\,\mathrm{km\,s}^{-1}}\right)^{2.41}\left(\frac{\nh}{\mathrm{cm}^{-3}}\right)\,\mathrm{erg\,cm}^{-2}\,\mathrm{s}^{-1}\,.
\end{equation}
The photoionized region ahead of the shock also contributes to the total \hb emission. The corresponding flux can be expressed as \citep[equation~4.4 of][]{dopita1996}
\begin{equation}
\label{eq:fhb_prec}
F_{\mathrm{H}\beta}^\mathrm{prec}=9.85\times10^{-6}\left(\frac{\Vs}{100\,\mathrm{km\,s}^{-1}}\right)^{2.28}\left(\frac{\nh}{\mathrm{cm}^{-3}}\right)\,\mathrm{erg\,cm}^{-2}\,\mathrm{s}^{-1}\,.
\end{equation}
This is an upper limit to the contribution by the shock precursor, as the ISM can run out before the end of the H-Str\"omgren column ahead of the shock (especially for the fastest shocks).

Equations~\eqref{eq:ftot_shock}--\eqref{eq:fhb_prec} allow one to express the ratio of total (shock+precursor) \hb flux to  total radiative flux generated by the shock in terms of the quantities \Vs and \nh. This ratio can also be regarded as the ratio of total \hb luminosity to total luminosity emitted by the shock, i.e., we write
\begin{equation}
\label{eq:fluxlum_shocks}
\frac{F_{\mathrm{H}\beta}^\mathrm{shock}+F_{\mathrm{H}\beta}^\mathrm{prec}}{F_\mathrm{tot}^\mathrm{shock}}=
\frac{L_{\mathrm{H}\beta}^\mathrm{shock}+L_{\mathrm{H}\beta}^\mathrm{prec}}{L_\mathrm{tot}^\mathrm{shock}}\,.
\end{equation}
Since we have assumed that the entire flux of mechanical energy through the shock is radiated, we can replace $L_\mathrm{tot}^\mathrm{shock}$ by the power injected into the ISM (equation~\ref{eq:edot}) and derive from this the total \hb luminosity of the shock. 

Unfortunately, analogues to equations~\eqref{eq:ftot_shock}--\eqref{eq:fhb_prec} (derived by \citealt{dopita1996} using {\footnotesize MAPPINGS\,II}) are not available for the more recent calculations of \citet[][based on \mappings]{alarie2019}. Instead, \citet{alarie2019} provide tables of $F_{\mathrm{H}\beta}^\mathrm{shock}$ and $F_{\mathrm{H}\beta}^\mathrm{prec}$ versus shock velocity $\Vs$ and density $\nh$, while equation~\eqref{eq:ftot_shock}, which states that $F_\mathrm{tot}^\mathrm{shock}$ is equal to the flux of mechanical energy passing through the shock, can be assumed not to depend on the photoionization code. We therefore combine the results of \citet{alarie2019} with the expression of $F_\mathrm{tot}$ as a function of  $\Vs$ and $\nh$ from equation~\eqref{eq:ftot_shock} to compute the ratio on the left-hand side of equation~\eqref{eq:fluxlum_shocks}. Then, we consider the luminosity $L_{\mathrm{H}\beta}^\mathrm{shock}+L_{\mathrm{H}\beta}^\mathrm{prec}$ separately for radiative shocks powered by stellar winds and SN explosions, as the ejection velocities associated with SN explosions \citep[$\ga10,000\,$km\,s$^{-1}$; e.g.,][]{draine1993} are higher than those associated with stellar winds \citep[$\la3000\,$km\,s$^{-1}$; e.g.,][]{garcia2014, graefener2017}. The implied radiative-shock velocities are therefore likely to differ on average.

We compute the total \hb emission from radiative shocks driven by stellar winds as (dropping the dependence on \tprime for simplicity)
\begin{equation}
\label{eq:lhb_winds}
\begin{aligned}
L_{\mathrm{H}\beta}^\mathrm{shock}+L_{\mathrm{H}\beta}^\mathrm{prec}=&\,\,4.39\times10^{2}\,\effw \,\dot{E}_\mathrm{W}\,
\left(\frac{F_{\mathrm{H}\beta}^\mathrm{shock}+F_{\mathrm{H}\beta}^\mathrm{prec}}{\mathrm{erg\,cm}^{-2}\,\mathrm{s}^{-1}}\right)\,
\\&\times\left(\frac{\Vs}{100\,\mathrm{km\,s}^{-1}}\right)^{-3.0}\left(\frac{\nh}{\mathrm{cm}^{-3}}\right)^{-1}\,\mathrm{erg\,s}^{-1}\,,
\end{aligned}
\end{equation}
where the total flux $F_{\mathrm{H}\beta}^\mathrm{shock}+F_{\mathrm{H}\beta}^\mathrm{prec}$ for shocks with velocity \Vs propagating in a medium with density \nh is taken from \citet{alarie2019}. We obtain the rate $\dot{E}_\mathrm{W}$ of energy input from winds (equation~\ref{eq:edot}) by summing the individual contributions $\frac{1}{2}\dot{m}v_\infty^2$ of all stars, where $\dot{m}$ is the stellar mass-loss rate and $v_\infty$ the terminal wind velocity.\footnote{For simplicity, we compute this estimate for a population of single stars and ignore the influence of binary-star processes on energy injection by stellar winds into the ISM.}  For OB stars, we adopt the widely used relation $v_\infty = 2.65\,v_\mathrm{esc}$, where $v_\mathrm{esc}$ is the stellar escape velocity \citep[e.g.,][]{garcia2014}. For WR stars,  we take $v_\infty$ to be the minimum of $1.30\,v_\mathrm{esc}$ and 1800\,km\,s$^{-1}$ \citep[e.g.,][]{graefener2017}. For simplicity, we assume here that all of the kinetic energy injected by winds into the ISM is thermalized, i.e., we take $\effw=1.0$ \citep[e.g.,][]{rosen2022}.

Similarly, we express the total \hb emission from radiative shocks driven by SN explosions as 
\begin{equation}
\label{eq:lhb_sne}
\begin{aligned}
L_{\mathrm{H}\beta}^\mathrm{shock}+L_{\mathrm{H}\beta}^\mathrm{prec}=&\,\,1.39\times10^{46}\,\effsn\left(\frac{\rsn}{\mathrm{yr}^{-1}}\right)\,
\left(\frac{F_{\mathrm{H}\beta}^\mathrm{shock}+F_{\mathrm{H}\beta}^\mathrm{prec}}{\mathrm{erg\,cm}^{-2}\,\mathrm{s}^{-1}}\right)\,
\\&\times\left(\frac{\Vs}{100\,\mathrm{km\,s}^{-1}}\right)^{-3.0}\left(\frac{\nh}{\mathrm{cm}^{-3}}\right)^{-1}\,\mathrm{erg\,s}^{-1}\,.
\end{aligned}
\end{equation}
Constraints from the prototypical starburst galaxy M82 suggest $0.3\lesssim\effsn\lesssim1.0$ \citep{strickland2009,heckman2017}. Here, we adopt for simplicity $\effsn=1.0$, which should provide an upper limit on the shock contribution to line emission. 

Once the \hb luminosity is known, whether for wind- or SN-driven shocks, the luminosity of any other line X can be computed as 
\begin{equation}
\label{eq:lxalarie}
L_\mathrm{X}^\mathrm{shock}+L_\mathrm{X}^\mathrm{prec}=
L_{\mathrm{H}\beta}^\mathrm{shock}\,
\frac{F_\mathrm{X}^\mathrm{shock}}
{F_{\mathrm{H}\beta}^\mathrm{shock}}\,+\,
L_{\mathrm{H}\beta}^\mathrm{prec}\,
\frac{F_\mathrm{X}^\mathrm{prec}}
{F_{\mathrm{H}\beta}^\mathrm{prec}}\,,
\end{equation}
where the ratios $F_\mathrm{X}^\mathrm{shock}/F_{\mathrm{H}\beta}^\mathrm{shock}$ and $F_\mathrm{X}^\mathrm{prec}/F_{\mathrm{H}\beta}^\mathrm{prec}$ are available from the tables of \citet{alarie2019}.


\bsp	
\label{lastpage}
\end{document}